\newif\ifsingle
\singlefalse

\newif\ifproofs
\ifsingle
\documentclass[11pt,draftclsnofoot, onecolumn]{IEEEtran}		
\else		
\documentclass[10pt,conference]{IEEEtran}
\fi
\usepackage{multirow} 
\usepackage{soul}
\usepackage{times}
\usepackage{amsmath,dsfont}
\usepackage{amssymb,amsthm}
\usepackage{epsfig,verbatim}
\usepackage{subfigure}
\usepackage{setspace}
\usepackage{color}
\usepackage{cite}
\usepackage{epstopdf}
\usepackage{graphics}
\usepackage{accents}
\usepackage{acronym}
\usepackage{booktabs}
\usepackage{mathtools}
\usepackage{algorithm}
\usepackage{algorithmic}
\usepackage{enumitem}
\usepackage{tcolorbox}
\usepackage{makecell}

\usepackage{multirow} %multirow for format of table

\newcommand{\myVec}[1]{{\boldsymbol{#1}}}
\newcommand{\myMat}[1]{{\boldsymbol{#1}}}
\newcommand{\mySet}[1]{\mathcal{#1}}
\newcommand{\Comm}{^{\left(\rm c\right)}}
\newcommand{\Radar}{^{\left(\rm r\right)}}

\newcommand{\NoCite}{\textcolor{red}{[CITE]}}
\newcommand{\NoRef}{\textcolor{red}{[REF]}}

\newcommand{\Revise}[1]{\textcolor{black}{#1}}
\newcommand{\ReviseReview}[1]{\textcolor{black}{#1}}

\newtheorem{theorem}{Theorem}
\newtheorem{corollary}{Corollary}
\newtheorem{proposition}{Proposition}
\newtheorem{lemma}{Lemma}

\definecolor{NewColor}{rgb}{0,0,0} %{0.2,0,0.5}

\ifsingle

\else

\fi % ------------------------------------------

\acrodef{adc}[ADC]{analog-to-digital convertor}
\acrodef{dac}[DAC]{digital-to-analog convertor}
\acrodef{cs}[CS]{compressed sensing}
\acrodef{dtft}[DTFT]{discrete-time Fourier transform}
\acrodef{bpsk}[BPSK]{binary phase shift keying}
\acrodef{ber}[BER]{bit error rate}
\acrodef{ofdm}[OFDM]{orthogonal frequency division multiplexing}
\acrodef{csi}[CSI]{channel state information}
\acrodef{map}[MAP]{maximum a-posteriori probability}
\acrodef{snr}[SNR]{signal-to-noise ratio}
\acrodef{bs}[BS]{base station} 
\acrodef{mimo}[MIMO]{multiple-input multiple-output}
\acrodef{mse}[MSE]{mean-squared error}
\acrodef{pdf}[PDF]{probability density function}
\acrodef{rv}[RV]{random variable}
\acrodef{lti}[LTI]{linear time-invariant}
\acrodef{wss}[WSS]{wide-sense stationary}
\acrodef{psd}[PSD]{power spectral density}
\acrodef{ser}[SER]{symbol error rate} 
\acrodef{isi}[ISI]{intersymbol interference} 
\acrodef{lstm}[LSTM]{long short-term memory} 
\acrodef{em}[EM]{expectation minimization} 
\acrodef{tdd}[TDD]{time division duplexing} 
\acrodef{ut}[UT]{user terminal} 
\acrodef{awgn}[AWGN]{additive white Gaussian noise}
\acrodef{cgac}[CGAC]{Complex-gain analog combiner}
\acrodef{psoac}[PSOAC]{Phase-shifter-only analog combiner}
\acrodef{fpga}[FPGA]{field-programmable gate array}
\acrodef{gui}[GUI]{grapichal user interface}
\acrodef{dfrc}[DFRC]{dual function radar-communications}
\acrodef{jrc}[JRC]{joint radar and communication}
\acrodef{pri}[PRI]{pulse repetition interval}
\acrodef{gsm}[GSM]{generalized spatial modulation}
\acrodef{smx}[SMX]{spatial multiplexing MIMO}
\acrodef{lfm}[LFM]{linear frequency modulation}
\acrodef{tws}[TWS]{track while scan}
\acrodef{stt}[STT]{single target tracker}
\acrodef{mi}[MI]{mutual information}
\acrodef{ula}[ULA]{uniform linear array}
\acrodef{scr}[SCR]{signal-to-noise ratio}
\acrodef{fmcw}[FMCW]{frequency modulated continuous waveform}
\acrodef{far}[FAR]{frequency agile radar}
\acrodef{cpi}[CPI]{coherent processing interval}
\acrodef{im}[IM]{index modulation}
\acrodef{lpf}[LPF]{low pass filter}
\acrodef{crrp}[CRRP]{coarse resolution range profile}
\acrodef{hrrp}[HRRP]{high resolution range profile}
\acrodef{idft}[IDFT]{inverse discrete Fourier transform}
\acrodef{ml}[ML]{maximum likelihood}
\acrodef{sod}[SOD]{successive orthogonal decoding}
\acrodef{pm}[PM]{phase modulation}
\acrodef{lidar}[LIDAR]{light detection and ranging}
\acrodef{adas}[ADAS]{autonomous driving assistance systems}
\acrodef{v2v}[V2V]{vehicle-to-vehicle} 
\acrodef{v2i}[V2I]{vehicle-to-infrastructure}
\acrodef{v2n}[V2N]{vehicle-to-network}
\acrodef{v2c}[V2C]{vehicle-to-cloud}
\acrodef{v2p}[V2P]{vehicle-to-pedestrian}
\acrodef{v2x}[V2X]{vehicle-to-everything}
\acrodef{caesar}[CAESAR]{carrier agile phased array radar}
\acrodef{omp}[OMP]{orthogonal matching pursuit}
\acrodef{qpsk}[QPSK]{quadrature phase shift keying}
\acrodef{psk}[PSK]{phase shift keying}
\acrodef{rf}[RF]{radio frequency}
\acrodef{cer}[CER]{sub-carrier errer rate}
\acrodef{frac}[FRaC]{FMCW based joint radar-communications system}

\acrodef{gnss}[GNSS]{global navigation satellite system}
\acrodef{spacor}[SpaCoR]{spatial modulation based communication-radar}
\acrodef{majorcom}[MAJoRCom]{multi-carrier agile joint radar communication}
\acrodef{lista}[LISTA]{learned iterative shrinkage thresholding}
\acrodef{dof}[DOF]{degrees of freedom}
\acrodef{gpu}[GPU]{graphics processing unit}

\IEEEoverridecommandlockouts

\title{FRaC: FMCW-Based Joint Radar-Communications System via Index Modulation
}

\author{
	\IEEEauthorblockN{Dingyou Ma, Nir Shlezinger,  Tianyao Huang, Yimin Liu,  and Yonina C. Eldar
	} 
	\thanks{Parts of this work were presented in the 2020 IEEE International Conference on Communications Workshops (ICC Workshops) as the paper \cite{Ma2020ADFRC}.
	}
	\thanks{
		D. Ma, T. Huang, and Y. Liu are with the EE Department, Tsinghua University, Beijing, China (e-mail: mdy16@mails.tsinghua.edu.cn; \{huangtianyao, yiminliu\}@tsinghua.edu.cn).
		N. Shlezinger is with the School of ECE, Ben-Gurion University of the Negev, Be'er-Sheva, Israel (e-mail: nirshl@bgu.ac.il). 
		Y. C. Eldar is with the Faculty of Math and CS, Weizmann Institute, Rehovot, Israel (e-mail: yonina.eldar@weizmann.ac.il). (\textit{Corresponding author: Tianyao Huang})				
	}	
	\thanks{This work received funding from the National Natural Science Foundation of China under grant 61801258, from the European Union’s Horizon 2020 research and innovation program under grant No. 646804-ERC-COG-BNYQ, from Futurewei Technologies, and from the Air Force Office of Scientific Research under grant No. FA9550-18-1-0208.}	
	\vspace{-0.5cm}
}

\vspace{-0.75cm}

\begin{document}
	
	\maketitle
	\pagestyle{plain}
	\thispagestyle{plain}
	%--------------------------------------------------------------------------------------
	%	ABSTRACT
	%----------------------------------------------------------------------------------------
	% It's preferable not to use \ac{} in the abstract, only in the main body
	\begin{abstract} 
		Dual function radar communications (DFRC) systems are  attractive technologies for autonomous vehicles, which utilize  electromagnetic waves to constantly sense the environment while simultaneously communicating with neighbouring devices. An emerging approach to implement DFRC systems is to embed information in radar waveforms via index modulation (IM). Implementation of DFRC schemes in vehicular systems gives rise to strict constraints in terms of cost, power efficiency, and hardware complexity. In this paper, we extend IM-based DFRC systems to utilize sparse arrays and frequency modulated continuous waveforms (FMCWs), which are popular in automotive radar for their simplicity and low hardware complexity. The proposed FMCW-based radar-communications system (FRaC) operates at reduced cost and complexity by transmitting with a reduced number of radio frequency modules, combined with narrowband FMCW signalling.  This is achieved via array sparsification in transmission, formulating a virtual multiple-input multiple-output array by combining the signals in one coherent processing interval, in which the narrowband waveforms are transmitted in a randomized manner. %This  randomness, which enables multiple systems to simultaneously probe with minor interference, is also exploited by FRaC to convey information via IM, embedding the transmitted bits in the selection of carrier frequencies and antennas. 
		Performance analysis and numerical results show that the proposed radar scheme achieves  similar resolution performance compared with a wideband radar system operating with a large receive aperture, while requiring  less hardware overhead. For the  communications subsystem, FRaC achieves higher rates and improved error rates compared to dual-function signalling based on conventional  phase modulation. 
	\end{abstract}
	
	% 	\begin{IEEEkeywords}
	% 		Joint Radar-Communication System, Index Modulation, FMCW Radar
	% 	\end{IEEEkeywords}
	
	\acresetall 
	%----------------------------------------------------------------------------------------
	%	Introduction
	%----------------------------------------------------------------------------------------
	%	\vspace{-0.2cm}
	\section{Introduction} 
	%	\vspace{-0.15cm}
	
	Autonomous vehicles are envisioned to revolutionize transportation, and are thus the focus of growing interests both in academia and industry. Such self-driving cars, which are aware of their environment and neighbouring vehicles, are expected to decrease accidents, improve traffic efficiency, and reduce transportation cost. To avoid obstacles, plan routes and comply with traffic regulations, autonomous vehicles are required to constantly sense the environment. Therefore, self-driving cars are equipped with multiple sensors, including \ac{lidar}, camera, \ac{gnss}, and automotive radar. Among these sensors, automotive radar is a necessary component due to its  capability to detect distant objects in bad weather conditions and poor visibility. In addition to environment sensing, autonomous vehicles also need to exchange information with nearby cars and control centers in order to realize efficient coordination. 
	Consequently, future cars  will transmit electromagnetic waves for both radar and  wireless communications. 
	
	In traditional designs,  individual separate hardware modules are  designed  for each functionality. Considering the similarities of  radar and communications in hardware and signal processing, an alternative strategy is to \emph{jointly design} both functionalities as a \ac{dfrc} system. Such dual-function systems are the focus of extensive research attention over recent years \cite{  Ma2020Joint,  Liu2020Joint, Paul2017, Sturm2011Waveform,Liu2018MUMIMO, Liu2018Toward, Zheng2019radar,  Huang2020MAJoRCom,Huang2020Multi,Kumari2018IEEE80211ad, Chiriyath2017,Sturm2013Spectrally, Bica2019Multicarrier,Ma2021Spatial,Sahin2017Anovel,Hassanien2016Dual,Wang2019Codesign, Ma2021Bit}. Jointly implementing both systems in a common platform contributes to reducing the system cost, size, weight, and power consumption, as well as alleviating concerns for electromagnetic compatibility and spectrum congestion, making it an attractive technology for vehicular applications \cite{Ma2020Joint, mishra2019towards}. %Nonetheless, vehicular \ac{dfrc} systems are required to operate at reduced cost, power, and complexity, while supporting multiple cars to simultaneously operate in the same environment with minimal mutual interference.  
	%	Although the \ac{dfrc} systems have been studied for over a decade, when focused on the vehicular applications, many constraints and requirements should be considered.  
	
	Various strategies are proposed to enable the dual function operation as surveyed in~\cite{Ma2020Joint}. A common approach is to utilize separate coordinating signals for radar and communications~\cite{Sturm2013Spectrally, Bica2019Multicarrier,Ma2021Spatial} in a co-existing manner.  	
	An alternative strategy is to realize radar sensing based on conventional communication waveforms~\cite{Sturm2011Waveform,Kumari2018IEEE80211ad}. When radar is the primary user, another scheme embeds the information bits into traditional radar waveforms~\cite{Sahin2017Anovel,Hassanien2016Dual,Wang2019Codesign, Ma2021Bit}.The fourth method is to design an optimized waveform according to the objectives and constraints from the dual functionalities~\cite{Liu2018MUMIMO, Liu2018Toward,Liu2020Joint}. The pros and cons of each category were analyzed in~\cite{Ma2020Joint} according to the radar and communications requirements in vehicular applications. Nonetheless, no single \ac{dfrc} scheme can satisfy all scenarios of vehicular applications. %They either have a high hardware complexity, such as the communication waveform based approach and the optimized waveform design, or lead to a low data rate, such as the radar waveform based scheme. 
	Design of a \ac{dfrc} scheme which has a good trade off between performance and hardware complexity is still desired.
	
	%	Recently, the emerging \ac{im} technique is employed to design the \ac{dfrc} systems~\cite{Ma2021Spatial, Huang2020MAJoRCom}. \ac{im} is a promising communication technique, which embeds the transmit information into the indices of certain building blocks, such as the antenna elements~\cite{Wang2012Generalised}, sub-carriers~\cite{Basar2013Orthogonal}, and spreading codes~\cite{Kaddoum2016Generalized}.
	%This technique can be utilized to design the \ac{dfrc} systems. On the one hand, \ac{im} can facilitate the co-existence of radar and communications.	
	
	\textcolor{black}{Recent years have witnessed a growing interest in communications based on \ac{im} techniques due to their increased spectral and energy efficiency \cite{Basar2016Index}. \ac{im} schemes convey additional information in the indices of the building blocks of communication systems, such as the selection of transmit antennas~\cite{Wang2012Generalised,Younis2010Generalised}, subcarriers~\cite{Basar2013Orthogonal, Datta2016Generalized}, and spreading codes~\cite{Kaddoum2016Generalized}.\label{txt:IM}} The idea of \ac{im} has been introduced to design \ac{dfrc} systems, facilitating the co-existence of radar and communications.	
	The \ac{spacor} system, which allocates antenna elements of a phased array between radar and communication according to spatial \ac{im}, i.e., \ac{gsm}~\cite{Wang2012Generalised,Younis2010Generalised}, was proposed in~\cite{Ma2021Spatial}. By employing \ac{gsm},  \ac{spacor} achieves increased communication rate while acquiring the same angle resolution as using the full antenna array for radar. However, the cost and power consumption of \ac{spacor} is high for consumer-oriented vehicular applications, since separate \ac{rf} components and a costly phased array are required.

	\ac{im} can also be employed to increase the \ac{dof} of  radar waveform-based \ac{dfrc} approaches for information embedding~\cite{Huang2020MAJoRCom}. \ReviseReview{In radar waveform-based \ac{dfrc} schemes, the information can be embedded into the phase of the radar waveform~\cite{Sahin2017Anovel}, the sidelobe~\cite{Hassanien2016Dual}, the carrier frequency~\cite{Wang2019Codesign} or the basis of the radar sub-pulses~\cite{Wu2020Waveform}.\label{txt:CiteReviewer3}} 
	While these schemes lead to minimal degradation to radar performances, the communication rates are typically low because they have limited freedoms to embed information. The \ac{majorcom} system was recently proposed in~\cite{Huang2020MAJoRCom} to utilize \ac{im} to embed more information bits into spectral and spatially agile radar waveforms. However, \ac{majorcom} utilizes simple pulse waveforms and phased array antenna, which are suitable for traditional radar systems such as military applications, as they require a power amplifier with high peak transmit power and a costly phased array.	
	For automotive radar systems, the most commonly utilized scheme is  \ac{fmcw} \cite{AWR18432020}, due to its simplicity, low complexity, and established accuracy. This motivates the combination of \ac{im} with  \ac{fmcw} systems for vehicular  systems. 
	
	\color{black}
	In this paper, we propose an \ac{frac}, which combines \ac{im}-based \ac{dfrc} design with existing automotive radar techniques and considerations. 
	To satisfy the requirements  of automotive radar, \ac{frac} reduces the complexity and cost both in the system level and in the waveform level. 
	In the system level, \ac{frac} utilizes a \ac{mimo} array architecture with separate transmit and receive arrays. To further reduce the number of \ac{rf} modules, the \ac{mimo} transmitter utilizes a randomized sparse array rather than using a full array as in \ac{spacor} and \ac{majorcom}. This transmission architecture decreases the hardware complexity while achieving high angular resolution by formulating a large virtual array in the radar receiver. In the waveform level, unlike \ac{spacor} which transmits a wideband waveform and \ac{majorcom} that uses a narrowband simple pulse, \ac{frac} transmits narrowband \ac{fmcw} signals from the sparse array, which enables utilizing \acp{adc} with low sampling rates. After signal processing, the range resolution is synthesized as the resolution of a wideband waveform radar. 
	
	\ac{frac} thus extends \ac{majorcom} to utilize sparse \ac{mimo} arrays and \ac{fmcw} signalling with modulated pulses, resulting in an \ac{im}-based \ac{dfrc} system geared towards vehicular applications.  Compared with \ac{majorcom}, which only embeds information into the spatial and frequency \ac{im}, \ac{frac} also embeds additional bits through \ac{pm}. During each transmission, 
	the carrier frequencies and the transmit antenna elements are randomly selected and assigned  according to the \ac{im} mapping rule.  
	The digital message communicated to a remote receiver is embedded in the selection of the  transmit antenna subset, the selection of carriers, and the \ac{pm} symbols. 
	
	\color{black}
	
	To study the performance of \ac{frac}, we first analyze the ambiguity function, then  identify the relationship between the maximum number of recoverable targets and the waveform parameters using phase transition thresholds. Theoretical analysis and simulation results show that \ac{frac} achieves improved performance for both radar and communications systems: The radar subsystem achieves the same range, velocity and angle resolutions as that  of a  wideband \ac{fmcw} radar system, which is much more costly and utilizes an equivalent large receiving aperture as the virtual aperture combined in~\ac{frac}. The communications subsystem of \ac{frac} achieves increased rates and decreased \ac{ber}  compared with the scheme which solely utilizes  \ac{pm} to embed the message as in~\cite{Sahin2017Anovel}. Furthermore, compared with the wideband \ac{fmcw} radar, \ac{frac} uses less \ac{rf} modules, and decreases the sampling rate  at the radar receiver, which reduces the hardware cost. Moreover, utilizing frequency and spatial agility facilitates operation in congested environments \cite{Huang2020MAJoRCom}. These benefits make \ac{frac} a promising candidate for future vehicular applications. 
	
	The rest of this paper is organized as follows: Section~\ref{sec:Model} presents the system model of the proposed \ac{dfrc} scheme, and compares the system model of \ac{frac}, \ac{majorcom} and \ac{spacor} in a high level. The designs of the radar and communications receiver are introduced in Section~\ref{sec:RadarCommRx}. Performance analysis of the radar subsystem is provided in Section~\ref{sec:PerformanceRadar}. 
	We numerically evaluate the performance of the radar and communications subsystems in Section~\ref{sec:Sims}. Finally, Section~\ref{sec:Conclusions} provides concluding remarks. 
	
	The following notations are used throughout the paper: Boldface lowercase and uppercase letters denote vectors and matrices, respectively.  %$\binom{\cdot}{\cdot}$ is the binomial coefficient. 
	We denote the transpose, conjugate,  Hermitian transpose, and integer floor operation as $\left(\cdot\right)^{\rm T}$,  	$\left(\cdot\right)^{\ast}$,  $\left(\cdot\right)^{\rm H}$, and $\lfloor \cdot \rfloor$,  respectively,  %The complex normal distribution with mean $\mu$ and variance $\sigma^2$ is expressed as $\mathcal{CN}\left(\mu, {\sigma}^2\right)$, while 
	and $\mathbb{E}\left\{\cdot\right\}$ is the expected value of a random argument.   The set of complex numbers is $\mathcal{C}$.

	\begin{figure}
		\centerline{\includegraphics[width= 0.8\columnwidth]{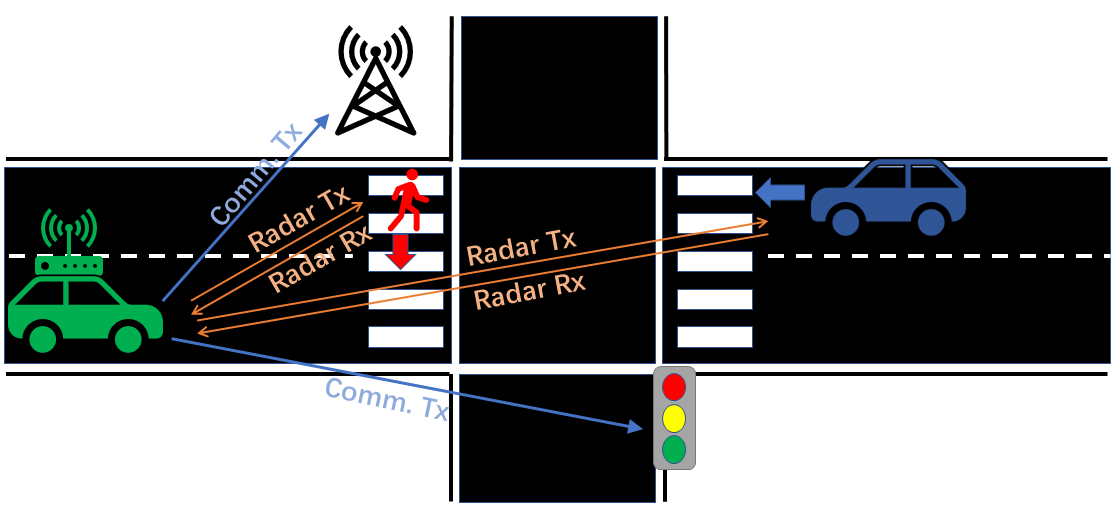}}
		\vspace{-0.2cm}
		\caption{An illustration of the \ac{dfrc} system for vehicular applications.} 
		\label{fig:SystemModel}
	\end{figure}
	
	%----------------------------------------------------------------------------------------
	% System Model
	%----------------------------------------------------------------------------------------	
	\vspace{-0.1cm}
	\section{System  Model}
	\label{sec:Model}
	\vspace{-0.1cm}
	\color{black}
	In this section we review the  \ac{dfrc} system model for which the proposed \ac{im}-based \ac{frac} is designed. To that aim, we first present a generic formulation of \ac{im}-based \ac{dfrc} systems in Subsection~\ref{subsec:GenericModel}. We show how this model specializes into the previously proposed  \ac{spacor} \cite{Ma2021Spatial} and \ac{majorcom} \cite{Huang2020MAJoRCom} \ac{dfrc} systems, and identify their shortcomings in the context of vehicular applications. Then, in Subsection~\ref{subsec:TxStrategy} we use the generic \ac{im}-based \ac{dfrc} system model to formulate the transmission model of the proposed \ac{frac} in Subsection~\ref{subsec:TxStrategy}, and elaborate on its differences from \ac{spacor} and \ac{majorcom}.

	%One  illustration of the \ac{dfrc} system is shown in Fig.~\NoRef. To reliably sense the environment and interact with the surroundings, autonomous vehicles are usually equipped with automotive radar and vehicular communication devices. Rather than implementing both functionalities on separate hardwares, we propose a  radar waveform-based \ac{dfrc} system, where the transmitter utilizes radar waveforms for both sensing and communication.
	
	%To formulate the system model of \ac{frac}, we begin by describing the  transmitted \ac{dfrc} waveform and the information embedding approach in Subsection~\ref{subsec:TxStrategy}. %Then, we introduce the  information embedding strategy extending the scheme to a \ac{dfrc} system in Subsection~\ref{subsec:InfoEmbed}. 
	%The received signal models and the processing algorithms of the radar and communication subsystems are discussed in Subsection~\ref{subsec:RadarRx} and Subsection~\ref{subsec:CommuRx}, respectively. %Finally,  discussions on \ac{frac} are provided in Subsection~\ref{subsec:Discussion}. %The design guidelines of the system parameters are given in Subsection~\ref{subsec:ParameterDesign}, and followed by the discussion in Subsection~\ref{subsec:Discussion}.
	
	%----------------------------------------------------------------------------------------
	% Generic model
	%----------------------------------------------------------------------------------------	
	\vspace{-0.11cm}
	\subsection{\ac{im}-Based \ac{dfrc} System Model}
	\vspace{-0.11cm}
	\label{subsec:GenericModel}
	\ac{dfrc} systems consist of a \ac{dfrc} transmitter, a radar receiver and communications receivers. Typically, the \ac{dfrc} receiver and the radar receiver are located on the same device, e.g., a self-driving car probing the environment, while the communication receiver, such as a road-side unit, is remotely located \cite{Ma2020Joint}.  
	Since the differences of the \ac{dfrc} schemes are mainly in the design of the \ac{dfrc} transmitter, we focus on presenting the generic model of the \ac{dfrc} transmitter. 
	
\begin{figure}
		\centerline{\includegraphics[width= \columnwidth]{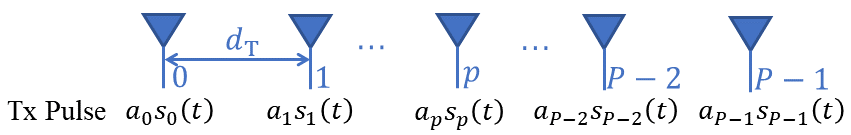}}
		\vspace{-0.2cm}
		\caption{The generic model of the \ac{im}-based \ac{dfrc} transmitter.} 
		\label{fig:GenericModel}
	\end{figure}

	The  generic \ac{dfrc} transmitter is depicted in Fig.~\ref{fig:GenericModel}. The transmitter is equipped with  a \ac{ula} with $P$ elements, the adjacent element distance of which is denoted by $d_{\rm T}$. During the transmission of each pulse, the waveform $s_{p}\left(t\right)$ is multiplied with the weight $a_{p}$, and then transmitted from the $p$th element of the array. The waveform $s_{p}\left(t\right)$ is chosen from the waveform set $\mathcal{S}$. 
	%while the multiplied weight $a_p$ is chosen from the set $\mathcal{A}$. 
	The fact that the transmitter implements both radar and communication implies that the transmitted waveforms $\{a_p s_p(t)\}$ are utilized for both sensing as well as conveying information to the remote receiver. 
	
	When different transmit waveforms and weights are assigned, special cases of the \ac{dfrc} schemes can be represented using the generic model. In \ac{im}-based \ac{dfrc} systems, the transmission parameters, such as the setting of $\{a_p\}$, are used to convey information to the remote receiver. Two recent   \ac{dfrc} systems which implement such forms of \ac{im} are \ac{spacor}  and \ac{majorcom}:
	%	Consequently, we can compare the existing schemes and the proposed \ac{frac} system in a unified  framework. To that aim, we first represent the \ac{spacor} and \ac{majorcom} using this high level model, then propose the high level model of \ac{frac} considering the requirements from vehicular applications. 
	
	%	\ac{im} is a communications schemes in which information is embedded in the selection of the parameters of the transmission systems \cite{Basar2016Index}. \ac{im} techniques were shown to be highly suitable for realizing \ac{dfrc} systems, due to their ability to facilitate  co-coexistence of the radar and communication functionality, as well or increase the \ac{dof} for information embedding. We formulate a generic system model of the \ac{im}-based \ac{dfrc} schemes in a high level. The 

	\subsubsection{SpaCoR} \ac{spacor} proposed in \cite{Ma2021Spatial} implements the \ac{dfrc} transmitter illustrated in Fig.~\ref{fig:GenericModel} while incorporating \ac{im} in the form of \ac{gsm}. In particular, \ac{spacor} utilizes spectrally distinct waveforms for radar and communications, such that $\mySet{S} = \left\{s^{\left(\rm r\right)}\left(t\right)\right\} \cup \mySet{S}^{\left(\rm c\right)}$, where $s^{\left(\rm r\right)}\left(t\right)$ is the radar waveform, and $\mySet{S}^{\left(\rm c\right)}$ is the set of the communication waveforms, i.e., $|\mySet{S}^{\left(\rm c\right)}|=J$ for a $J$th order digital constellation. \ac{im} is realized by setting whether $s_p(t)$ should be $s^{\left(\rm r\right)}\left(t\right)$ or taken from $\mySet{S}^{\left(\rm c\right)}$.
	
	During each pulse, $K$ antenna elements are utilized for radar probing and $P-K$ elements are allocated to wireless communications. The radar waveforms are beamformed in the selected phased array elements, i.e., the weights $\{a_p\}$ are set to steer the beam in a desired direction following phased array principles~\cite[Ch. 8.2]{skolnik2001introduction}.
	%the set $\mySet{A}$ is set to $\mySet{A} = \Big\{1, \big\{e^{\frac{j2\pi f_c p d_T  \sin\theta_0}{c}}\big\}_{p=0}^{P-1}\Big\}$, 
	%where $\theta_0$ is the desired beamborming direction, $f_c$ is the carrier frequency of $s\Radar\left(t\right)$ and $c$ is the speed of light. The selection of the multiplied weight is as following: When $s_{p}\left(t\right) \in \mySet{S}^{\left(c\right)}$, $a_p$ is set to $1$; When $s_{p}\left(t\right) = s^{\left(r\right)}\left(t\right)$, $a_p$ is set to $e^{\frac{j2\pi f_c p d_T  \sin\theta_0}{c}}$. 	
	The information is conveyed both by the communication waveforms and the spatial \ac{im}. Assume that the cardinality of $\mySet{S}^{\left(\rm c\right)}$ is $J$.   
	The total number of bits conveyed in each pulse is $\log_{2}J + \lfloor\log_2\binom{P}{K}\rfloor$.
	
	\subsubsection{MAJoRCom} While \ac{spacor} utilizes a single radar waveform and spectrally distinct communication signals, \acs{majorcom} proposed in \cite{Huang2020MAJoRCom} utilizes radar waveforms only,   chosen from an orthogonal waveform set. As a result, the communication message is conveyed only via \ac{im}, which includes the selection of the orthogonal waveforms to be transmitted (spectral \ac{im}) as well as in their division among the antenna elements (spatial \ac{im}).
	
	In \acs{majorcom}, the transmit waveform set $\mySet{S}$ is composed of $M$ waveforms orthogonal in frequency, i.e., $\mySet{S} = \{s_{f_m}\Radar\left(t\right)\}_{m=0}^{M-1}$ where   $s_{f_m}\Radar\left(t\right)$ denotes a radar waveform with carrier frequency $f_m$. During the transmission of each pulse, $K$ waveforms are first chosen from $\mySet{S}$, after which each waveform is transmitted from a subset of the array, e.g., from $P_K = P/K$ elements. Both these selections are dictated by the transmitted message. The total number of bits conveyed in each pulse is $\lfloor \log_{2}\binom{P}{K}\rfloor + \lfloor \log_{2} \frac{P!}{\left(P_K !\right)^{K}} \rfloor$.  
	The weights $\{a_p\}$ are set to achieve phased-array beamsteering as in \ac{spacor}. 
	%	The transmit information of \ac{majorcom} is embedded into spatial and frequency index modulation. 

	\smallskip
	\ac{spacor} and \ac{majorcom} are  both special cases of the \ac{dfrc} transmitter model illustrated in Fig.~\ref{fig:GenericModel}, which utilize \ac{im} to enhance spatial allocation  efficiency (as in \ac{spacor}) and to enable communication functionality with pure radar waveforms (as in \ac{majorcom}).  Nonetheless, these designs both rely on phased array radar with the full antenna array transmissions, and are thus geared towards military applications, which utilize high transmit power. However, these cost and power constraints associated with these designs may not be acceptable to the consumer-oriented vehicular applications. Furthermore, the weighting coefficients $\{a_p\}$ are used solely for beamsteering, and are not exploited to convey additional information and further increase the spectral efficiency. This motivates the design of an \ac{im}-based \ac{dfrc} system which builds upon the generic formulation in Fig.~\ref{fig:GenericModel}, while being geared towards automotive systems. This is achieved by integrating array sparsification, using conventional \ac{fmcw} waveforms, and exploiting the ability to set the weights $\{a_p\}$ in order to increase spectral efficiency, as detailed in the following section.

	\color{black}
	%----------------------------------------------------------------------------------------
	% System Model of FRaC
	%----------------------------------------------------------------------------------------	
	\vspace{-0.11cm}
	\subsection{Transmission Model of FRaC}
	\vspace{-0.11cm}
	\label{subsec:TxStrategy}
	The \ac{frac} system consists of a \ac{dfrc} transmitter equipped  on the autonomous vehicle, and several communication receivers, which can be passengers, other vehicles,  road-side units, or wireless base stations, as illustrated in Fig.~\ref{fig:SystemModel}. In this subsection, we only introduce the transmission model of the \ac{dfrc} transmitter. The  radar and communications  receivers are presented in Section~\ref{sec:RadarCommRx}.	
	%We next elaborate on how the \ac{dfrc} waveform is generated, and how the information is embedded into it.
	
	\smallskip
	\subsubsection{Dual Function Waveform Transmission}
	\label{subsubsec:DualWaveform}
\begin{figure}
		\centerline{\includegraphics[width= 0.7\columnwidth]{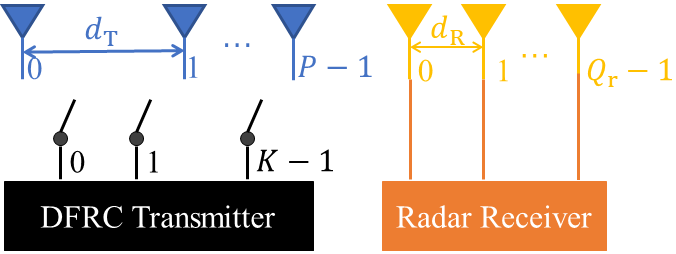}}
		\vspace{-0.2cm}
		\caption{\ac{dfrc} transmitter and radar receiver schematic illustration.} 
		\label{fig:DFRCTX}
	\end{figure}

	The schematic architecture of the \ac{dfrc} transmitter and the radar receiver is shown in Fig.~\ref{fig:DFRCTX}. The \ac{dfrc} transmitter utilizes a  \ac{ula} with $P$ elements to transmit the dual function waveform, while the radar receiver is equipped with a \ac{ula} of $Q_{\rm r}$ elements. The inter-element distance of the radar receive array is denoted by $d_{\rm R}$. The transmit array and the receive array forms a \ac{mimo} radar architecture, i.e., $d_{\rm T} = Q_{\rm r}d_{\rm R}$. 
	%For each pulse, the \ac{pri} is $T_{0}$ and the pulse width is $T_{p}$. 
	To decrease the cost and hardware complexity,  only $K$ antenna elements are activated in the \ac{mimo} transmitter during each pulse, and $K \le P$ waveforms with different carrier frequencies are transmitted from the $K$ active elements. The active antenna elements may change between different pulses, which is controlled by the \ac{rf} switch, as shown in Fig.~\ref{fig:DFRCTX}. The transmit waveforms combine \acs{fmcw} signalling via the setting of the waveform set $\mySet{S}$ with \ac{pm} for conveying a digital message via the weighting coefficients $\{a_p\}$. Substituting the expression of \ac{fmcw} to the high level model of \ac{frac} described in Subsection~\ref{subsec:GenericModel}, the composed waveforms of $\mySet{S}$ are expressed as 
	\vspace{-0.1cm}
	\begin{equation}
	s_{f_m}\Radar\left(t\right) = s\left(t\right) e^{j2\pi \left(f_c + m\Delta f\right)t},
	\vspace{-0.1cm}
	\label{eqn:TxWave}
	\end{equation}
	where $f_c$ is the start of the carrier frequency, $\Delta f$ is frequency step. \ReviseReview{In particular, $s\left(t\right)$ denotes a baseband \ac{fmcw}, which is expressed as
		\vspace{-0.1cm}
		\begin{equation}
		s\left(t\right) = \mathrm{rect}\left(\frac{t}{T_{p}}\right)e^{j\kappa\pi t^2},\ \ 0\le t\le T_{0},
		\vspace{-0.1cm}
		\label{eqn:BasebandFMCW}
		\end{equation}
		where $\mathrm{rect}\left(t\right) = 1$ for $0\leq t < 1$, $\kappa$ is the frequency modulation rate of the \acs{fmcw} waveform, $T_0$ is the duration of one \ac{fmcw} pulse, i.e., the \ac{pri}, and $T_p\le T_0$ is the pulse width. In each \ac{pri}, the \acs{fmcw} waveform is transmitted when $ 0 \le t  \le T_p$, and no signal is transmitted when $T_p< t \le T_0$.\label{txt:T0andTp} The bandwidth of $s\left(t\right)$, denoted  $B_{\rm sub}$, is given by the product of the frequency modulation rate $\kappa$ and the transmit duration, i.e., $B_{\rm sub} := \kappa T_{p}$~\cite{Patole2017}.\label{txt:BandwidthFMCW}}
	
	During each transmission, $K$ \acs{fmcw} waveforms with different carrier frequencies are first chosen from $\mySet{S}$. Then the selected waveforms are simultaneously multiplied with $K$ \ac{pm} symbols. According to \eqref{eqn:TxWave}, the carrier frequencies of the transmit waveforms are chosen from the frequency set $\mySet{F} := \left\{f_{c} + m \Delta f| m \in \mySet{M}\right\}$, where $\mySet{M} := \left\{ 0,1,\cdots, M-1\right\}$. To guarantee that the transmit waveforms are spectrally orthogonal, we set $\Delta f = B_{\rm sub}$. One radar \ac{cpi} consists of $N$ periodically transmitted pulses. In the $n$th radar pulse, the transmitted waveform assigned to the $k$th active element is given by
	\vspace{-0.11cm}
	\begin{equation}
	x_{n,k}\left(t\right) = s\left(t- nT_{0}\right) e^{j2\pi f_{n,k} t}\cdot e^{j\phi_{n,k}},
	\label{eqn:FMCW1}
	\vspace{-0.11cm}
	\end{equation}
	where $f_{n,k} := f_{c} + m_{n,k}\Delta f$ is the carrier frequency chosen from $\mySet{F}$, $m_{n,k}\in \mySet{M}$ is the index of $f_{n,k}$ in $\mySet{F}$, and $\phi_{n,k}\in \mySet{J}$ is the phase modulated on the \acs{fmcw} waveform. \textcolor{black}{Consequently, when the $p$th antenna element transmits at carrier frequency $f_{n,k}$ during the $n$th pulse, the transmitter sets $a_p = e^{j\phi_{n,k}}$.}

	%$\myMat{\Omega}_{m}\in \left\{0, 1\right\}^{K\times N}$ is the waveform selection matrix which has a single non-zero entry in each row, and $\myMat{P}_{m}$ is a $K \times K$ permutation matrix. The selected waveform vector $\myVec{x}_{m}\left(t\right)$ is transmitted from the $K$ activated antenna elements. 
	
	%Assume $Q$ targets \Verify{Add Target Model, e.g. Swerling I}are located in the far field of the antenna array. The range, velocity and direction of the $p$th targets are denoted as $r_{p}$, $v_{p}$ and $\theta_{p}$, respectively. With the far field and narrowband assumptions, the signal received at the $p$th target is expressed as
	%	\begin{align}
	%		y_{p}\Radar\left(t\right) = 
	%	\end{align}
	%	
	%	Denote $\myMat{\Lambda}_{m} \in \left\{0, 1\right\}^{K\times L_{T}}$ as the antenna selection matrix.  
	%	
	%\vspace{-0.2cm}
	\smallskip
	\subsubsection{Information Embedding}
	\label{subsubsec:InfoEmbed}
	%\vspace{-0.1cm}		
	%Here, we introduce the information embedding strategy of the \ac{dfrc} system. 
	In \ac{frac}, the transmit bits are conveyed by both the \ac{pm} symbols and spatial-frequency \ac{im} in the selection of the carrier frequency and the active antenna elements.
	%The idea of using \ac{im} to design \ac{dfrc} systems are also exploited in \ac{majorcom} and \ac{spacor}. The difference of \ac{majorcom} and \ac{frac} is that \ac{majorcom} only embeds information through \ac{im}, while \ac{frac} exploits both \ac{pm} and \ac{im}. For \ac{spacor}, the information bits are conveyed by traditional communication waveforms and spatial \ac{im}, while in \ac{frac} the information bits are conveyed on the dual function waveform, and no communication waveform is transmitted.} %Unlike only modulating the phase of the \ac{fmcw} waveform to convey information, additional bits are conveyed in the form of \ac{im} via the selection of carrier frequencies, the selection of active transmit elements, and through waveform permutation. %To understand how many uncoded bits can be conveyed in each \ac{pri} of the proposed scheme, .
	Embedding the message bits in the transmitted waveform is based on the following steps:
	First, the $K$  \ac{pm} symbols in \eqref{eqn:FMCW1} are set. As each \ac{pm} symbol takes $J$ possible values, the number of bits conveyed by \ac{pm} is $N_{\rm PM} = K\log_{2} J$. Then, $K$ carrier frequencies  are selected from $\mySet{F}$, and are assigned to  the baseband signals. This selection is based on the digital message, and since there are $\binom{M}{K}$ possible combinations of the selected carrier frequencies, a total of  $\lfloor\log_2\binom{M}{K}\rfloor$ bits are conveyed in this selection. Next, $K$ antenna elements are chosen from the transmit antenna array with $P$ elements, i.e., $\binom{P}{K}$ possible selections, where again the information bits are used to determine which elements are chosen. Finally, the up-converted waveforms are assigned to the selected transmit antennas, which enables to convey $K!$ different symbols in the permutation of the waveforms. Hence, the total number of bits  conveyed via \ac{im}, denoted by  $N_{\rm IM}$, is given by $N_{\rm IM}=\lfloor\log_2\binom{M}{K}\rfloor  + \lfloor  \log_2\binom{P}{K}\rfloor +  \lfloor \log_{2}K!\rfloor $, and the total number of bits conveyed in each \ac{pri} is $N_{\rm Total} = N_{\rm PM} + N_{\rm IM}$.
	
\begin{figure}
		\centerline{\includegraphics[width= 0.8\columnwidth]{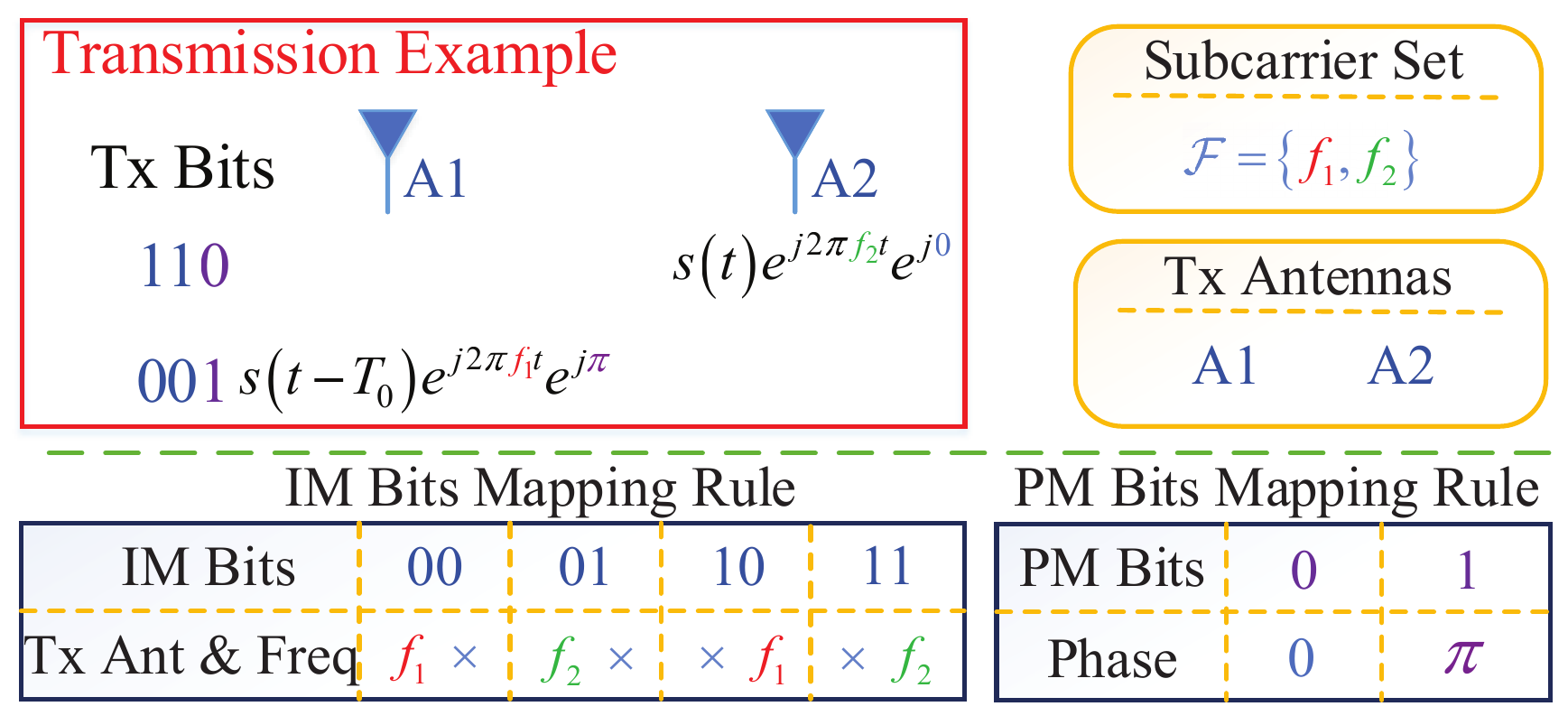}}
		\vspace{-0.3cm}
		\caption{An example of \ac{frac}. The bits in blue fonts are conveyed via \ac{im}. The purple bit is assigned to a \acs{pm} symbol. %The mapping rules of the \acs{im} bits and the \acs{pm} bits are shown. 
			In the \ac{im} mapping rule, "$\times$" implies an unused element. For example, "$f_{1} \ \times$" means that carrier $f_1$ is transmitted from element $\mathrm{A1}$ while antenna $\mathrm{A2}$ is unused.} 
		\label{fig:Example}
	\end{figure}

	Based on the above steps, the generation of the transmit waveforms and their assignment among the transmit antennas are determined by the conveyed data bits. 
	In particular, the transmit bits are first divided into \ac{im} bits, which are used for selecting the carriers and the spatial assignment of the waveforms, and \ac{pm} bits, which are modulated into the \ac{pm} symbols. 
	An example of \ac{frac} is illustrated in Fig.~\ref{fig:Example}. In this example, the cardinality of $\mySet{F}$ is $M = 2$. During each \ac{pri}, $K = 1$ carrier is selected and mixed with the baseband signal $s\left(t\right)$. The \ac{pm} symbols are  generated from a \ac{bpsk} constellation. In the first pulse, the conveyed bits are $110$, where the \acs{im} bits are $11$ and the \ac{pm} bit is $0$. The mapping of the bits is shown in Fig.~\ref{fig:Example}, where the \ac{fmcw} waveform is multiplied with $e^{j0}$, then mixed with carrier $e^{j2\pi f_{2}t}$, and transmitted from antenna $\mathrm{A2}$. %The mapping in the second pulse is similar to the first pulse.

	\color{black}
	%----------------------------------------------------------------------------------------
	% Discussion
	%----------------------------------------------------------------------------------------
	\smallskip
	\subsubsection{Comparison to \ac{spacor} and \ac{majorcom}}
	\ac{frac} implements \ac{im}-based \ac{dfrc} operation in a manner which is geared towards vehicular systems. This is reflected in its differences from the previously proposed  \ac{spacor} and \ac{majorcom}, which also follow the model detailed in Subsection~\ref{subsec:GenericModel}. In particular, \ac{frac} utilizes sparse \ac{mimo} arrays, allowing it to operate with less RF modules than antenna elements, while \ac{spacor} and \ac{majorcom} utilize phased array transmissions from the complete antenna array. Furthermore, \ac{frac} focuses on the usage of \ac{fmcw} signals, which are commonly utilized in automotive radar systems, while embedding additional information in the form of \ac{pm}. 
	An additional fundamental difference of \ac{frac} from \ac{spacor} and \ac{majorcom} lies in the radar receiver array.  The  transmission and reception of \ac{majorcom} and \ac{spacor} utilize the same array through time division duplexing, which is suitable for identifying distant targets as in military applications. \ac{frac} uses distinct receive and transmit arrays, facilitating the identification of nearby targets, as required in automotive applications. 
	
	The comparison of \ac{frac} to \ac{spacor} and \ac{majorcom}  in terms of antenna architecture,  angle resolution,  information embedding strategy,  data rates, and target applications, are summarized in TABLE~\ref{tab:CompareSchemes}. The  transmission and reception of \ac{majorcom} and \ac{spacor} utilize the same array through time division duplexing. The normalized angle resolutions are computed for half-wavelength element spacing, where the corresponding measures for \ac{spacor} and \ac{majorcom} are analyzed in~\cite{Ma2021Spatial} and~\cite{Huang2020Multi}, respectively, while the angle resolution of \ac{frac} is derived in Subsection~\ref{subsec:RadarReso} of this paper.  We numerically demonstrate in Subsection~\ref{subsec:RadarSimu} that this improved angular resolution of \ac{frac}  is directly translated into target recovery gains compared to \ac{majorcom}.
	
	\begin{table*}
		\vspace{-0.25cm}
		\centering
		\caption{Comparisons of \acs{spacor}, \acs{majorcom} and \acs{frac}.} 
		\vspace{-0.2cm}
		\label{tab:CompareSchemes}
		\footnotesize
		\begin{tabular}{|c||c|c|c|}
			\hline
			\hline
			\textbf{Characteristic} & \textbf{\acs{spacor}} & \textbf{\acs{majorcom}} & \textbf{\acs{frac}} \\
			\hline
			\hline
			Antenna architecture & Phased array & Phased array & Randomized sparse \acs{mimo} array \\
			\hline
			Normalized angle resolution & $\frac{1}{P}$ & $\frac{1}{P}$ & $\frac{1}{PQ_{\rm r}}$\\
			\hline
			Information embedding & Comm. symbol + spatial \acs{im} & Spatial and frequency \acs{im} & \acs{pm} + spatial and frequency \acs{im} \\
			\hline
			\acs{im} bits & $ \lfloor\log_2\binom{P}{K}\rfloor$ & $\lfloor \log_{2}\binom{P}{K}\rfloor + \lfloor \log_{2} \frac{P!}{\left(P_K !\right)^{K}} \rfloor$ & $\lfloor\log_2\binom{M}{K}\rfloor  + \lfloor  \log_2\binom{P}{K}\rfloor +  \lfloor \log_{2}K!\rfloor$\\
			\hline
			Total bits & $\log_{2}J + \lfloor\log_2\binom{P}{K}\rfloor$ & $\lfloor \log_{2}\binom{P}{K}\rfloor + \lfloor \log_{2} \frac{P!}{\left(P_K !\right)^{K}} \rfloor$ & $\lfloor\log_2\binom{M}{K}\rfloor  + \lfloor  \log_2\binom{P}{K}\rfloor +  \lfloor \log_{2}K!\rfloor + \log_2 J$ \\
			\hline
			Use cases & Military & Military & Commercial \\
			\hline	
			\hline	
		\end{tabular}  
		\vspace{-0.4cm}
	\end{table*}

	\color{black}
	%----------------------------------------------------------------------------------------
	% Radar and Communications Receivers
	%----------------------------------------------------------------------------------------
	\vspace{-0.2cm}
	\section{Radar and Communications Receivers}
	\label{sec:RadarCommRx}
	\vspace{-0.1cm}
	The received signal models and the processing algorithms of the radar and communication subsystems are discussed in Subsection~\ref{subsec:RadarRx} and Subsection~\ref{subsec:CommuRx}, respectively.
	%----------------------------------------------------------------------------------------
	% Radar Receiver
	%----------------------------------------------------------------------------------------
	\vspace{-0.2cm}
	\subsection{Radar Receiver}
	\label{subsec:RadarRx}
	\vspace{-0.1cm}
	To formulate the radar receiver processing, we first present the received signal model, after which we introduce a \ac{cs} based detection algorithm.
	\subsubsection{Radar Received Signal Model}
	Let $p_{n,k}$ be the index of the $k$th transmit element in the $n$th \ac{pri}. Assume that $L$ ideal point targets are located in the far field of the transmit antenna array with ranges $\{r_{l}\}$, velocities $\{v_{l}\}$ and angles $\{\theta_{l}\}$, $l = 0,1,\dots,L-1$. Under the far field assumption and the ``stop and go"  model~\cite{Huang2020Multi}, the round-trip delay between the $p_{n,k}$th transmit element and the $q_{\rm r}$th receive element is $\tau_{n,k,q_{\rm r}}^{l} = \frac{2\left(r_{l} + nv_{l}T_{0}\right)}{c} - \frac{\left(p_{n,k}d_{\rm T} + q_{\rm r}d_{\rm R}\right)\sin\theta_{l}}{c}$ for the echo from the $l$th target, where $c$ is the speed of light. The received signal at the $q_{\rm r}$th  element is 
	\vspace{-0.2cm}
	\begin{equation}
	\tilde{y}\Radar_{n,q_{\rm r}}\left(t\right) = \sum_{l=0}^{L-1}\alpha_{l} \sum_{k=0}^{K-1}x_{n,k}\left(t- \tau_{n,k,q_{\rm r}}^{l}\right) + \breve{w}_{n,q_{\rm r}}\Radar\left(t\right),
	\label{eqn:RadarRx1}
	\vspace{-0.2cm}
	\end{equation}
	where $\alpha_{l}$ is the reflective factor of the $l$th target, and  $\breve{w}_{n,q_{\rm r}}\Radar\left(t\right)$ denotes additive white Gaussian noise.
	
	To separate the transmit waveforms in the receive element, the received signal  $\tilde{y}\Radar_{n,q_{\rm r}}\left(t\right)$ is simultaneously mixed with the transmit waveforms $x_{n,k}\left(t\right)$ via the traditional de-chirp processing\cite{Patole2017}. Then, the output of each mixer is fed into a \ac{lpf} whose cutoff frequency is set to $B_{\rm sub}$. The separated signal transmitted from the $p_{n,k}$th  element is  $\tilde{y}\Radar_{n,k,q_{\rm r}}\left(t\right) = \mathrm{LPF}\big(\tilde{y}\Radar_{n,q_{\rm r}}\left(t\right)\cdot x_{n,k}^{\ast}\left(t\right)\big)$ \cite{Winkler2007Range}, which \ReviseReview{approximates} \label{txt:Equal2Approx}
	\begin{align}
	\!\!\!&\tilde{y}\Radar_{n,k,q_{\rm r}}\left(t\right) 
	% 	= \mathrm{LPF}\left(\tilde{y}\Radar_{n,q_{\rm r}}\left(t\right)\cdot x_{n,k}^{\ast}\left(t\right)\right)  \notag\\
	% 	\!\!\!
	=\sum_{l=0}^{L-1}\tilde{\alpha}_{l} e^{-j2\pi \kappa \left[\frac{2r_{l}}{c}\! +\! \frac{2v_{l}nT_{0}}{c} \!- \!\frac{\left(p_{n,k}Q_{\rm r} \!+ \!q_{\rm r}\right)d_{\rm R}\sin\theta_{l}}{c}\right]\left(t \!-\! n T_{0}\right)}\notag\\ 
	%&=  \beta e^{-j2\pi \frac{2\kappa \Delta f m_{n,k}}{c}} \cdot e^{ - j2\pi\frac{2\kappa r}{c}t} \notag\\
	%& \times e^{ - j2\pi \frac{2vT_{0}f_{n,k}}{c}n} \cdot e^{-j2\pi f_{n,k}\frac{\sin\theta}{c}\left(p_{n,k}Q + q\right)} \notag\\
	&  \times \!\!e^{\!-j2\pi \left(m_{n,k}\!\frac{2r_{l}\Delta f}{c} \!+\! \xi_{n,k} f_{v}^{l} n \!+\! \xi_{n,k} f_{\theta}^{l}  \left(Q_{\rm r}p_{n,k} \!+\! q_{\rm r}\right)\right)} \!\!+ \!\tilde{w}\Radar_{n,k,q_{\rm r}}\!\!\left(t\right)\!,
	\vspace{-0.1cm}
	\label{equ:RadarRx1}
	\end{align}
	where %$\mathrm{LPF}\left(\cdot\right)$ is the \acl{lpf},
	%	$\left(\cdot\right)^{\ast}$ is the complex conjugate operation, %\footnote{\textcolor{red}{Nir Comment - I thought you defined $\alpha_l$, i.e., it depends on the index $l$, but here you use $\alpha$ and the whole expression is just lost in the definition of $\beta_l[g]$ below.}},
	$\tilde{\alpha}_{l} := \alpha_{l} e^{-j2\pi \frac{2r_{l}f_{c}}{c}}$, %$f_{r}^{l} := \frac{2 r_{l} \Delta f}{c}$ is the normalized range frequency, 
	$f_{v}^{l} := \frac{2v_{l}T_{0}f_c}{c}$ is the normalized velocity frequency, $f_{\theta}^{l} := \frac{f_c d_{\rm R}\sin\theta_{l}}{c}$ is the normalized spatial frequency, and $\xi_{n,k} := \frac{f_{n,k}}{f_c}$ is the relative factor between $f_{n,k}$ and $f_{c}$. Since the sequence $x_{n,k}^{\ast}\left(t\right)$ has  constant unit amplitude within a pulse, the equivalent noise $\tilde{w}\Radar_{n,k,q_{\rm r}}\!\!\left(t\right):= \mathrm{LPF}\big(\breve{w}_{n,q_{\rm r}}\Radar\left(t\right)\cdot x_{n,k}^{\ast}\left(t\right)\big)$ is a band-limited  Gaussian signal. %\footnote{\textcolor{red}{Nir Comment - how come you have noise in \eqref{equ:RadarRx1} but no noise in \eqref{eqn:RadarRx1}? and how come the noise is white? It is obtained after filtering and mixing withe a deterministic waveform, I do not see any justification for it to be stationary not to mention white... }}\VerifyN{Denote $\acute{w}_{n,k,q}\Radar{\left(t\right)}: = \breve{w}_{n,q}\Radar\left(t\right) \cdot x_{n,k}^{\ast}\left(t\right)$, which is the mixing of $\breve{w}_{n,q}\Radar{\left(t\right)}$ and $x_{n,k}^{\ast}\left(t\right)$. We claim that $\acute{w}_{n,k,q}\Radar{\left(t\right)}$ is an additive white Gaussian noise. We prove this conclusion by calculating its expectation and auto-correlation functions. As $\breve{w}_{n,q}\Radar\left(t\right)$ is the thermal noise, it is modeled as the zero mean white Gaussian noise. Assume the spectral density of $\breve{w}_{n,q}\Radar{\left(t\right)}$ is $S_{\breve{w}}\left(f\right) = \frac{N_{0}}{2}$, and the auto-correlation of $R_{\breve{w}}\left(\tau\right) = \frac{N_{0}}{2}\delta\left(\tau\right)$. The expectation of $\acute{w}_{n,k,q}\Radar\left(t\right)$ is  $\mySet{E}\left(\acute{w}_{n,k,q}\Radar\left(t\right)\right) = \mySet{E}\left(\breve{w}_{n,q}\Radar{\left(t\right)} \cdot x_{n,k}^{\ast}\left(t\right)\right) = x_{n,k}^{\ast}\left(t\right) \cdot \mySet{E}\left(\breve{w}_{n,q}\Radar\left(t\right)\right) = 0$. The auto-correlation function of $\acute{w}_{n,k,q}\Radar\left(t\right)$ is $R_{\acute{w}}\left(t_{1}, t_{2}\right) = \mySet{E}\left(\breve{w}_{n,q}\Radar{\left(t_{1}\right)} \cdot x_{n,k}^{\ast}\left(t_{1}\right) \cdot \left[\breve{w}_{n,q}\Radar{\left(t_{2}\right)}\right]^{\ast} \cdot x_{n,k}\left(t_{2}\right)\right) = \mySet{E}\left(\breve{w}_{n,q}\Radar{\left(t_{1}\right)}\cdot \left[\breve{w}_{n,q}\Radar{\left(t_{2}\right)}\right]^{\ast}\right) \cdot x_{n,k}^{\ast}\left(t_{1}\right) \cdot x_{n,k}\left(t_{2}\right) = \frac{N_{0}}{2} \delta\left(\tau\right)$. This is because $\mySet{E}\left(\breve{w}_{n,q}\Radar{\left(t_{1}\right)}\cdot \left[\breve{w}_{n,q}\Radar{\left(t_{2}\right)}\right]^{\ast}\right) \cdot x_{n,k}^{\ast}\left(t_{1}\right) \cdot x_{n,k}\left(t_{2}\right) = \frac{N_{0}}{2}\delta\left(\tau\right)\cdot x_{n,k}^{\ast}\left(t_{1}\right) \cdot x_{n,k}\left(t_{2}\right) = 0$, when $t_{1}\ne t_{2}$, i.e. $\tau \ne 0$. And when $t_{1} = t_{2}$, i.e., $\tau = 0$,  $\mySet{E}\left(\breve{w}_{n,q}\Radar{\left(t_{1}\right)}\cdot \left[\breve{w}_{n,q}\Radar{\left(t_{2}\right)}\right]^{\ast}\right) \cdot x_{n,k}^{\ast}\left(t_{1}\right) \cdot x_{n,k}\left(t_{2}\right) = R_{\breve{w}}\left(0\right)\cdot x_{n,k}^{\ast}\left(t_{1}\right) \cdot x_{n,k}\left(t_{2}\right)= \delta\left(0\right) \cdot x_{n,k}^{\ast}\left(t_{1}\right) \cdot x_{n,k}\left(t_{1}\right) = \frac{N_{0}}{2}\delta\left(0\right) \exp{j0} = \frac{N_{0}}{2}\delta\left(0\right)$. Therefore $\acute{w}_{n,k,q}\Radar{\left(t\right)}$ is still an additive white Gaussian noise. As $\tilde{w}\Radar_{n,k,q}\left(t\right)$ is the output of $\acute{w}\Radar_{n,k,q}\left(t\right)$ after low pass filtering, $\tilde{w}\Radar_{n,k,q}\left(t\right)$ is an additive band-limited white Gaussian noise. } 
	
	\ReviseReview{After individually separated, the received signal is uniformly sampled with rate $F_s\Radar = \frac{1}{T_{s}\Radar}$, where $T_{s}\Radar$ is the sampling interval. In particular, the sampling rate $F_{s}\Radar$ is set to~\cite[Eqn. 1]{ProgramChirpTI} 
		\vspace{-0.1cm}
		\begin{equation}
		F_{s}\Radar = \frac{2r_{\rm max}B_{\rm sub}}{cT_{0}},
		\vspace{-0.1cm}
		\label{eqn:SampRateFMCW}
		\end{equation}
		where $r_{\rm max}$ is the maximum detection range. In vehicular applications, $r_{\rm max}$ typically ranges from tens to several hundred meters. Therefore $\frac{2r_{\rm max}}{c}$ is usually in the order of several microseconds, which is much less than $T_{0}$, whose order is of tens to hundreds of microseconds~\cite[Tab. 1]{ProgramChirpTI}. Consequently, the sampling rate $F_s\Radar$ is much less than the bandwidth $B_{\rm sub}$.\label{txt:SamplingRate}}
	
	The number of sample points in each \ac{pri} is $G = \lfloor \frac{T_{0}}{T_{s}\Radar} \rfloor$, and the sample time instances are $t = nT_{0} + \tilde{g}T_{s}\Radar$, where $\tilde{g} \in \left\{ 0,1, \cdots, G-1\right\}$. Substituting the time instances into \eqref{equ:RadarRx1}, the sampled signal is given by
	\begin{flalign}
	& \tilde{y}\Radar_{n,k,q_{\rm r}}\left[\tilde{g}\right] =\sum_{l=0}^{L-1} \tilde{\alpha}_{l} e^{-j2\pi \kappa \left[\frac{2r_{l}}{c} + \frac{2v_{l}nT_{0}}{c} - \frac{\left(p_{n,k}Q_{\rm r} + q_{\rm r}\right)d_{\rm R}\sin\theta_{l}}{c}\right]\tilde{g}T_{s}\Radar} \notag  \\ 
	\!\!\!\!& \times \!\!e^{\!-\!j2\pi m_{n\!,k}\!\frac{2r_{l}\Delta f}{c} \!-\! j2\pi \xi_{n,k} f_{v}^{l} n \!-\! j2\pi \xi\!_{n\!,k}\! f_{\theta}^{l}  \left(Q\!_{\rm r}p_{n\!,k} \!+\! q_{\rm r}\right)} \!\!+ \!\tilde{w}^{\left(\!r\!\right)}_{n,k,q_{\rm r}}\!\!\left[\tilde{g}\right]\!,\!\!\!\!\!
	\vspace{-0.1cm}
	\label{equ:RadarRx2}
	\end{flalign}
	where $\tilde{w}_{n,k,q_{\rm r}}\left[\tilde{g}\right] := \tilde{w}_{n,k,q_{\rm r}}\left(nT_{0} + \tilde{g}T_{s}\right)$ is the discrete-time Gaussian noise. In \eqref{equ:RadarRx2}, the terms $\frac{2v_{l}nT_{0}}{c}$ and $\frac{\left(p_{n,k}Q_{\rm r} + q_{\rm r}\right)d_{\rm R}\sin\theta_{l}}{c}$ represent the range migration between different pulses and different receive antenna elements, respectively. 
	\Revise{In \ac{frac}, the coarse range resolution is determined by the bandwidth of the narrowband waveform, and equals $\frac{c}{2B_{\rm sub}}$. Here, we assume that the targets move in low speed, i.e., ${v_l N T_0} \ll \frac{c}{2B_{\rm sub}}$, and assume the waveform is narrowband, i.e. ${Q_{\rm r}P d_{\rm R}} \ll \frac{c}{2B_{\rm sub}}$  as in~\cite{Cohen2018summer}. Applying these assumptions to \eqref{equ:RadarRx2}, the migration between different pulses and different antenna elements are neglected.} %These migrations can be eliminated through range profile alignment~\cite{Zhang2011High}.
	After neglecting the migrations, the sampled signal is rewritten as 
	\vspace{-0.25cm}
	\begin{equation}
	\begin{aligned} 
	&\tilde{y}\Radar_{n,k,q_{\rm r}}\left[\tilde{g}\right] =\sum_{l=0}^{L}\tilde{\alpha}_{l} e^{-j2\pi  \frac{2\kappa r_{l} T_{s}\Radar}{c}\tilde{g}}\cdot e^{-j2\pi m_{n,k}\frac{2r_{l}\Delta f}{c}} \\ 
	& \quad \times e^{- j2\pi \xi_{n,k} f_{v}^{l} n - j2\pi \xi_{n,k} f_{\theta}^{l}  \left(Q_{\rm r}p_{n,k} + q_{\rm r}\right)} \!\!+ \!\tilde{w}\Radar_{n,k,q_{\rm r}}\left[\tilde{g}\right].
	\label{equ:RadarRx3} 
	\end{aligned}
	\vspace{-0.2cm}
	\end{equation}
	In radar detection, the task is to recover the target parameters $\left\{r_l, v_l, \theta_l\right\}_{l=0}^{L-1}$  from the received signal \eqref{equ:RadarRx3}, as discussed next.
	
	\subsubsection{Radar Processing}	
	Radar receive processing consists of two stages:  The received radar data of each pulse is first processed by pulse compression, generating a \ac{crrp}. Then, the parameters of the radar targets are recovered with  enhanced resolution from all the samples in one \ac{cpi} via \ac{cs}-based recovery. 
	%	In the sequel, the detailed algorithms are presented.
	
	The coarse range profile is generated after pulse compression through \ac{idft}, yielding $y_{n,k,q_{\rm r}}\Radar\left[g\right] = \mathrm{IDFT}(\tilde{y}_{n,k,q_{\rm r}}\Radar\left[\tilde{g}\right]) = \sum_{\tilde{g} = 0}^{G - 1}\tilde{y}_{n,k,q_{\rm r}}\Radar\left[\tilde{g}\right]e^{\frac{j2\pi\tilde{g}g}{G}}$ which equals
	\vspace{-0.25cm}
	\begin{equation}
	\begin{aligned} 
	y_{n,k,q_{\rm r}}\Radar\left[g\right] %&= \mathrm{IDFT}\left(\tilde{y}_{n,k,q_{\rm r}}\Radar\left[\tilde{g}\right]\right) \\
	&= \sum_{l=0}^{L-1}\beta_{l}\left[g\right] e^{-j2\pi m_{n,k}\frac{2r_{l}\Delta f}{c}}\cdot e^{- j2\pi \xi_{n,k} f_{v}^{l} n}  \\
	&\times e^{ - j2\pi \xi_{n,k} f_{\theta}^{l}  \left(Q_{\rm r}p_{n,k} 
		+ q_{\rm r}\right)} + w\Radar_{n,k,q_{\rm r}}\left[g\right],
	\label{equ:RadarRx4} 
	\vspace{-0.25cm}
	\end{aligned}
	\end{equation}
	where $\mathrm{IDFT}\left(\cdot\right)$ is the \ac{idft} operation over $G$ samples,
	\vspace{-0.25cm}
	\begin{equation}
	\beta_{l}\left[g\right] \!:=\! \tilde{\alpha}_{l}\! \cdot \! \mathrm{IDFT}\!\left(\!\!e^{-j2\pi \frac{2\kappa r_{l} T_{s}\Radar}{c}\tilde{g}}\!\!\right) \!= \tilde{\alpha}_{l} \!\cdot\!\! \sum_{\tilde{g}=0}^{G-1}\!e^{j2\pi\tilde{g}\left(\!\frac{g}{G}-\frac{2\kappa r_{l} T_{s}\Radar}{c}\!\right)},\notag
	\vspace{-0.2cm}
	\end{equation}
	and the additive noise $w\Radar_{n,k,q_{\rm r}}\left[g\right] := \mathrm{IDFT}(\tilde{w}\Radar_{n,k,q}\left[\tilde{g}\right])$ is still white and Gaussian as \ac{idft} is a unitary transformation.
	
	The output signal $y_{n,k,q_{\rm r}}\Radar\left[g\right]$  is referred to as the $g$th \ac{crrp} with range  resolution  $\frac{c}{2B_{\rm sub}}$.  %	In the proposed radar subsystem, %the baseband \ac{fmcw} waveforms are multiplied with different phase modulation symbols and mixed up with $K$ selected carriers. 
	%	the transmission scheme leads to a resource compression in time, frequency and antenna elements. Similar kinds of radar systems are studied, such as  \ac{far}\cite{Huang2014TAES,Huang2018Analysis,Wang2019Analysis}, SUMMeR\cite{Cohen2018summer}, TenDSuR\cite{Na2019Tendsur}, etc.
	In the sequel, we introduce a \ac{cs} based method to recover the high-range-resolution profiles, velocity, and angle of the target  utilizing this \ac{crrp}. 
	
	The parameters of the radar targets are recovered by processing the samples collected from the same coarse range cell. Samples from other coarse range cells can be processed identically and individually. Therefore, we focus here on one coarse range cell, and assume that the targets are located in the $g$th  cell without loss of generality, i.e., $r_{l}\in \big[ \frac{\left(2g - 1\right)c}{4\Delta f}, \frac{\left(2g+1\right)c}{4\Delta f}\big)$. 	
	In \eqref{equ:RadarRx4}, the term $e^{-j2\pi m_{n,k}\frac{2r_{l}\Delta f}{c}}$ also equals $e^{-j2\pi m_{n,k} \frac{2\delta_{r}\Delta f}{c}}$, where $\delta_{r} = r_{l} - \frac{gc}{2\Delta f}$, which means that the maximum unambiguous range is $\frac{c}{2\Delta f}$.  In the unambiguous range interval $\big[ \frac{\left(2g - 1\right)c}{4\Delta f}, \frac{\left(2g+1\right)c}{4\Delta f}\big)$, the normalized range frequency is denoted by  $f_{r}^{l} := \frac{2\delta_{r}\Delta f}{c}$. Substituting $f_{r}^{l}$ in \eqref{equ:RadarRx4}, we get
	\vspace{-0.25cm}
	\begin{equation}
	\begin{aligned} 
	y_{n,k,q_{\rm r}}\Radar=& \sum_{l=0}^{L-1}\beta_{l} e^{-j2\pi m_{n,k}f_{r}^{l}}\cdot e^{- j2\pi \xi_{n,k} f_{v}^{l} n} \\
	&\qquad\times e^{ - j2\pi \xi_{n,k} f_{\theta}^{l}  \left(Q_{\rm r}p_{n,k} + q_{\rm r}\right)} + w\Radar_{n,k,q_{\rm r}}.
	\label{eqn:RadarRx5} 
	\end{aligned} 
	\vspace{-0.2cm}
	\end{equation}
	The task of radar detection is to recover the range, velocity and angle of the targets, which  can be recovered by estimating the values of $\{f_{r}^{l}\}$, $\{f_{v}^{l}\}$ and $\{f_{\theta}^{l}\}$, respectively. 
	
	\Revise{To recover the radar targets, $f_r$, $f_v$ and $f_\theta$ are first discretized by intervals equal to the resolutions of the normalized range, velocity, and angle frequencies, which equal  $\frac{1}{M}$, $\frac{1}{N}$, and $\frac{1}{PQ_{\rm r}}$, respectively, according to the analysis in Subsection~\ref{subsec:RadarReso}. 
		Thus the grid sets of discretized $f_r$, $f_v$ and $f_\theta$ are denoted by $\mySet{R}:= \left\{\frac{{m}}{{M}}| {m} = 0, 1,\cdots, {M}-1\right\}$, $\mySet{V}:= \left\{\frac{{n}}{{N}}| {n} = 0, 1,\cdots, {N}-1\right\}$, ${\Theta}:= \left\{\frac{{q}}{{Q}}| {q} = 0, 1,\cdots, {Q}-1\right\}$, respectively.}  %The grid sets of the discretized range, velocity and angle frequencies are denoted by $\mySet{R}:= \left\{\frac{{m}}{{M}}| {m} = 0, 1,\cdots, {M}-1\right\}$, $\mySet{V}:= \left\{\frac{{n}}{{N}}| {n} = 0, 1,\cdots, {N}-1\right\}$, ${\Theta}:= \left\{\frac{{q}}{{Q}}| {q} = 0, 1,\cdots, {Q}-1\right\}$, respectively. 
	The received signals form into a data cube, denoted by $\myMat{Y} \in \mySet{C}^{N\times K \times Q_{\rm r}}$, where $\big[\myMat{Y}^{\Radar}\big]_{n,k, q_{\rm r}}:= y_{n,k,q_{\rm r}}^{\Radar}$. Assuming the targets are located on these grids, the target scene can be indicated by  $\myMat{B}\in \mySet{C}^{N\times M \times Q}$ with
	entries 
	\vspace{-0.2cm}
	\begin{equation}
	\left[\myMat{B}\right]_{n,m, q} := \left\{
	\begin{aligned}
	&\beta_{l}, \mathrm{if\ exists}\  \left( f_v^l, f_r^l, f_\theta^l\right) = \left( \frac{n}{N}, \frac{m}{M}, \frac{q}{Q}\right),\\
	&0, \ \qquad \mathrm{otherwise}.
	\end{aligned}		
	\right.\notag
	\vspace{-0.2cm}
	\end{equation}
	To formulate
	the radar target recovery  as a sparse recovery problem, let $\myVec{y}{\Radar}$ and $\myVec{b}$ be the vectorized representations of $\myMat{Y}{\Radar}$ and $\myMat{B}$, respectively, i.e., $\big[\myVec{y}\Radar\big]_{nKQ_{\rm r} + kQ_{\rm r} + q_{\rm r}} := \big[\myMat{Y}{\Radar}\big]_{n,k, q_{\rm r}}$ and $\myVec{b}_{nMQ + mQ + q} = \left[\myMat{B}\right]_{n, m, q}$. Following \eqref{eqn:RadarRx5}, it holds that 
	\vspace{-0.1cm}
	\begin{equation}
	\myVec{y}\Radar = \myMat{A}\myVec{b} + \myVec{w}\Radar,
	\label{eqn:RadarObseEqn}
	\vspace{-0.15cm}
	\end{equation}
	where $\myMat{A}\in \mySet{C}^{NKQ_{\rm r}\times {N}{M}{Q}}$ is the observation matrix, the entries of which are given by
	\vspace{-0.15cm}
	\begin{equation}
	\begin{aligned}
	&\left[\myMat{A}\right]_{{n}KQ_{\rm r} + kQ_{\rm r} + q_{\rm r}, \tilde{n}{M}{Q}+{m}{Q}+{q}}  \\
	&= e^{-j2\pi m_{n,k} \frac{{m}}{{M}}}\cdot e^{- j2\pi \xi_{n,k} \frac{\tilde{n}}{{N}}n - j2\pi \xi_{n,k} \frac{{q}}{{Q}}  \left(Q_{\rm r}p_{n,k} + q_{\rm r}\right)},
	\end{aligned}
	\vspace{-0.15cm}
	\end{equation}
	where $n, \tilde{n} \in \left\{0, 1, \cdots, N-1\right\}$. %\Revise{In addition to vectorizing the received signals as a vector, the received signal can also be arranged in a tensor form, and processed according to the method proposed  in~\cite{Na2019Tendsur}, which is left for future work.}
	
	The range, velocity and angle can be estimated by recovering $\myVec{b}$ from the observations \eqref{eqn:RadarObseEqn}. Assume the $({n}{M}{Q}+{m}{Q}+{q})$th entry of $\myVec{b}$ is non-zero, the range, velocity and angle of the corresponding target can be calculated as $r = \frac{c}{2\Delta f}\left(\frac{{m}}{{M}} - \frac{1}{2}\right) +\frac{gc}{2\Delta f}$, $v = \frac{c}{2T_{0}f_{c}}\left(\frac{{n}}{{N}} - \frac{1}{2}\right)$, and $\theta = \arcsin{\left\{\frac{c}{f_c d_{\rm R}}\left(\frac{{q}}{{Q}} - \frac{1}{2}\right)\right\}}$, respectively, where $v \in \left[\frac{-c}{4T_0 f_c}, \frac{c}{4T_0 f_c}\right)$ and $\theta \in \left[-\arcsin\big({\frac{c}{2f_c d_{\rm R}}}\big), \arcsin\big({\frac{c}{2f_c d_{\rm R}}}\big)\right)$. 
	%\footnote{\textcolor{red}{How exactly are the radar target parameters obtained from $\myVec{b}$? You should detail this.}}
	Solving \eqref{eqn:RadarObseEqn} is an under-determined problem as the number of rows is less than the number of columns, i.e,. $NKQ_{\rm r} \leq {N}{M}{Q}$. Due to the sparsity of $\myVec{b}$,  whose entries and sparsity pattern encapsulate the radar target parameters,  it can be recovered by solving
	\vspace{-0.15cm}
	\begin{equation}
	\min\limits_{\myVec{b}} {\Vert \myVec{b} \Vert}_{0},\  \mathrm{subject\ to}\ {\Vert {\myVec{y}\Radar - \myMat{A}\myVec{b}} \Vert} \le \epsilon,
	\label{eqn:RadarEquCS}
	\vspace{-0.15cm}
	\end{equation}	
	where $\epsilon$ is related to the noise level. 
	\ReviseReview{The optimization problem \eqref{eqn:RadarEquCS} can be solved by \ac{cs} algorithms, such as greedy approaches and $\ell_{1}$ relaxation-based optimization \cite{eldar2012compressed, de2019compressed}.  
		One can increase the speed of solving \eqref{eqn:RadarEquCS} by utilizing hardware accelerators, e.g., \acp{gpu}. Furthermore, recent advances in deep learning for \ac{cs} have shown that model-based and structured neural networks can be trained to rapidly solve problems of the form of  \eqref{eqn:RadarEquCS} \cite{shlezinger2020model}, e.g., via deep unfolding of sparse recovery algorithms~\cite{Fu2021Structured}. These indicate that identifying the targets by solving \eqref{eqn:RadarEquCS} can be carried out in real time.\label{txt:RealTime}}

	%----------------------------------------------------------------------------------------
	%	Communication Reciever ----------------------------------------------------------------------------------------
	\vspace{-0.2cm}
	\subsection{Communications Receiver}
	\label{subsec:CommuRx}
	\vspace{-0.1cm}	
	Next, we discuss how the digital message is recovered from the received \ac{dfrc} waveform. We begin by introducing the signal model, and then we present the \ac{ml} rule as well as a low complexity decoding method. 
	\subsubsection{Received Communication Signal}
	\Revise{%Compared with \ac{majorcom} which considers a flat fading channel and a single receive antenna, w
		We consider a receiver with $Q_{\rm c}$ antennas which is synchronized with the \ac{dfrc} transmitter. Each receiver observes the output of a noisy multipath channel with $I$ taps whose coefficients remain fixed during the radar \ac{cpi}. The impulse response of the channel relating the $p$th transmit antenna and the $q_{\rm c}$th receiver antenna is given by $\myVec{h}_{p,q_{\rm c}} := \big[h_{p,q_{\rm c}}^{0}, h_{p,q_{\rm c}}^{1},  \cdots, h_{p,q_{\rm c}}^{I-1} \big]^{\rm T}$.} \ReviseReview{In the proposed scheme, the channel knowledge can be typically obtained using some preliminary pilot transmission phase, which can be achieved in a \ac{dfrc} system by using some periodically pre-defined waveform patterns for this purpose. The detailed derivation of channel estimation procedures and its analysis are left for future work.\label{txt:CSIEstimate}}  
	
	After down conversion by being mixed with the carrier of frequency $f_{c}$, the signal is sampled with rate $F_s\Comm := \frac{1}{T_{s}\Comm}$, where $T_{s}\Comm$ is the sampling interval. 
	The samples received at the $q_{\rm c}$th antenna in the $n$th pulse are stacked as the vector 
	%\vspace{-0.2cm}
	%	\begin{equation}
	$\myVec{y}_{n,q_{\rm c}}\Comm := \big[y_{n,q_{\rm c}}\Comm\left[0\right], y_{n,q_{\rm c}}\Comm\left[1\right], \cdots, y_{n,q_{\rm c}}\Comm\left[U-1\right]\big]^{\rm T}$,
	%\vspace{-0.2cm}
	%	\end{equation}
	where $U := \lfloor\frac{T_{p}}{T_{s}\Comm}\rfloor$ is the number of samples received in one pulse. The $u$th entry of $\myVec{y}_{n,q_{\rm c}}\Comm$ is given by 
	\vspace{-0.2cm}
	\begin{equation}
	y_{n,q_{\rm c}}\Comm\left[u\right] \!=\! \sum_{k=0}^{K-1}\sum_{i=0}^{I-1} {h}_{p_{n,k},q_{\rm c}}^{i}{s}_{m_{n,k}}\!\left[u \!-\! i\right]\!e^{j\phi_{n,k}} \!+\! w\Comm_{n,q_{\rm c}}\!\left[u\right],
	\label{equ:CommRx1}
	\vspace{-0.2cm}
	\end{equation}
	where $u \in \left\{0, 1, \cdots, U-1\right\}$, $i\in \left\{0, 1, \cdots, I-1\right\}$; $p_{n,k}$ is the index of the $k$th transmit antenna of the $n$th pulse;  $w\Comm_{n}\left[u\right]$ is white Gaussian noise with variance $\sigma_{\rm c}^{2}$; and $\myVec{s}_{m_{n,k}} := \big[{s}_{m_{n,k}}[0], {s}_{m_{n,k}}[1], \cdots, {s}_{m_{n,k}}[U-1]\big]^{\rm T}$ 
	is the discrete-time baseband transmitted waveform whose $u$th entry is $s_{m_{n,k}}\left[u\right] := s_{m_{n,k}}\left(uT_{s}\Comm\right)$, where 
	$s_{m_{n,k}}\left(t\right) := s\left(t\right)e^{j2\pi m_{n,k} \Delta f t}$. 
	
	Using \eqref{equ:CommRx1}, $\myVec{y}_{n,q_{\rm c}}\Comm $ is rewritten  as
	\vspace{-0.15cm}
	\begin{equation}
	\myVec{y}_{n,q_{\rm c}}\Comm = \myMat{\Psi}_{q_{\rm c}}\myVec{e}_{n} + \myVec{w}\Comm_{n,q_{\rm c}},
	\label{eqn:EqInOut1}
	\vspace{-0.15cm}
	\end{equation}
	where $\myVec{e}_{n}\in \mySet{E} \subset (\mySet{J}\cup \{0\}) ^{PM}$ denotes the transmit symbol vector of the $n$th pulse in the \ac{dfrc} transmitter, and $\mySet{E}$ is the set of transmit symbol vectors. The selection of carrier frequencies, the subset of transmit antennas, the waveform permutation pattern, and the modulated phases can be obtained from $\myVec{e}_{n}$. The structure of $\myVec{e}_{n}$ is given by
	%	\vspace{-0.2cm}
	%	\begin{equation}
	$\myVec{e}_{n} = \big[(\myVec{e}_{n}^{1})^{\rm T}, (\myVec{e}_{n}^{2})^{\rm T}, \cdots, 
	(\myVec{e}_{n}^{M})^{\rm T}\big]^{\rm T}$,
	%	\vspace{-0.2cm}
	%	\end{equation}
	where each $P\times1$ vector $\myVec{e}_{n}^{m}$  is either all zero  or  has one non-zero entry at the $p$th entry if there exist $m_{n,k} = m$ and $p_{n,k} = p$. The value of this nonzero entry is $[e_{n}^{m}]_{p} = e^{j\phi_{n,k}}$. 
	The vector $\myVec{w}\Comm_{n,q_{\rm c}} := \big[w\Comm_{n,q_{\rm c}}\left[0\right], \cdots, w\Comm_{n,q_{\rm c}}\left[U-1\right]\big]^{\rm T}$ represents the additive noise. The matrix $\myMat{\Psi}_{q_{\rm c}} \in \mySet{C}^{U\times PM}$ is comprised of the set of $U\times P$ matrices $\{\myMat{\psi}_{q,m}\}$ via   $\myMat{\Psi}_{q_{\rm c}} := \left[\myMat{\psi}_{q_{\rm c},0}, \myMat{\psi}_{q_{\rm c},1}, 
	\cdots, \myMat{\psi}_{q_{\rm c},M-1}\right]$. Each sub-matrix $\myMat{\psi}_{q_{\rm c},m}$ is given by $\myMat{\psi}_{q_{\rm c},m} := \big[\myMat{\psi}_{q_{\rm c},m}^{0},  \myMat{\psi}_{q_{\rm c},m}^{1},  \cdots, 
	\myMat{\psi}_{q_{\rm c},m}^{P-1}\big]$, 
	where $\myMat{\psi}_{q_{\rm c},m}^{p} := \myMat{H}_{p,q_{\rm c}}\myVec{s}_{m}$, and $\myMat{H}_{p,q_{\rm c}}\in \mySet{C}^{U\times U}$ denotes the channel response matrix between the $p$th transmit antenna element and the $q_{\rm c}$th receive antenna of the communications receiver, 
	%\vspace{-0.15cm}
	\begin{equation}
	\!\!\!\!{{\myMat{H}}_{p,q_{\rm c}}} \!: = \!\left[ {\begin{array}{*{20}{c}}
		{{h_{p,q_{\rm c}}^{0}}}&0&0&0& \cdots &0\\
		{{h_{p,q_{\rm c}}^{1}}}&{{h_{p,q_{\rm c}}^{0}}}&0&0& \cdots &0\\
		\vdots & \vdots & \ddots &0& \cdots &0\\
		{{h_{p,q_{\rm c}}^{I-1}}}&{{h_{p,q_{\rm c}}^{I-2}}}& \cdots &{{h_{p,q_{\rm c}}^{0}}}& \cdots &0\\
		\vdots & \vdots & \vdots & \vdots & \ddots &0\\
		0& \cdots &{{h_{p,q_{\rm c}}^{I-1}}}&{{h_{p,q_{\rm c}}^{I-2}}}& \cdots &{{h_{p,q_{\rm c}}^{0}}}
		\end{array}} \right]\!.	
	\label{equ:CommuChannelMtx}	
	%\vspace{-0.15cm}
	\end{equation}
	
	The signal received in all the antennas is stacked as the vector 
	%$\myVec{y}_{n}\in \mySet{C}^{UQ_{\rm c}}$,  i.e.,
	%		\begin{equation}
	$\myVec{y}_{n}\Comm := \big[\big(\myVec{y}\Comm_{n, 0}\big)^{\rm T}, \big(\myVec{y}\Comm_{n, 1}\big)^{\rm T}, \cdots, \big(\myVec{y}\Comm_{n, Q_{\rm c} - 1}\big)^{\rm T}\big]^{\rm T}$.
	%		\label{eqn:RecModel}
	%		\end{equation}
	By defining  $\myMat{\Psi}:=\big[\myMat{\Psi}_{0}^{\rm T}, \myMat{\Psi}_{1}^{\rm T}, \cdots, \myMat{\Psi}_{Q_{\rm c} - 1}^{\rm T}\big]^{\rm T}$, and $\myVec{w}_{n}:= \big[\myVec{w}_{n,0}^{\rm T}, \myVec{w}_{n,1}^{\rm T}, \cdots, \myVec{w}_{n,Q_{\rm c}-1}^{\rm T}\big]^{\rm T}$, the received signal  is given by 
	\vspace{-0.2cm}
	\begin{equation}
	\myVec{y}\Comm_{n} = \myMat{\Psi} \myVec{e}_{n} + \myVec{w}_{n}.
	\vspace{-0.2cm}
	\end{equation}
	The received signal model in \eqref{eqn:EqInOut1} is used next to formulate the decoding algorithm for recovering $\myVec{e}_{n}$ from $\myVec{y}_{n}\Comm$.
	
	\subsubsection{Communications Decoder}
	%In order to decode the transmitted message, the receiver should detect the selected carrier frequencies, the active transmit antenna elements, the waveform permutation patterns, and the modulated phases, which can all be obtained through the estimation of $\myVec{e}_{n}$ in \eqref{eqn:EqInOut1}.
	\Revise{%Compared with \ac{majorcom} which only conveys information bits through \ac{im}, \ac{frac} embeds message both through \ac{im} and \ac{pm}. 
		In order to decode the transmitted message, the receiver should detect the selected carrier frequencies, the active transmit antenna elements, the waveform permutation patterns, and the modulated phases, which can all be obtained through the estimation of $\myVec{e}_{n}$ in \eqref{eqn:EqInOut1}.} %Thus the index combinations of sub-carriers and antenna elements, as well as the modulated phases are all encapsulated in ${\myVec{e}}_{n}$.} 
	Here, we first introduce the \ac{ml} rule for recovering $\myVec{e}_{n}$. Since \ac{ml} recovery may be computationally prohibitive, we also propose a \ac{sod} algorithm with reduced complexity building upon the inherent orthogonality of the  \ac{mimo} radar waveforms.
	
	\emph{\ac{ml} Algorithm:} Since the receiver has full \ac{csi}, i.e., knowledge of the matrix $\myVec{\Psi}$ and the distribution of $\myVec{w}^{\left(c\right)}$, it can compute the error probability minimizing \ac{ml} decoder, i.e.,
	\vspace{-0.2cm}
	\begin{equation}
	\label{equ:ML}
	\hat{\myVec{e}}_{n} = \mathop {\arg \max}\limits_{\myVec{e}_{n} \in \mySet{E}} p\left(\myVec{y}^{\left(c\right)}_{n}|\myVec{e}_{n},\myMat{\Psi}\right).
	\vspace{-0.2cm}
	\end{equation}
	As the noise obeys a proper-complex white Gaussian distribution, 
	it holds that \eqref{equ:ML} specializes to the minimum distance detector 
	\vspace{-0.15cm}
	\begin{equation}
	\label{equ:ML2}
	\hat{\myVec{e}}_{n} =  \mathop {\arg \min}\limits_{\myVec{e}_{n} \in \mySet{E}} \|\myVec{y}^{\left(c\right)}_{n}-\myMat{\Psi}\myVec{e}_{n}\|^2_{2}.
	\vspace{-0.15cm}
	\end{equation}
	Recovering ${\myVec{e}}_{n}$ via \eqref{equ:ML} generally involves searching over the set $\mySet{E}$ whose cardinality is $2^{N_{\rm Total}}$. To facilitate decoding, we next propose the \ac{sod}-based method with reduced complexity by leveraging the orthogonality of the transmit waveforms.
	
	\emph{\acs{sod} Algorithm:} The high computational complexity of the \ac{ml} detector follows from the need to search over the entire set $\mySet{E}$. To decrease the complexity of detection, we propose a \acl{sod} detector, which decodes in two steps: First, the carrier frequencies of each transmit waveform are estimated through matched filtering. Then, the transmit antenna elements and the symbols modulated on the waveforms are detected, as detailed next.
	
	Define $\myVec{\psi}_m^p:= \big[\big(\myVec{\psi}_{0,m}^{p}\big)^{\rm T}, \big(\myVec{\psi}_{1,m}^{p}\big)^{\rm T}, \cdots, \big(\myVec{\psi}_{Q_c - 1,m}^{p}\big)^{\rm T} \big]^{\rm T}$, which is the $\big(\left(m-1\right)P + p\big)$th column of $\myMat{\Psi}$. The received signal $\myVec{y}_n\Comm$ is first matched with $MP$ normalized filters. Each filter is given by  $\tilde{\myVec{\psi}}_{m}^{p} := \frac{\myVec{\psi}_{m}^{p}}{{\Vert\myVec{\psi}_{m}^{p}\Vert_{2}}}$.  Define $\myMat{\psi}_{m}:= \left[\myMat{\psi}^{\rm T}_{0,m}, \myMat{\psi}^{\rm T}_{1,m},\cdots, \myMat{\psi}^{\rm T}_{Q_{\rm c}-1,m}\right]^{\rm T}$. 
	The output of each matched filter is denoted by  ${g}_{m}^{p} := \big(\tilde{\myVec{\psi}}_{m}^{p}\big)^{\rm H}\myVec{y}_{n}^{\left(c\right)}$.
	As the transmit waveforms are orthogonal to each other when $m \ne m{'}$, i.e.,  $\big(\tilde{\myVec{\psi}}_{m}^{p}\big)^{\rm H}\myVec{\psi}_{m{'}} = \myVec{0}$, the output is expressed as
	\vspace{-0.15cm}
	\begin{equation} 
	\begin{aligned}
	{g}_{m}^{p} =  \big(\tilde{\myVec{\psi}}_{m}^{p}\big)^{\rm H}\myMat{\psi}_{m}\myMat{e}_{n}^{m} + \big(\tilde{\myVec{\psi}}_{m}^{p}\big)^{\rm H}\myVec{w}^{\left(c\right)}_{n}.
	\end{aligned}
	\vspace{-0.15cm}
	\end{equation}
	Let $g_{m}^{\rm max}$ be the maximum amplitude output, i.e., $g_{m}^{\rm max} :=\max_p{g}_{m}^{p}$. To identify the used carriers, the matched outputs $g_{m}^{\rm max}$ are sorted by descending order according to their amplitudes. %, i.e. sorted according to the received power of each waveform $s_{m}\left(t\right)$. 
	Denoting the $K$ largest outputs that belong to the transmit carrier set as $\{ {g}_{m^{'}_{0}}^{\rm max},{g}_{m^{'}_{1}}^{\rm max}, %\cdots, \myVec{g}_{m{'}_{k}}, 
	\cdots, {g}_{m^{'}_{K-1}}^{\rm max}\}$, the estimated carriers,  denoted  $\{\hat{m}_{n,k}\}$,  are obtained via $\hat{m}_{n,k} = m^{'}_{k}$.
	
	After determining the transmit frequencies, the receiver needs to further estimate the transmit antenna element and the phase symbols modulated on each waveform. Once the carrier frequencies of the transmit waveforms have been found, detection of the transmit antenna elements and the \ac{pm} symbols is carried out by searching over the  subset $\mySet{E}_{\rm sub} \subset \mySet{E}$, which is composed of all $\myVec{e}_{n}$ corresponding to the transmission patterns that transmit waveforms with the detected carrier frequencies. The estimation is expressed as
	\vspace{-0.2cm}
	\begin{equation}
	\begin{aligned}
	\hat{\myVec{e}}_{n} = \mathop{\arg \min} \limits_{\myVec{e}_{n} \in \mySet{E}_{\rm sub}} {\Vert \myVec{y}_{n}\Comm - \myVec{\Psi}\myVec{e}_{n} \Vert}_{2}^{2}.
	\end{aligned}
	\vspace{-0.2cm}
	\end{equation}
	Neglecting the computation in the first step, the computational complexity is reduced by a factor of ${2^{\lfloor \log_{2}\binom{M}{K}\rfloor}}$ compared to  the \ac{ml} rule, as the cardinality of $\mySet{E}$ is ${2^{\lfloor \log_{2}\binom{M}{K}\rfloor}}$ times the cardinality of $\mySet{E}_{\rm sub}$.

	\iffalse
	%----------------------------------------------------------------------------
	% Guidelines of the Parameter Design
	%---------------------------------------------------------------------------------------
	%\subsection{Guidelines of the Parameter Design}
	%\label{subsec:ParameterDesign}
	
	%----------------------------------------------------------------------------
	% Discussion
	%---------------------------------------------------------------------------------------
	\subsection{Discussion}
	\label{subsec:Discussion}
	Compared to military applications, vehicular applications have more strict constraints in the power consumption, hardware complexity and cost. To design a \ac{dfrc} system for vehicular application, these consideration should be take into consideration. In \ac{frac}, a spare \ac{mimo} array and narrowband waveforms are utilized, which reduces the number of \ac{rf} modules, decreases the sampling rate and the data volume in the radar receiver. To improve the abilities for conveying information, additional information bits are conveyed on the index modulation. 
	
	\textcolor{red}{Not finished.}
	\fi 
	
	% TODO NIR CONTINUE FROM HERE
	%-----------------------------------------------------------------------------	-----------
	%	Performance analyisis
	%----------------------------------------------------------------------------------------
	%\vspace{-0.2cm}
	\section{Performance Analysis of Radar Subsystem}
	%\vspace{-0.1cm}
	\label{sec:PerformanceRadar}
	We now analyze the performance of \ac{frac}. The information embedding mechanism detailed in Subsection~\ref{subsec:TxStrategy} utilizes a combination of \ac{pm} and \ac{im}, and thus its fundamental limits can be obtained from the existing communication literature on \ac{im}, e.g., \cite{Younis2018Information}. For this reason, we focus our theoretical analysis on the radar subsystem, while the  communications performance is numerically evaluated in Subsection~\ref{subsec:CommSimu}.  In the following, we first characterize the radar ambiguity function in Subsection~\ref{subsec:RadarReso}, which shows that the range/velocity/angle resolutions of \ac{frac} are similar to that of radar systems transmitting wideband waveforms and receiving with the equivalent aperture size as the virtual aperture formulated by  \ac{frac}. Then, we study the relationship between the maximum number of recoverable targets and the waveform parameters using the theory of phase transition in Subsection~\ref{subsec:PhaseThreshold}.
	
	%\vspace{-0.2cm}
	\subsection{Ambiguity Function}
	%\vspace{-0.1cm}
	\label{subsec:RadarReso}
	The ambiguity function  is a useful measure for characterizing the radar resolution~\cite[Ch. 4]{Richards2014fundamentals}. For mathematical convenience,  we adopt the relative narrow  bandwidth assumption
	as in~\cite{Huang2018Analysis}, i.e., $\xi_{n,k}\approx 1$, such that \eqref{eqn:RadarRx5} becomes 
	\vspace{-0.15cm}
	\begin{equation}
	\begin{aligned}
	y_{n,k,q_{\rm r}}\Radar=& \sum_{l=0}^{L-1}\beta_{l} e^{-j2\pi m_{n,k}f_{r}^{l}}\cdot e^{- j2\pi n f_{v}^{l} } \\
	&\qquad\qquad\times e^{ - j2\pi    \left(Q_{\rm r}p_{n,k} + q_{\rm r}\right)f_{\theta}^{l} } + w\Radar_{n,k,q_{\rm r}}.
	\label{eqn:RadarRx5Approx}
	\end{aligned}
	\vspace{-0.15cm}
	\end{equation}	
	The ambiguity function is defined as the correlation of the noiseless received signal whose parameters are $\{\beta_l = 1, f_{r}, f_{v}, f_{\theta}\}$ and the reference signal whose parameters are $\{\beta_l = 1, f_{r}^{\rm ref}, f_{v}^{\rm ref}, f_{\theta}^{\rm ref}\}$.  Thus, the expression of the ambiguity function, denoted by  $\chi\left(\delta \! f_r, \delta \! f_v, \delta \! f_\theta\right)$, is written as 
	\vspace{-0.15cm}
	\begin{equation}
	\begin{aligned} 
	\chi\left(\delta \! f_r, \delta \! f_v, \delta \! f_\theta\right) &= \sum_{n=0}^{N-1}\sum_{k=0}^{K-1}\sum_{q=0}^{Q_{\rm r}-1} e^{-j2\pi m_{n,k} \delta \! f_r} \\ 
	&\qquad \times e^{- j2\pi n \delta \! f_v } \cdot e^{ - j2\pi  \left(Q_{\rm r}p_{n,k} + q\right) \delta \! f_\theta },
	\label{eqn:AmbFun} 
	\vspace{-0.15cm}
	\end{aligned}
	\end{equation}
	where $\delta \! f_r \!:= \!f_{r} - f_{r}^{\rm ref}$, $\delta \! f_v \!:=\! f_{v} - f_{v}^{\rm ref}$, and $\delta \! f_\theta \!:=\! f_{\theta} - f_{\theta}^{\rm ref}$. 
	
	Since the active antenna elements and  the selected sub-carriers are determined by the transmitted messages, the  ambiguity function is a random quantity which takes a different realization on each \ac{pri}. 
	\Revise{To study the performance of the radar subsystem, we evaluate the stochastic properties of the ambiguity function of \ac{frac}, as done in~\cite{Huang2018Analysis, hu2014randomized}. It is emphasized that these previous studies  considered different agile radar transmission schemes, while focusing on two-dimensional ambiguity functions, i.e., the range-velocity ambiguity function~\cite{Huang2018Analysis} and the delay-spatial ambiguity function~\cite{hu2014randomized,Ma2021Spatial}. In the sequel, we extend this approach to study the stochastic properties for a three-dimensional ambiguity function, accounting for how \ac{frac} combines random selections in both the sub-carriers and the active antenna elements.}
	
	We first characterize the expected ambiguity function, which is approached by the averaged ambiguity function over a large number of \acp{cpi}. The expected ambiguity function is given in the following theorem:
	\begin{theorem}
		\label{thm:ExpctatedAmbFun}
		The absolute value of the expected ambiguity function \eqref{eqn:AmbFun} of \ac{frac} is 
		\vspace{-0.12cm}
		\begin{equation}
		\begin{aligned} 
		&\left|\mathbb{E}\big\{\chi\left(\delta \! f_r, \delta \! f_v, \delta \! f_\theta\right) \big\}\right|  \\
		&\!\!\!\! = \!\frac{K}{MP} \left|\frac{\sin\left(M\pi \delta \! f_r\right)}{\sin\left(\pi \delta \! f_r\right)}\right| \!\cdot\! \left|\frac{\sin\left(N\pi \delta \! f_v\right)}{\sin\left(\pi \delta \! f_v\right)} \right| \!\cdot \! \left|\frac{\sin\left(PQ_{\rm r}\pi \delta \! f_\theta\right)}{\sin\left(\pi \delta \! f_\theta\right)}\right|. \!\!\!
		\label{eqn:ExpctatedAmbFun}
		\end{aligned}
		\vspace{-0.12cm}
		\end{equation}
	\end{theorem}
	\begin{IEEEproof}
		The proof is given in Appendix~\ref{app:Proof1}. 	
	\end{IEEEproof}	
	The expectation in \eqref{eqn:ExpctatedAmbFun} is carried out with respect to the random indices of the random antenna elements and sub-carriers, i.e., $\{m_{n,k}\}$ and $\{p_{n,k}\}$. These indices are determined by the transmit bits, which are assumed to be i.i.d.. It follows from the law of large number that as the number of \ac{pri} grows, 
	the average transmit beam pattern approaches its expected value with probability one \cite[Ch. 8.4]{papoulis2002probability}.
	Consequently, after a large number of \acp{pri}, the magnitude of the average transmit beam pattern coincides with \eqref{eqn:ExpctatedAmbFun}.  Furthermore, we numerically demonstrate in Section~\ref{sec:Sims} that the expected ambiguity function of \ac{frac} is similar to the ambiguity function computed for a finite number of pulses (the exact parameters are listed in TABLE~\ref{tab:ParameterValue}), which can be observed in Fig.~\ref{fig:AmbguityFun}. In typical vehicular applications, the number of pulses can be even larger than the simulated settings~\cite[Tab. 1]{ProgramChirpTI}. Therefore, the expected ambiguity function can be a useful tool to indicate the radar resolution in practical \ac{dfrc} applications. %also in the horizon of a finite number of pulses.
	
	The resolutions are calculated as half the width of the first two null points in each dimension. Thus, from \eqref{eqn:ExpctatedAmbFun}, we obtain that $\Delta f_r = \frac{1}{M}$, $\Delta f_v = \frac{1}{N}$, and $\Delta f_{\theta} = \frac{1}{PQ_{\rm r}}$. We next transfer $\Delta f_r$, $\Delta f_{v}$ and $\Delta f_{\theta}$ to the range resolution, velocity resolution and angle resolution, denoted by $\Delta r$, $\Delta v$ and $\Delta \theta$, respectively. According to the definition of $\Delta f_{r}$, $\Delta f_{v}$ and $\Delta f_{\theta}$, we obtain 
	\vspace{-0.15cm}
	\begin{equation}
	\label{eqn:Resolution}
	\left\{
	\begin{aligned}
	\Delta r &= \frac{c \Delta f_r}{2\Delta f} = \frac{c}{2M\Delta f},\\
	\Delta v &= \frac{c \Delta f_v}{2T_{0}f_{c}} =  \frac{\lambda }{2NT_{0}},\ \text{and} \\
	\Delta \theta & = \arcsin\left(\frac{c \Delta f_\theta}{f_{c}d_{\rm R}}\right) = \arcsin\left(\frac{\lambda }{PQ_{\rm r}d_{\rm R}}\right),
	\end{aligned}
	\right.	
	\vspace{-0.15cm}
	\end{equation}
	where $\lambda := \frac{c}{f_c}$ is the wavelength. From \eqref{eqn:Resolution}, we find that the resolutions of \ac{frac} are equal to that of wideband \ac{mimo} radar, whose bandwidth is $M\Delta f$, and utilizes $PQ_{\rm r}$ antenna elements~\cite[Ch. 2.10]{Wehner1995High}.  %\Verify{can we cite a reference on this? can we state that the ambiguity function in Theorem~\ref{thm:ExpctatedAmbFun} is the same up to a constant factor of $\frac{K}{MP}$, which encapsulates the usage of a subset of the array and the available band?}. 
	This indicates that the randomization in the antenna selection and the frequency division, which \ac{frac} utilizes to embed communication messages and reduce hardware complexity, also contributes to its resolution.
	
	\vspace{-0.15cm}
	\subsection{Phase Transition Threshold}
	\label{subsec:PhaseThreshold}
	\vspace{-0.1cm}
	%In \ac{frac}, the \ac{dfrc} waveform is sparse in the frequency domain and antenna domain.\VerifyT{This sentence is misleading.} 
	For a specific scenario, the parameters of a \ac{dfrc} system should be carefully designed in order to satisfy both the radar requirements, such as target recovery performance, and the  communication demands, e.g., achievable rate. Here, we analyze the relationship between the waveform parameters and the radar target recovery performance, particularly, focusing on the maximum number of recoverable targets. We utilize  phase transition analysis, which emerges in many convex optimization problems, and have been adopted to characterize  sparse recovery  in~\cite{amelunxen2014living}. This motivates the usage of phase transition for analyzing \ac{frac}, whose radar receiver operation is formulated as a \ac{cs} problem  \eqref{eqn:RadarEquCS}. 
	
	In \ac{cs}, phase transition  threshold divides the plane of parameters into regions where  recovery succeeds and fails with high probability. Although existing results of phase transition are focused on Gaussian matrices, numerical simulations have been utilized to show that such an analysis also reflects on radar systems whose recovery can be expressed as \ac{cs} with structured measurement matrices~\cite{Li2021Phase}. %Moreover, approximations of the transition curves have been derived by elementary functions, which facilitates predicting the recovery performance based on this simplified analysis~\cite{Li2021Phase}. 
	In the sequel, we study the phase transitions of \ac{frac}, which extends \ac{far} considered in \cite{Li2021Phase} to include spatial agility and \ac{fmcw} signalling. By doing so, we characterize the  number of recoverable targets as a function of the  waveform parameters.
	
	% Recently, phase transition is utilized to analyze the recovery performance of the \ac{far} in terms of the waveform parameters and the target scenario in~\NoCite. Finding analytical expression of these curves in an active area. For standard sparse recovery, bounds on the phase transition curve of $l_1$ norm minimization under standard Gaussian matrices were established in \NoCite. Generalization to block sparse recovery and for complex number form were given in \NoCite. However, there approximate results assume the observation matrices to be large, and have complicated form, resulting in difficulty to apply in practice. A more concise and tight bound, which has no requirement on the size of the observation matrix, was given in \NoCite using integral geometry. Nevertheless, it is confined to standard sparse recovery under real-valued Gaussian matrices. 
	
	%We begin by calculating the phase transition threshold under the Gaussian matrix condition, then demonstrate the applicability of the derived phase transition curve to the proposed radar subsystem through simulations in Subsection~\ref{subsec:RadarSimu}.
	
	We begin by providing some preliminaries on the phase transition in the context of  \ac{cs}. Consider an under-determined problem $\myVec{y} = \myMat{\Omega}\myVec{x}$, where $\myVec{y}\in \mySet{C}^{N_1}$, $\myMat{\Omega}\in \mySet{C}^{N_1\times N_2}$ is a complex Gaussian matrix, and $\myVec{x} \in \mySet{C}^{N_2}$ is a sparse vector with $N_s$ nonzero entries. \ac{cs} recovers $\myVec{x}$ by solving the following optimization problem
	\vspace{-0.15cm}
	\begin{equation}
	\begin{aligned}
	\min\limits_{\myVec{x}} {\Vert \myVec{x} \Vert}_{0},\  \mathrm{subject\ to}\ {\myVec{y} = \myMat{\Omega}\myVec{x}},
	\label{eqn:CSGaussianl0}
	\end{aligned}
	\vspace{-0.15cm}
	\end{equation}
	which is typically relaxed into an
	%Since the $l_0$ norm minimization in \eqref{eqn:CSGaussianl0} is NP-hard, under certain conditions \eqref{eqn:CSGaussianl0} can be exactly solved by
	$\ell_1$ norm minimization~\cite{eldar2012compressed}, 
	\vspace{-0.15cm}
	\begin{equation}
	\begin{aligned}
	\min\limits_{\myVec{x}} {\Vert \myVec{x} \Vert}_{1},\  \mathrm{subject\ to}\ {\myVec{y} = \myMat{\Omega}\myVec{x}}.
	\label{eqn:CSGaussian}
	\end{aligned}
	\vspace{-0.15cm}
	\end{equation}
	
	Problem \eqref{eqn:CSGaussian} is convex, and thus its recovery limits can be characterized using the phase transition theory. The phase transition threshold of~\eqref{eqn:CSGaussian} indicates the maximum sparse degree, denoted by $N_s^{\ast}$, for a given $N_1$ and $N_2$. This threshold is the demarcation point that separates the successful recovery and failing recovery with high probability. Namely, when  $N_s \le N_s^{\ast}$, then $\myVec{x}$ can be recovered with high probability, while when $N_s > N_s^{\ast}$, the probability of exact recovery dramatically decreases. 	
	The phase transition threshold of solving \eqref{eqn:CSGaussian} can be computed via the following lemma:
	\begin{lemma}
		\label{lem:CSGaussian}
		The sparse  $\myVec{x}$ can be exactly recovered via \eqref{eqn:CSGaussian} with high probability, when $N_s\le N_s^{\ast}$, where $N_s^{\ast}$ is related to $N_1$ and $N_2$  via
		\vspace{-0.1cm}
		\begin{equation}
		\begin{aligned}
		\!\!\!\! N_1  \!\!=\!\! \mathop{\inf} \limits_{\beta \ge 0}\frac{1}{2}\Big\{ \!N_s^{\ast}\!\left(2 \!+\! \beta^{2}\right) \!\!+\! \left(N_2 \!-\! N_s^{\ast}\right)\!\!\int_{\beta}^{\infty}\!\!\!\!\!\!\left(u \!-\! \beta\right)^2 \!\phi_2\left(u\right){\rm d}u\!\Big\}\!,\!\!
		\label{eqn:PhaseTransCSGaussian}		 		
		\end{aligned}
		\vspace{-0.1cm}
		\end{equation}
		where $\phi_2\left(u\right) = ue^{-u^2/2}$.
	\end{lemma}
	\begin{IEEEproof}
		Lemma~\ref{lem:CSGaussian} follows from combing \cite[Pro. 3]{Li2021Phase} and \cite[Eqn. 24]{Li2021Phase}.
	\end{IEEEproof}
	
	We next specialize Lemma~\ref{lem:CSGaussian} to the target recovery procedure of \ac{frac}. In particular, we consider a noiseless scenario, as commonly assumed in such studies~\cite{amelunxen2014living, foucart2013sparse,donoho2006high}, %\NoCite\VerifyT{Cite some typical approaches from \cite{Li2021Phase}},
	i.e., the noise term in \eqref{eqn:RadarObseEqn} is omitted. In this case, the target recovery problem formulated in \eqref{eqn:RadarEquCS}, tackled via $\ell_1$ norm minimization,  
	% 	turns into
	% 	\begin{align}
	% 	\min\limits_{\myVec{b}} {\Vert \myVec{b} \Vert}_{0},\  \mathrm{subject\ to}\ {\myVec{y}\Radar = \myMat{A}\myVec{b}}.
	% 	\label{eqn:RadarEqnCS1}
	% 	%\vspace{-0.15cm}
	% 	\end{align}	
	% 	Problem \eqref{eqn:RadarEqnCS1} can be solved by $l_1$ norm minimization~\cite{eldar2012compressed}, which 
	is given by the convex problem
	\vspace{-0.15cm}
	\begin{equation}
	\begin{aligned}
	\min\limits_{\myVec{b}} {\Vert \myVec{b} \Vert}_{1},\  \mathrm{subject\ to}\ {\myVec{y}\Radar = \myMat{A}\myVec{b}}.
	\label{eqn:RadarEqnCS2}
	\end{aligned}
	\vspace{-0.15cm}
	\end{equation}
	%Since \eqref{eqn:RadarEqnCS2} is a convex optimization problem, the target recovery performance can be analyzed using the phase transition theory when the target parameters are recovered by \eqref{eqn:RadarEqnCS2}. 
	%In this analysis, the range frequency, velocity frequency, and angle frequency are discretized according to the resolutions. %, which yield  $\delta f_r = \frac{1}{M}$, $\delta f_{v} = \frac{1}{N}$ and $\delta f_\theta = \frac{1}{PQ_{\rm r}}$ according to \eqref{eqn:ExpctatedAmbFun}. 
	%Thus the number of discretized grids in each dimension are $\tilde{M} = M$, $\tilde{N} = N$ and $\tilde{Q} = PQ_{\rm r}$, and the measurement matrix $\myMat{A} \in \mathcal{C}^{NKQ_{\rm r}\times NMPQ_{\rm r}}$. 
	
	The main difference between \eqref{eqn:CSGaussian}, used in Lemma~\ref{lem:CSGaussian}, and \eqref{eqn:RadarEqnCS2} which represents \ac{frac}, is in the structure of measurement matrices.  
	Although phase transition results such as  Lemma~\ref{lem:CSGaussian} assume  Gaussian measurements, it was numerically shown that the analysis for Gaussian matrices is also accurate for structured measurements which arise in \ac{far}~\cite{Li2021Phase}. %Inspired by this, we also exploit the theory of phase transition to analyze the radar performance of \ac{frac}. 
	
	The phase transition threshold of \eqref{eqn:RadarEqnCS2} indicates the maximum number of exactly recovered targets, denoted by $L^{\ast}$, for given waveform parameters, i.e., pulse number $N$, number of active antennas $K$, amount of available transmit antennas $P$, number of sub-carriers $M$, and amount of radar receive antennas $Q_{\rm r}$. %This threshold is the demarcation point that separates the successful recovery and failing recovery with high probability. 
	In particular, when the number of targets obeys $L \le L^{\ast}$, then they can be exactly recovered with high probability, while when $L > L^{\ast}$, the probability of exact recovery  dramatically decreases. The relationship between this threshold and the \ac{dfrc} waveform $\left\{N, K, M, P, Q_{\rm r}\right\}$ is obtained from Lemma~\ref{lem:CSGaussian} as stated in the following corollary:
	%Inspired by~\cite{Li2021Phase}, which directly utilized the phase transition theory based on Gaussian to analyze the performance of \ac{far}, Currently, there is no theoretical evidence that Proposition~\ref{pro:PhaseTransition1} holds for the measurement matrix which has a structure given by  \eqref{eqn:MeasureMatrix}.	
	\begin{corollary}
		\label{cor:specialcase}
		For a given $\left\{N, K, M, P, Q_{\rm r}\right\}$, the phase transition threshold  $L^{\ast}$ for \eqref{eqn:RadarEqnCS2} with Gaussian measurements $\myMat{A}$ satisfies
		\vspace{-0.15cm}
		\begin{equation}
		\begin{aligned} 
		NKQ_{\rm r} & = \mathop{\inf} \limits_{\beta \ge 0}\frac{1}{2}\bigg\{ L^{\ast}\left(2 + \beta^{2}\right) \\
		&  \qquad + \left(NMPQ_{\rm r} \!-\! L^{\ast}\right)\!\!\int_{\beta}^{\infty}\!\!\!\!\left(u \!-\! \beta\right)^2 \phi_2\left(u\right){\rm d}u\bigg\}.
		\end{aligned}
		\label{eqn:PhaseTransition1}
		\vspace{-0.15cm}
		\end{equation}
	\end{corollary}
	% 	Given parameters of the \ac{dfrc} waveform, i.e., , the phase transition threshold $L^{\ast}$ can be obtained by substituting the parameters $N_1$, $N_2$, and $N_s$ in \eqref{eqn:PhaseTransCSGaussian} by $NKQ_{\rm r}$, $NMPQ_{\rm r}$, and $L^{\ast}$, respectively. Consequently, the relationship between the maximum number of recoverable targets and the waveform parameters follows
	% 	\begin{align} 
	% 	NKQ_{\rm r} & = \mathop{\inf} \limits_{\beta \ge 0}\frac{1}{2}\bigg\{ L^{\ast}\left(2 + \beta^{2}\right) \notag\\
	% 	&  \qquad + \left(NMPQ_{\rm r} \!-\! L^{\ast}\right)\!\!\int_{\beta}^{\infty}\!\!\!\!\left(u \!-\! \beta\right)^2 \phi_2\left(u\right){\rm d}u\bigg\}.
	% 	\label{eqn:PhaseTransition1}
	% 	\end{align}
	%where $\phi_2\left(u\right) = ue^{-u^2/2}$.\VerifyT{No need to use the subscript 2. No need to define it again since it appears in Lemma 1.} 
	The threshold characterized in \eqref{eqn:PhaseTransition1} is numerically shown to be tight in Subsection~\ref{subsec:RadarSimu}. Due to the tightness, Corollary \ref{cor:specialcase}  can be used for guiding waveform design. 
	
	%	Substituting $\myVec{y}$, $\myMat{\Omega}$ and $\myVec{x}$ in  \eqref{eqn:PhaseTransition1_2} by $\myVec{y}^{\left(r\right)}$, $\myMat{A}$, and  $\myMat{b}$, respectively, we transfer \eqref{eqn:PhaseTransition1_2} to \eqref{eqn:RadarEqnCS2}. Thus \eqref{eqn:PhaseTransition1} is proved by substituting the parameters $N_1$, $N_2$, and $N_s$ in \eqref{eqn:PhaseTransition1_2} by $NKQ_{\rm r}$, $NMPQ_{\rm r}$, and $L^{\ast}$, respectively.

	%Specified to our analysis, the phase transition threshold indicates the maximum number of exactly recovered targets, denoted by $L^{\ast}$, for given waveform parameters, i.e., the pulse number $N$, the number of active antennas $K$, the number of sub-carriers $M$, and the number of radar receive antennas $Q_{\rm r}$. This threshold is the demarcation point that separates the successful recovery and failing recovery with high probability. It means when the number of targets obeys $L \le L^{\ast}$, the radar targets can be exactly recovered with high probability, while when $L > L^{\ast}$, the probability of exact recovery  dramatically decreases.

	The phase transition threshold  $L^{\ast}$ is numerically computed using \eqref{eqn:PhaseTransition1} for a given set of \ac{dfrc} parameters $\left\{N, K, M, P, Q_{\rm r}\right\}$. To facilitate the waveform design, a more explicit  relationship between $L^{\ast}$ and the waveform parameters is required. To this aim, we next approximate \eqref{eqn:PhaseTransition1} under a quantitative assumption on the relationship between the parameters in the following proposition: 
	\begin{proposition}
		\label{pro:PhaseTransition2}
		When $\frac{NMPQ_{\rm r}}{L^{\ast}}\gg 1$, \eqref{eqn:PhaseTransition1} is approximated by 
		\vspace{-0.12cm}
		\begin{equation}
		\begin{aligned}
		NKQ_{\rm r} \approx 2L^{\ast} + \frac{L^{\ast}\beta_{\ast}^2}{2},
		\label{pro:phasethreshold}
		\end{aligned}
		\vspace{-0.12cm}
		\end{equation}
		where $\beta_{\ast}$ is the solution of the following equation
		\vspace{-0.1cm}
		\begin{equation}
		\begin{aligned}
		\ln\left(\beta_{\ast}^{2} +1\right) = \ln \frac{NMPQ_{\rm r} - L^{\ast}}{L^{\ast}} - \frac{\beta_{\ast}^{2}}{2}.
		\label{eqn:betaast2}
		\end{aligned}
		\vspace{-0.1cm}
		\end{equation}
	\end{proposition}
	\begin{IEEEproof}
		This is a direct consequence of Prop. 6 in \cite{Li2021Phase} with some notation changes. 
	\end{IEEEproof} 
	%The condition  $\frac{NMPQ_{\rm r}}{L^{\ast}}\gg 1$ means that the observed scene\Verify{I disagree here, This does not mean that the scene is sparse because the threshold $L^\ast$ is not the number of targets in the scene. Is there any intuitive understanding we can provide on when this approximation holds? What does the product $NMPQ_{\rm r}$ represent in the \ac{cs} problem?} is relatively sparse with respect to the overall number of measurements. %\VerifyT{Please verify.} It is typically easy to satisfy, which can be observed in the simulation results of Subsection~\ref{subsec:RadarSimu}.
	The condition $\frac{NMPQ_{\rm r}}{L^{\ast}}\gg 1$, for which the simplified relationship in \eqref{pro:phasethreshold} is formulated, is numerically shown to faithfully represent typical \ac{dfrc} parameter setting, as shown in Subsection~\ref{subsec:RadarSimu}. This implies that  \eqref{pro:phasethreshold}, which is simple to compute compared to \eqref{eqn:PhaseTransition1}, can often be used to obtain the number of recoverable radar targets of \ac{frac}.  
	In this regime where the threshold $L^{\ast}$ is much smaller than $NMPQ_{\rm r}$, Proposition \ref{pro:PhaseTransition2} also reveals how the number of detectable targets scales when the \ac{dfrc} parameters grow arbitrarily large. This asymptotic scaling of $L^{\ast}$ is stated in the following corollary:
	\begin{corollary}
		\label{cor:PhaseTransition1}
		When $\frac{NMPQ_{\rm r}}{L^{\ast}}\gg 1$, the maximum  number of recoverable targets $L^{\ast}$ has the order of $\mathcal{O}\left(\frac{NKQ_{\rm r}}{\ln\left(NMPQ_{\rm r}\right)}\right)$.
	\end{corollary}
	\begin{IEEEproof}
		The proof is given in Appendix~\ref{app:Proof4}.
	\end{IEEEproof} 
	From Corollary \ref{cor:PhaseTransition1} it follows that the maximum number of recoverable targets $L^{\ast}$ grows when increasing the active transmit antenna elements $K$, the number of pulses  in one \ac{cpi} $N$, and the number of elements in radar receive array $Q_{\rm r}$. It also indicates that increasing  the  total number of sub-carriers $M$ and/or  the total  number of transmit antenna elements  $P$ reduces $L^{\ast}$. %\VerifyT{It would be more readable, if we repeat here the meaning of these parameters.} 
	This is because increasing $N$ and $Q_{\rm r}$ effectively leads to more observations being acquired, which enables to recover more targets. When $K$ grows, the radar waveform is transmitted over more sub-carriers and using more antenna elements, thus improving the maximum number of recoverable targets. However, when increasing $M$ and $P$, the transmit waveform utilizes a smaller portion of the bandwidth and \ac{ula}, which decreases the number of recoverable targets.
	
	\ReviseReview{From the above analysis and the throughput analysis of \ac{im} provided in~\cite{Datta2016Generalized}, we observe that there exists a tradeoff between the performances of radar and communications, depending on the waveform parameters. Increasing the number of transmit antennas $P$ improves the angle resolution of radar as well as the number of bits conveyed by spatial \ac{im}, while decreasing the maximum number of recoverable targets. For fixed total frequency band $B$, increasing $M$, i.e., dividing more sub-carriers, improves the number of bits conveyed by frequency \ac{im}, while also decreasing the maximum number of recoverable targets. The combined analysis indicates how to set $K$: As long as $K \leq \min \big\{\lfloor\frac{M}{2}\rfloor,\lfloor \frac{P}{2}\rfloor\}$,  increasing $K$ improves both the message cardinality  and the number of recoverable targets. Nonetheless, it also  increases the hardware complexity, as discussed in Subsection~\ref{subsec:RadarSimu}.\label{txt:ParameterDiscussion} }
	
	%\textcolor{red}{We need to be able to relate these observations to the cardinality of the conveyed message and to the results from the ambiguity function analysis. These can appear in a discussion which I think is missing. }
	
	%	When specialized to the minimum distance detector, e.g., when the conditional distribution in \eqref{eqn:ML} is Gaussian, the complexity is of the order of $\mathcal{O}\left(2^{\beta} M_T^c M_R^c\right)$ products, as evaluating the distance for each tested $\mathbf{x} \in \mySet{X}$ involves $M_T^c M_R^c$ multiplications. This implies that the computational complexity grows with the dimensionality of the system, i.e., the number of antenna elements, as well as with the cardinality of the search space. Nonetheless, when $\beta$ is large, one can implement symbol detection with lower computational burden  using existing reduced complexity \acs{gsm} decoding methods, see e.g.,~\cite{Wang2012Generalised}.
	
	\iffalse
	\color{red}
	\subsection{Performance of Communication Subsystem}
	\label{subsec:PermComm}
	We now characterize the achievable rate of the  communications subsystem of \ac{frac}. The 
	analysis allows us to numerically compare the communication capabilities of the communication subsystem of \ac{frac} with the communication system which only exploits \ac{pm} for data conveying in Section~\ref{sec:Sims}. 
	
	\fi
	
	\color{black}
	%-----------------------------------------------------------------------------	-----------
	%	Numerical Evaluations
	%----------------------------------------------------------------------------------------
	\vspace{-0.2cm}
	\section{Numerical Evaluations}
	\vspace{-0.15cm}
	\label{sec:Sims}
	We next numerically evaluate the radar and communication performance of \ac{frac}. For convenience, we summarize  the main parameters used in the experimental study in TABLE~\ref{tab:ParameterValue}.

	\vspace{-0.2cm}
	\subsection{Radar Subsystem Evaluation}
	\label{subsec:RadarSimu}
	\vspace{-0.1cm}
	To demonstrate the radar performance, we first verify the ambiguity function of \ac{frac}. In particular, we show that the range/velocity/angle resolutions of \ac{frac} follow the analysis in Subsection~\ref{subsec:RadarReso}, and achieve the same resolutions as the radar transmitting a wideband waveform and with an equivalent virtual aperture. Then, we test the resolution of the radar subsystem by recovering three closely located targets, and calculating the hit rate curve to show the detection performance with different noise levels. Next, the phase transition curves of the radar subsystem are theoretically calculated, and compared with the simulated phase transition threshold with different waveform parameters. Finally, the hardware overheads of \ac{frac} and the compared radar system are provided. Due to the page limitations, the ability of \ac{frac} to decrease mutual interference, which was shown for other \ac{im}-based \ac{dfrc} systems, e.g., \cite{Huang2020MAJoRCom}, is not studied here, is left for future work.  %in~\cite{Huang2020Multi}\Verify{have you carried out such experiments? if so, we can submit it as supplementary material to further convince the reviewers}. 

	\begin{table}  
		\caption{Simulation Parameters}
		\vspace{-0.2cm}
		\small
		\begin{center}  
			\begin{tabular}{|c|c||c|c|}  
				\hline 
				\hline 
				\textbf{Parameter} & \textbf{Value} & \textbf{Parameter} & \textbf{Value}\\
				\hline
				$f_c$ &  $\mathrm{77\ GHz}$ &
				$B $ &  $100\ \mathrm{MHz}$ \\
				\hline
				$N$ &  32 &
				$M$ &  8\\			
				\hline
				$K$ & 1 &
				$P$ &  4\\	
				\hline
				$Q_{\rm r}$ & 2 &
				$T_{0}$ &  $60.88\ \mu \mathrm{s}$\\
				\hline
				$T_{p}$ & $50\  \mu\mathrm{s}$ &
				$F_s\Radar$ &  $ 416.68\ \mathrm{kHz}$\\
				\hline
				\hline		
			\end{tabular}  
		\end{center} 
		\vspace{-0.2cm}
		\label{tab:ParameterValue} 
	\end{table} 
	
	In the sequel, the ambiguity function, recovery performance, and number of detectable targets are evaluated. Unless stated otherwise, we use the waveform parameters listed in TABLE~\ref{tab:ParameterValue}. %In the evaluation of the phase transition, the \ac{dfec} parametes $\{N, M, P, Q_r, K\}$ are detailed in TABLE~\ref{tab:PhaseTransition}, while those other parameters still follow  TABLE~\ref{tab:ParameterValue}. 
	\subsubsection{Ambiguity Function}
\begin{figure}
		\begin{minipage}[b]{0.495\linewidth}
			\centering
			\centerline{\includegraphics[width= \columnwidth]{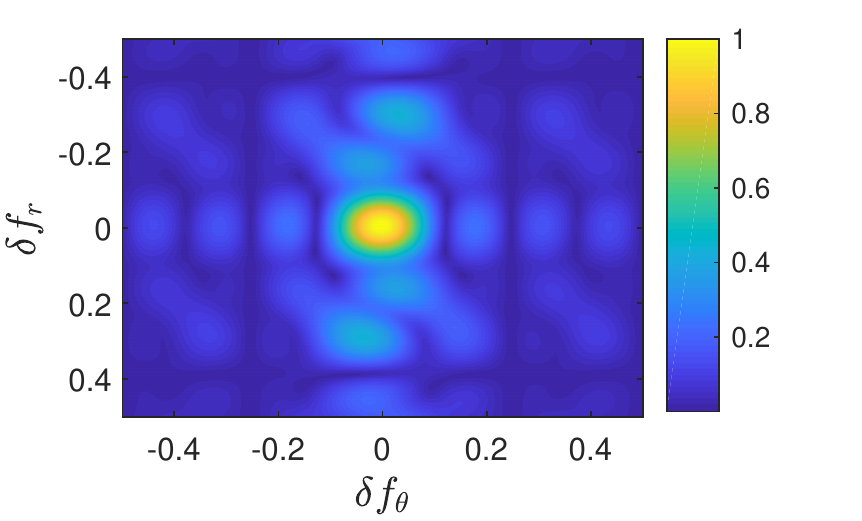}}
			\centerline{(a) $\left|\chi\left(\delta \! f_r, 0, \delta \! f_\theta\right)\right|$}	 
		\end{minipage}
		\begin{minipage}[b]{0.495\linewidth}
			\centering
			\centerline{\includegraphics[width=\columnwidth]{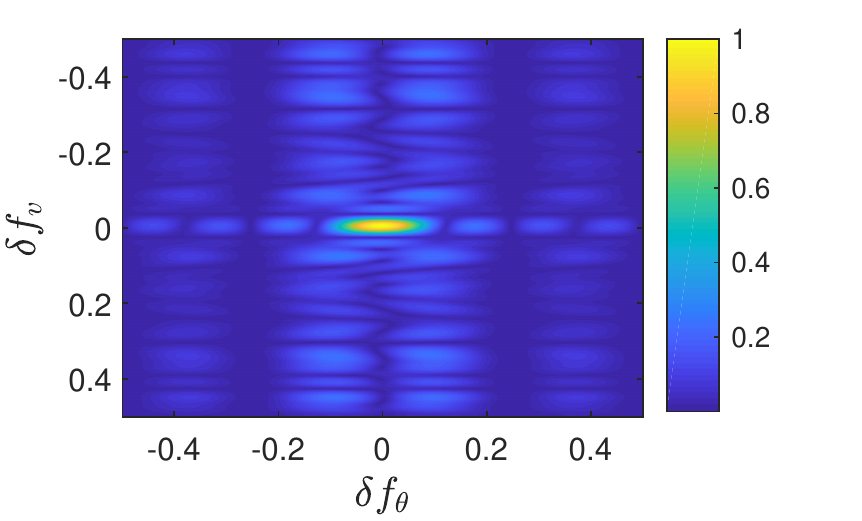}}
			\centerline{(b) $\left|\chi \left(0, \delta \! f_v, \delta \! f_\theta\right)\right|$}
		\end{minipage}
		
		\begin{minipage}[b]{0.495\linewidth}
			\centering	         \centerline{\includegraphics[width=\columnwidth]{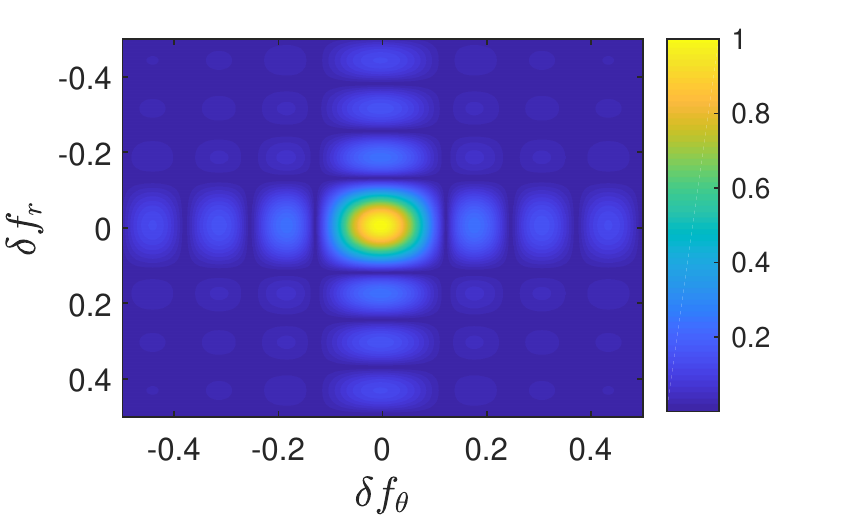}}
			\centerline{(c) $\left|\mathcal{E}\{\chi\left(\delta \! f_r, 0, \delta \! f_\theta\right)\}\right|$}	 
		\end{minipage}
		\begin{minipage}[b]{0.495\linewidth}
			\centering
			\centerline{\includegraphics[width= \columnwidth]{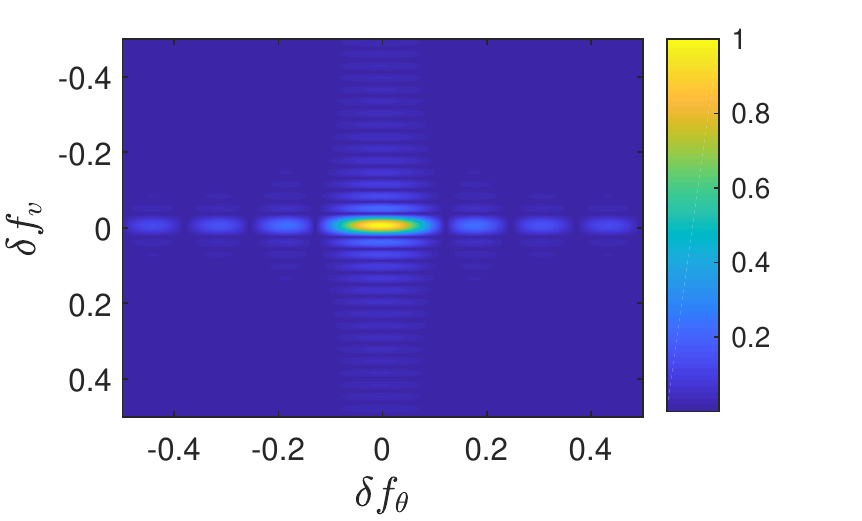}}		
			\centerline{(d) $\left|\mathcal{E}\{\chi \left(0, \delta \! f_v, \delta \! f_\theta\right)\}\right|$}	 
		\end{minipage}
		
		\begin{minipage}[b]{0.495\linewidth}
			\centering	         \centerline{\includegraphics[width=\columnwidth]{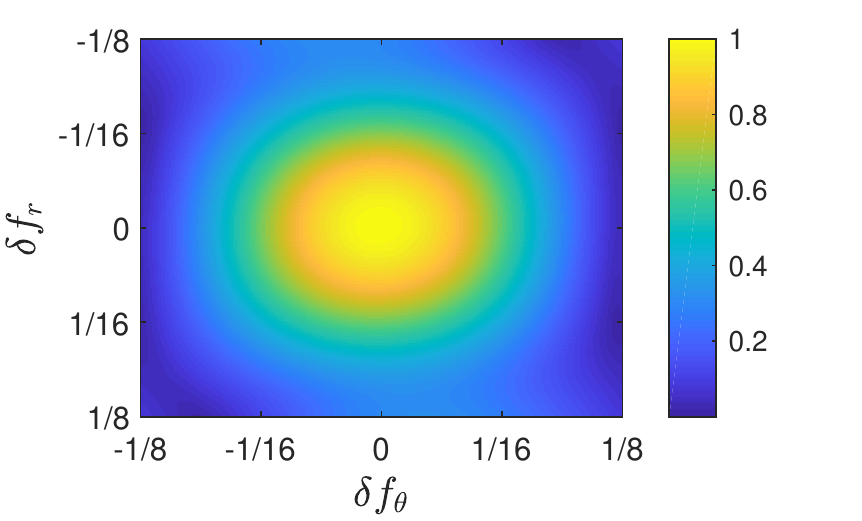}}
			\centerline{(e) Zoomed $\left|\chi\left(\delta \! f_r, 0, \delta \! f_\theta\right)\right|$}	 
		\end{minipage}
		\begin{minipage}[b]{0.495\linewidth}
			\centering
			\centerline{\includegraphics[width= \columnwidth]{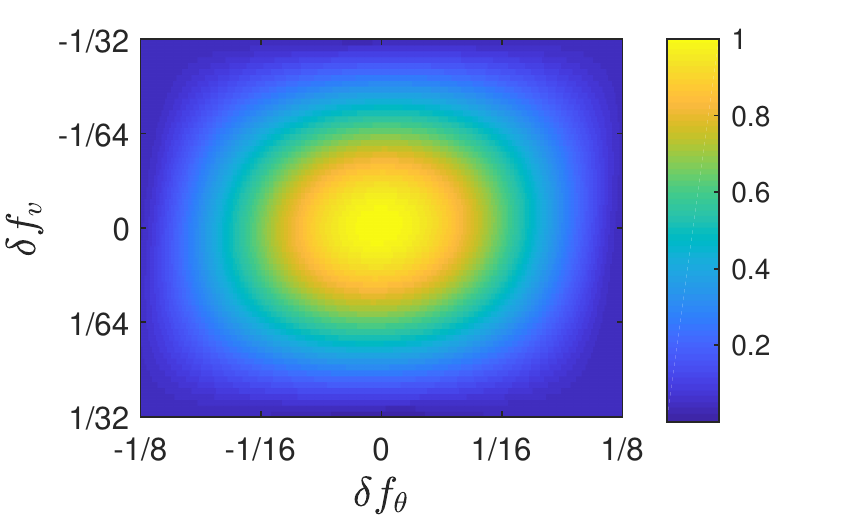}}		
			\centerline{(f) Zoomed $\left|\chi \left(0, \delta \! f_v, \delta \! f_\theta\right)\right|$}	 
		\end{minipage}
		\vspace{-0.4cm}
		\caption{Numerically evaluated ambiguity functions.} 
		\label{fig:AmbguityFun}
	\end{figure}

	We now empirically evaluate the ambiguity function of \ac{frac}, and compare it with the theoretical expected ambiguity function given by \eqref{eqn:ExpctatedAmbFun}. To show the characteristics of the ambiguity function of \ac{frac}, a single realization of the ambiguity function is calculated.  Since the ambiguity function has three arguments, i.e., $\delta \! f_r, \delta \! f_v, \delta \! f_\theta$, we only draw the amplitude of two cross sections at $\chi\left(\delta \! f_r, 0, \delta \! f_\theta\right)$ and $\chi \left(0, \delta \! f_v, \delta \! f_\theta\right)$ in Figs.~\ref{fig:AmbguityFun}(a)-\ref{fig:AmbguityFun}(b), respectively. For comparison, we depict the theoretical expected ambiguity function using \eqref{eqn:ExpctatedAmbFun} in Figs.~\ref{fig:AmbguityFun}(c)-\ref{fig:AmbguityFun}(d).
	
	The similarity between the instantaneous ambiguity function  and its theoretically evaluated expectation is observed in Fig.~\ref{fig:AmbguityFun}. 
	We note that the ambiguity function  of \ac{frac} is thumbtack, and has a mainlobe centered at $\left(0, 0, 0\right)$. 
	The resolution is obtained from the width of the mainlobe. Comparing Figs.~\ref{fig:AmbguityFun}(a)-(b) and Figs~\ref{fig:AmbguityFun}(c)-(d), we find that the mainlobe width of both ambiguity functions are nearly the same, which demonstrates that the expected ambiguity function is  useful to characterize the resolution of radar subsystem. To further show the resolution performance, we further zoom in around the mainlobe of $\chi\left(\delta \! f_r, 0, \delta \! f_\theta\right)$ and $\chi \left(0, \delta \! f_v, \delta \! f_\theta\right)$ in Figs.~\ref{fig:AmbguityFun}(e)-\ref{fig:AmbguityFun}(f). From the zoomed ambiguity function, we note that $\Delta \! f_r \approx \frac{1}{M} \approx \frac{1}{8}$, $\Delta \! f_v \approx \frac{1}{N} \approx \frac{1}{32}$ and $\Delta \! f_\theta \approx  \frac{1}{PQ_{\rm r}} \approx \frac{1}{8}$, which are in line with the analysis in Subsection~\ref{subsec:RadarReso}. According to \eqref{eqn:Resolution}, we obtain the resolutions in range, velocity, and angle, via $\Delta r = \frac{c \Delta  f_r}{2\Delta f} = 1.5\ \mathrm{m}$, $\Delta v = \frac{c \Delta  f_v}{2T_{0}f_{c}} = 1\  \mathrm{m/s}$ and $\Delta \theta = \arcsin\left\{\frac{c \Delta f_\theta}{f_{c}d_{\rm R}}\right\} = 14.48^{\circ}$, respectively. These are the same resolutions of \ac{mimo} \ac{fmcw} radar with a bandwidth of $100$~MHz, and receive array of $8$ elements~\cite[Ch. 2.10]{Wehner1995High}. %\Verify{can we cite anyone on this statement?}. 
	
	\color{black}
	\subsubsection{Target Recovery}
	\label{txt:SchemeCompareRadar}
	We next evaluate the radar resolution by recovering three adjacent targets with unit reflective factors using the detection mechanism proposed in Subsection~\ref{subsec:RadarRx}. 
	The number of active elements are set to $K =2$. The target recovery performance of \ac{frac} is compared with that of the \ac{im}-based \ac{majorcom} scheme \cite{Huang2020MAJoRCom}. Since the transmission and reception of  \ac{majorcom} utilize the same array through time division duplexing, to adapt it to the vehicular scenario, the compared \ac{majorcom} is modified to utilize separate transmit array and receive array, which also formulate a \ac{mimo} architecture as \ac{frac}. For fair comparison, the same number of \ac{rf} modules are utilized in \ac{frac} and \ac{majorcom}. Thus, the number of receive antennas of \ac{majorcom} is the same as that of \ac{frac}, while the number of transmit antennas of \ac{majorcom} equals the number of active transmit elements in \ac{frac}.  The remaining parameters of \ac{majorcom}, such as the cardinality of the sub-carrier set, the total bandwidth, the \ac{pri}, and the number of pulses in one \ac{cpi}, are set the same as that of \ac{frac}. In the compared \ac{majorcom}, $K$ waveforms with different sub-carriers are separately assigned to the $K$ transmit elements in a permutation manner. A virtual array can also be combined in the modified \ac{majorcom}, the aperture of which equals $KQ_{\rm r}$. Therefore, the normalized spatial resolution of the compared \ac{majorcom} is $\frac{1}{KQ_{\rm r}}$, which is wider than that of \ac{frac} equaling $\frac{1}{PQ_{\rm r}}$.
	
\begin{figure}
		\begin{minipage}[b]{0.48\linewidth}
			\centering
			\centerline{\includegraphics[width= \columnwidth]{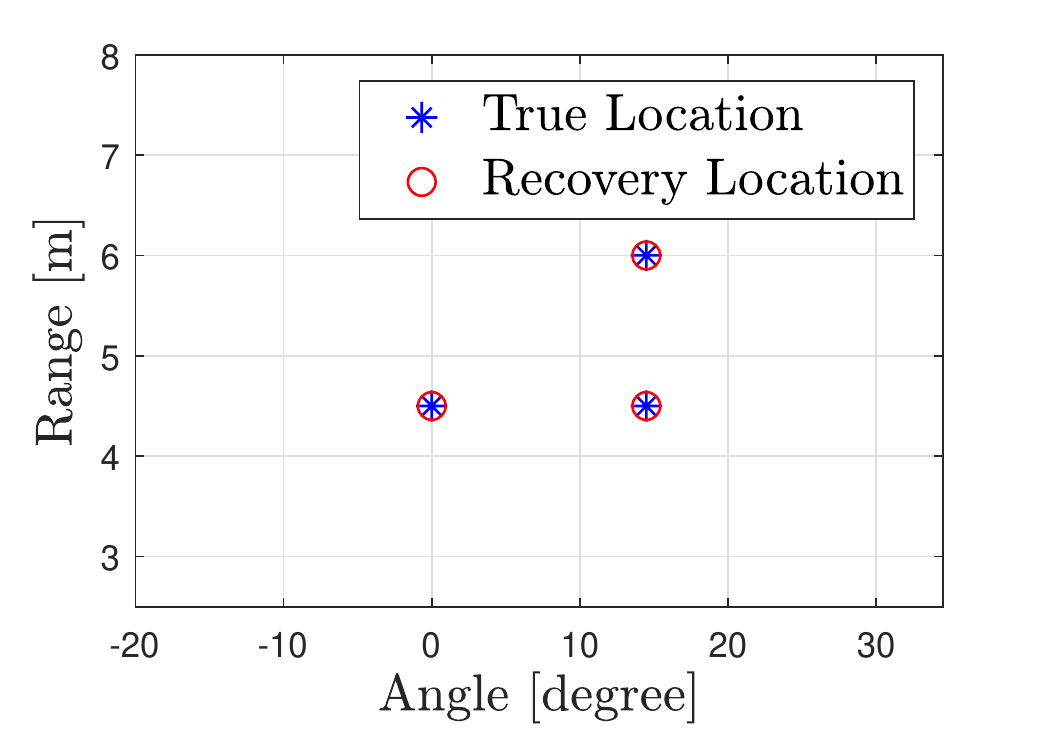}}
			\centerline{(a)}
			\vspace{-0.1cm}
		\end{minipage}
		\begin{minipage}[b]{0.48\linewidth}
			\centering
			\centerline{\includegraphics[width=\columnwidth]{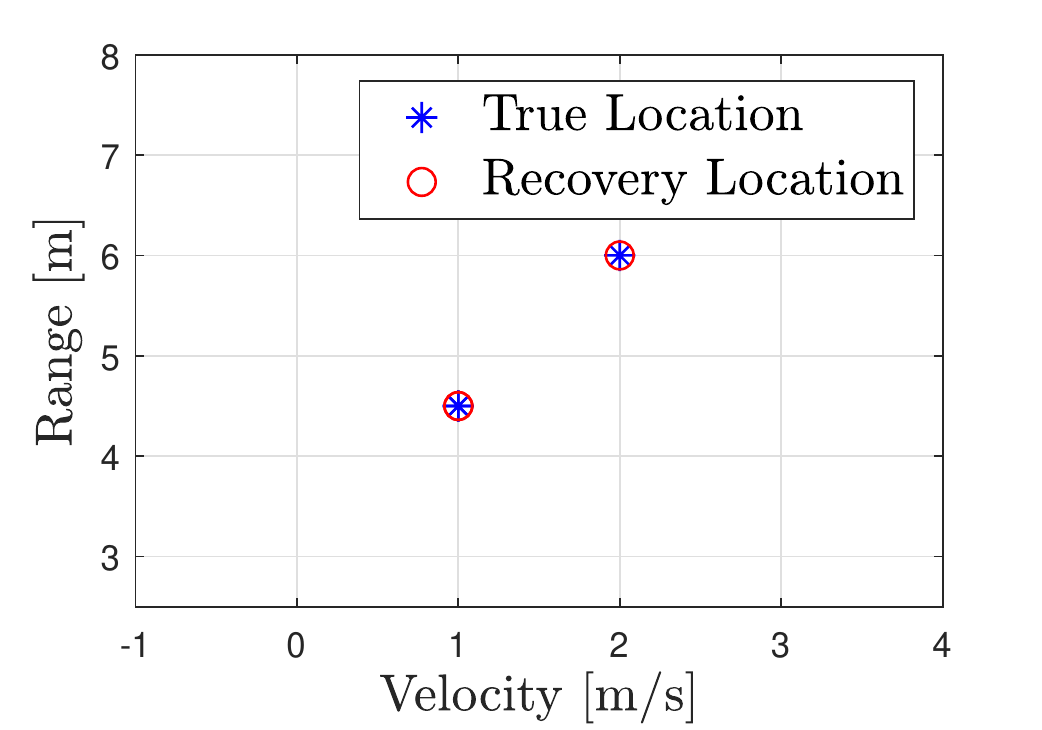}}
			\centerline{(b)}
			\vspace{-0.1cm}
		\end{minipage}
		\vspace{-0.2cm}
		\caption{Recovery results of \acs{frac}.} 
		\label{fig:TargetRecoryFRaC}
	\end{figure}

\begin{figure}
		\vspace{-0.5cm}
		\begin{minipage}[b]{0.48\linewidth}
			\centering
			\centerline{\includegraphics[width= \columnwidth]{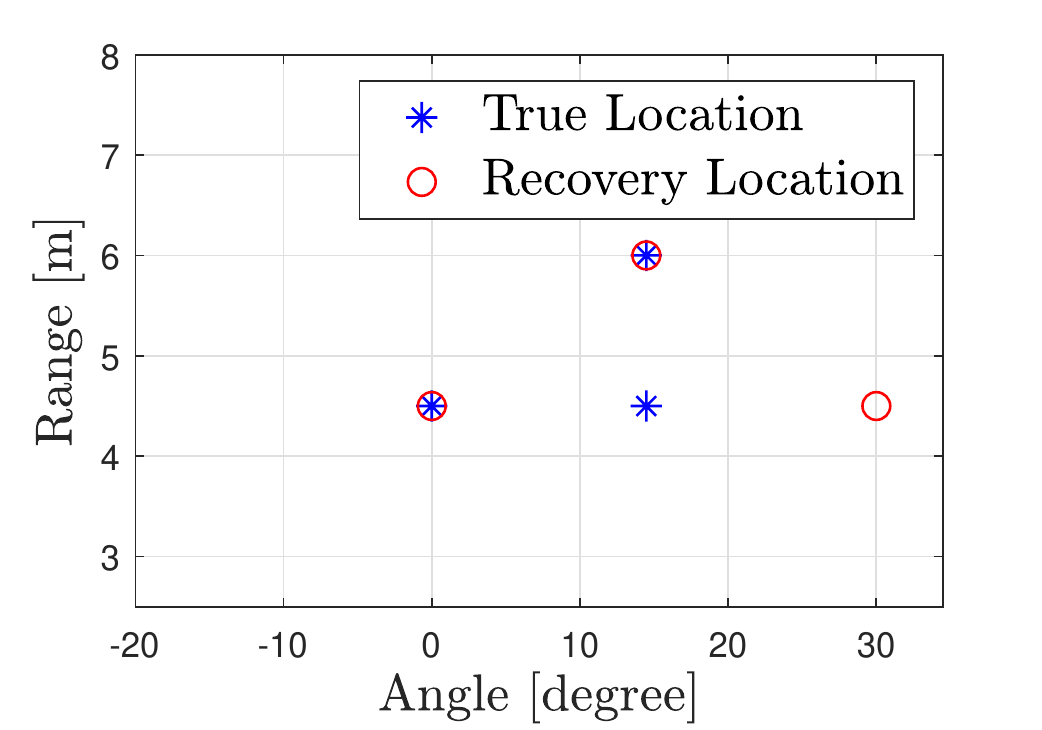}}
			\centerline{(a)}
			\vspace{-0.1cm}
		\end{minipage}
		\begin{minipage}[b]{0.48\linewidth}
			\centering
			\centerline{\includegraphics[width=\columnwidth]{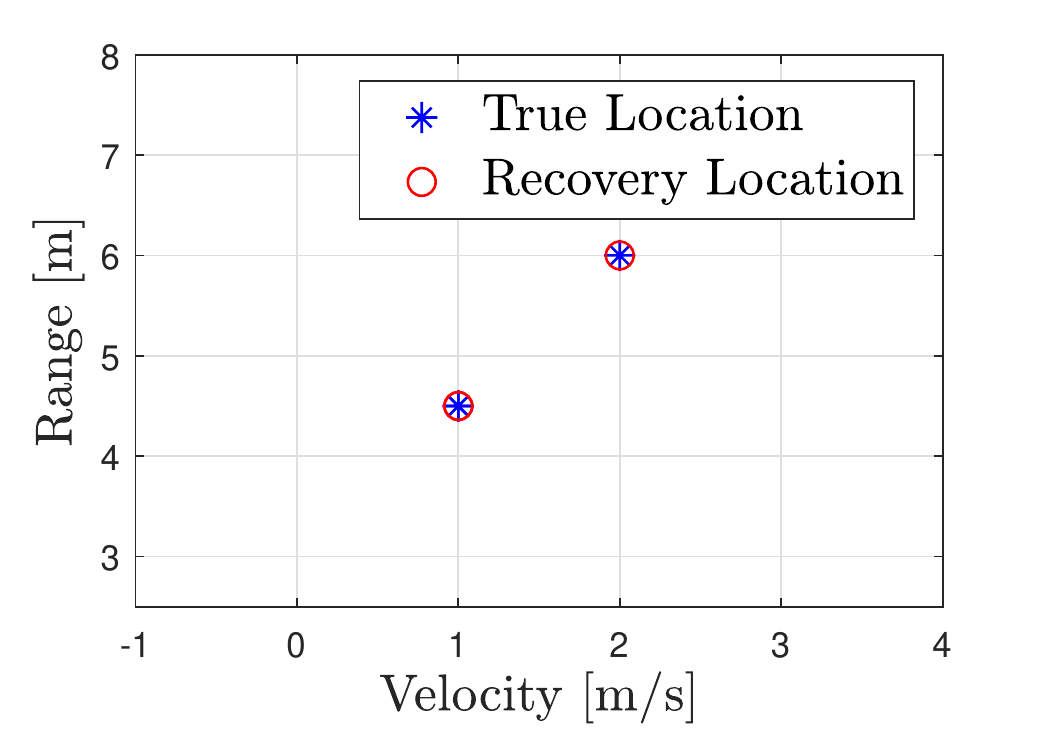}}
			\centerline{(b)}
			\vspace{-0.1cm}
		\end{minipage}
		\vspace{-0.2cm}
		\caption{Recovery results of \acs{majorcom}.} 
		\label{fig:TargetRecoryMAJoRCom}
	\end{figure}

	Following the ambiguity function experiments, we set the parameters of these targets to comply with the evaluated radar resolutions. In particular, we simulate three targets with parameters  $\left\{{r}_{1}, {v}_{1}, {\theta}_{1}\right\} = \left\{4.5\ \mathrm{m},  1\ \mathrm{m/s}, 0^{\circ} \right\}$, $\left\{{r}_{2}, {v}_{2}, {\theta}_{2}\right\} = \left\{4.5\ \mathrm{m}, 1\ \mathrm{m/s},   14.48^{\circ}\right\}$, and $\left\{{r}_{3}, {v}_{3}, {\theta}_{3}\right\} = \left\{6\ \mathrm{m}, 2\ \mathrm{m/s}, 14.48^{\circ} \right\}$, respectively. To demonstrate the radar resolution, this target map is recovered by \ac{frac} and \ac{majorcom} without adding noise. In \ac{frac}, the target locations are recovered utilizing the detection model in \eqref{eqn:RadarEquCS} using the \acl{omp} \ac{cs} algorithm  \cite{eldar2012compressed}. The recovery target locations compared to their true locations  are shown in Fig.~\ref{fig:TargetRecoryFRaC} and Fig.~\ref{fig:TargetRecoryMAJoRCom} for \ac{frac} and \ac{majorcom}, respectively. From Fig.~\ref{fig:TargetRecoryFRaC}, we observe \ac{frac} exactly recovers the targets such that they can be distinguished with the given parameter settings, demonstrating that the evaluated radar resolutions of \ac{frac} are indeed translated into accurate target recovery capabilities. From Fig.~\ref{fig:TargetRecoryMAJoRCom}(a), we find that \ac{majorcom} fails to recover the angle of Target3. This is because Target1 and Target2 have the same range and velocity values, and the angle distance of Target1 and Target2 is finer than the angle resolution of \ac{majorcom}. Thus, Target1 and Target2 can not be distinguished by \ac{majorcom}, which leads to a wrong recovery.
	
\begin{figure}
		\centerline{\includegraphics[width= 0.8\columnwidth]{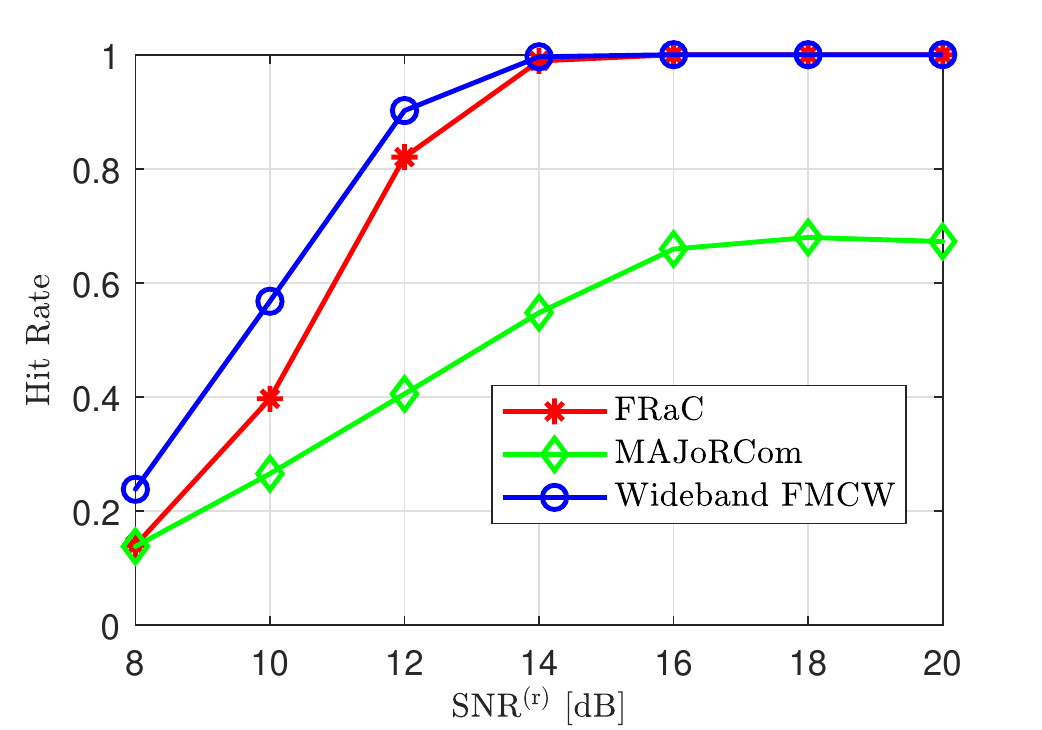}}
		\vspace{-0.4cm}
		\caption{Range-velocity-angle recovery hit rate versus different radar $\mathrm{SNR}\Radar$}.  %\textcolor{red}{can you change the legend to '\ac{frac}' and 'Wideband FMCW'?}.} 
		\label{fig:HitRate}
	\end{figure}

	To demonstrate the radar performance in different noise levels, we use hit rate as the performance criterion. A ``hit" is defined if the range-velocity-angle parameter of a scattering point is successfully recovered. Each hit rate is calculated over $20,000$ Monte Carlo trials by recovering the target scenario depicted in Fig.~\ref{fig:TargetRecoryFRaC}. In the simulation, the power of the received echoes from different targets are set to be equal and are normalized. The radar \ac{snr} is defined as $\mathrm{SNR}\Radar = \frac{NKQ_{\rm r}}{\sigma^{2}_{\rm r}}$, where $\sigma^{2}_{\rm r}$ is the variance of  $w_{n,k,q}\Radar$ in \eqref{eqn:RadarRx5}.  
	The hit rate of \ac{frac} is compared to that of \ac{majorcom} and a wideband \ac{fmcw} radar with the same number of elements $P\times Q_{\rm r}$, and the results are depicted in Fig.~\ref{fig:HitRate}. Observing Fig.~\ref{fig:HitRate}, we note that \ac{frac} has a  performance loss of $1$ dB in low $\rm{SNR}\Radar$ compared to the costly wideband \ac{fmcw} system, while achieving the same hit rate as the wideband radar for $\rm{SNR}\Radar $ higher than $14$ dB. For \ac{majorcom}, the hit rate does not reach probability one with the increasing of $\rm SNR\Radar$. This is because Target1 and Target2 can not be distinguished by \ac{majorcom} as observed in Fig.~\ref{fig:TargetRecoryMAJoRCom}(a), even in the absence of noise.  These results indicate that the notable hardware reduction and the natural operation as a \ac{dfrc} system of \ac{frac} result in only minor degradation in the radar accuracy compared to the wideband \ac{fmcw} \ac{mimo} benchmark, and outperforms the \ac{majorcom} utilizing the same number of \ac{rf} modules for scenarios of close targets.
	
\begin{figure}
		\begin{minipage}[b]{0.485\linewidth}
			\centering
			\centerline{\includegraphics[width= \columnwidth]{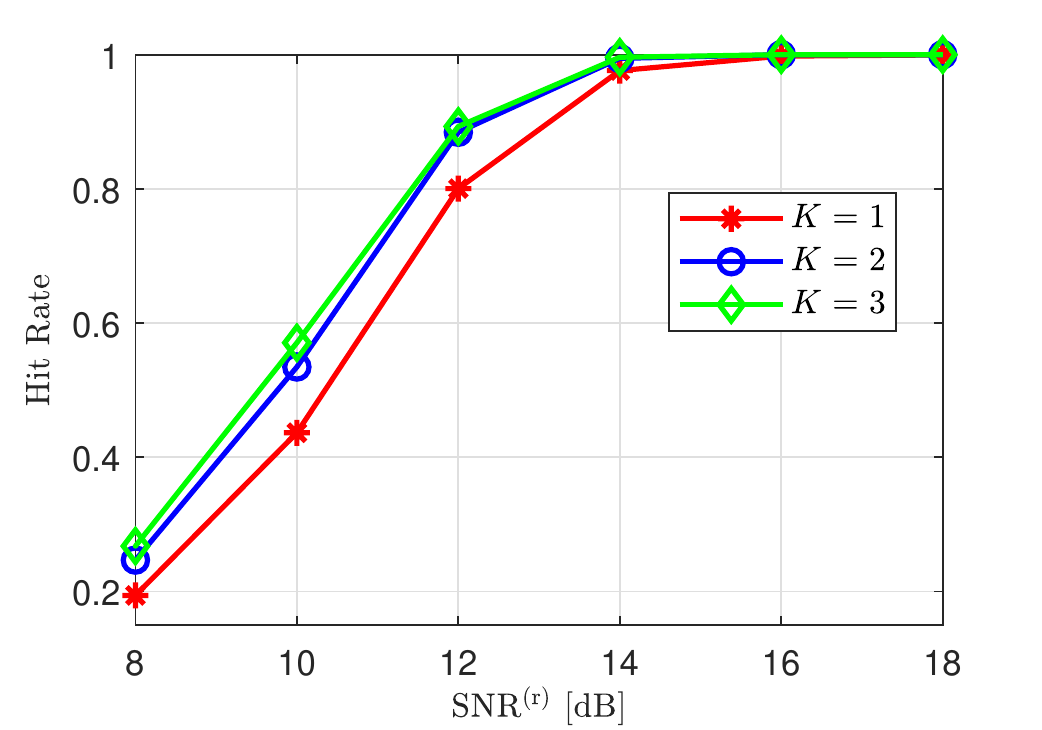}}
			\centerline{(a) }	 
		\end{minipage}
		\begin{minipage}[b]{0.485\linewidth}
			\centering
			\centerline{\includegraphics[width=\columnwidth]{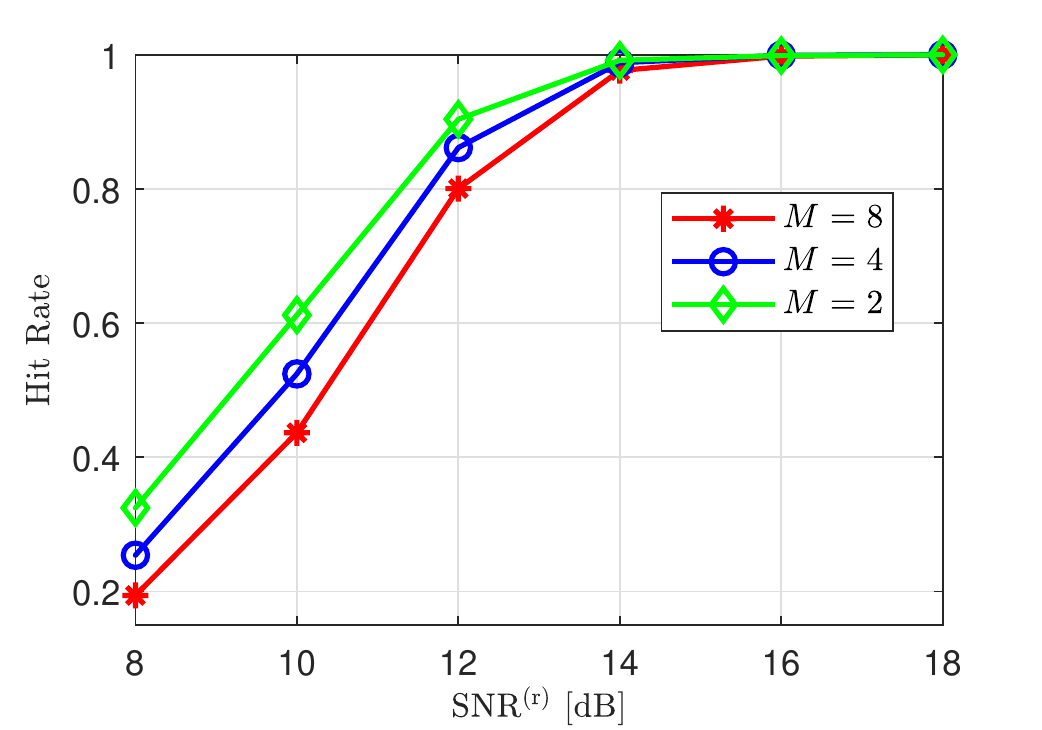}}
			\centerline{(b) }
		\end{minipage}
		\vspace{-0.3cm}
		\caption{Hit rate versus system parameters. (a) shows hit rate versus $K$ and $\rm SNR\Radar$. (b) shows hit rate versus $M$ and $\rm SNR\Radar$ versus $M$ and $\rm SNR\Radar$.} 
		\label{fig:HitRateVsPar}
	\end{figure}

	\textcolor{black}{We next simulate the target recovery performance versus different system parameters. To that aim, the hit rate performance is evaluated over $20,000$ randomly generated target maps with three targets. The hit rates versus $K$, i.e., the number of active antenna elements or the transmit subcarriers, are shown in Fig.~\ref{fig:HitRateVsPar}(a). Observing Fig.~\ref{fig:HitRateVsPar}(a), we note that the hit rates increase with the increase of $K$, where the more substantial improvement is observed when $K$ increases from one to two. This is because more sub-carriers and transmit elements are utilized, thus improves the performance. The hit rates versus $M$, i.e., the number of sub-carriers in the carrier set, are depicted in Fig.~\ref{fig:HitRateVsPar}(a). From Fig.~\ref{fig:HitRateVsPar}(a), we find that the hit rates increase with the decreasing of $M$. This stems from the fact that when $M$ is decreased, the waveform transmitted in each pulse occupies more frequency resource, thus improving the radar performance.\label{txt:HitRateVsPar}} 
	
	\color{black}
	\subsubsection{Phase Transition Threshold}
	Here, we verify the theoretical phase transition threshold derived in Subsection~\ref{subsec:PhaseThreshold}, by comparing the simulated phase transition thresholds with the theoretical quantities computed via \eqref{eqn:PhaseTransition1}. To obtain the simulated phase transition threshold, the targets are recovered without noise according to \eqref{eqn:RadarEqnCS2} by solving a convex optimization problem. For a given number of targets and waveform parameters, $1,000$ Monte Carlo trials are carried out to recover the radar targets, which are randomly generated on the grids. An exact recovery is defined when the average $\ell_2$ distance between the recovered target vector $\hat{\myVec{b}}$ and the true %\Verify{do you compute the $\ell_2$ or the average $\ell_2$ norm? the former depends on the dimensionality of $b$ which appears to be changing here.}
	$\myVec{b}$ is at most $10^{-4}$. The simulated phase transition threshold is observed from the behavior of recovery probability curve versus the number of targets. When the number of targets grows, the probability of exact recovery starts to drastically decrease at some point, which is regarded as the simulated phase transition threshold. In particular, the  recovery probability treated as the  phase transition threshold is $\sim 0.6$.
	
\begin{figure}
		\begin{minipage}[b]{0.485\columnwidth}
			\centering
			\centerline{\includegraphics[width= \columnwidth]{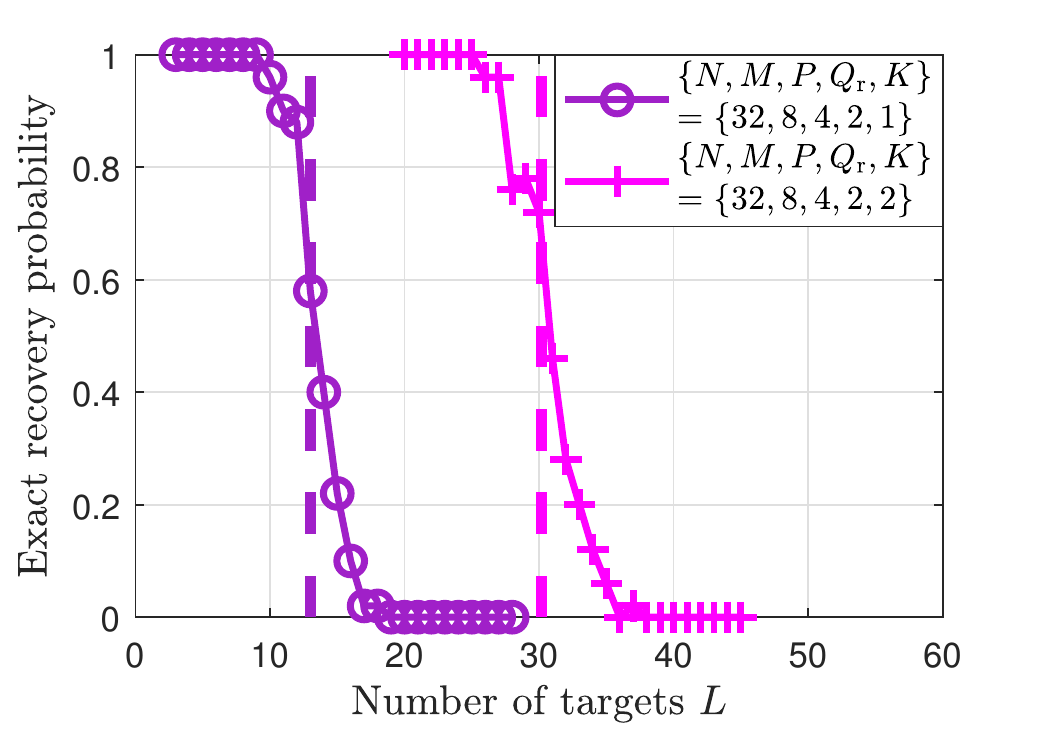}}
			\centerline{(a) Varying $K$}	 
		\end{minipage}		
		\begin{minipage}[b]{0.485\columnwidth}
			\centering
			\centerline{\includegraphics[width=\columnwidth]{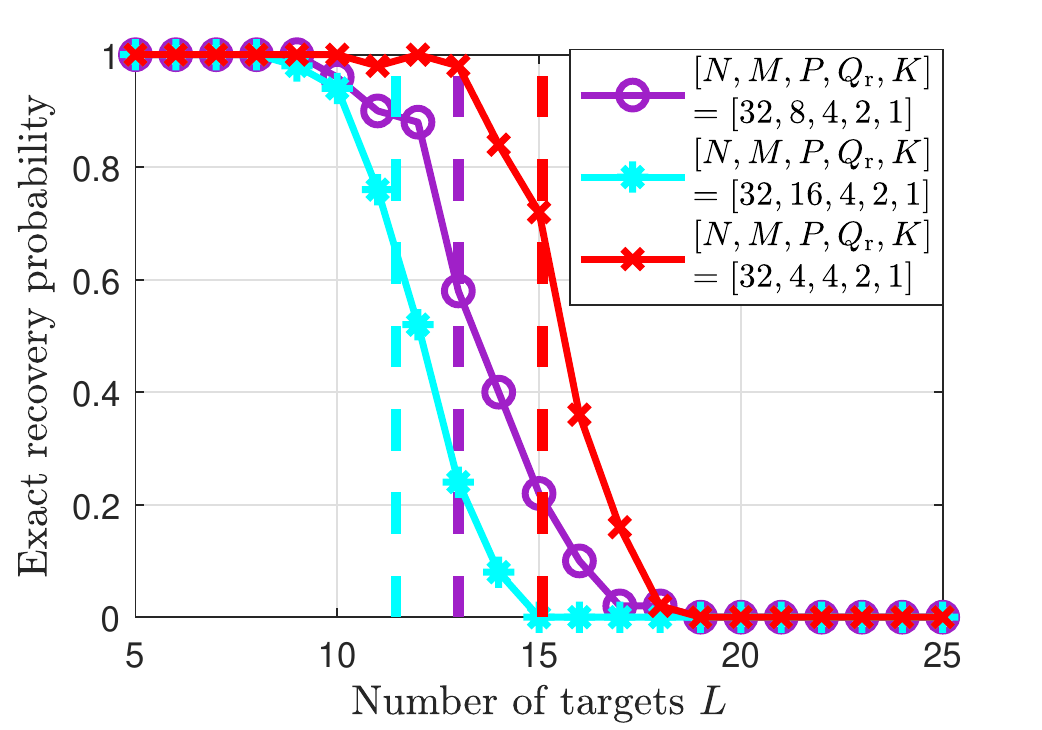}}
			\centerline{(b) Varying $M$}
		\end{minipage}
		
		\begin{minipage}[b]{0.485\columnwidth}
			\centering	         \centerline{\includegraphics[width=\columnwidth]{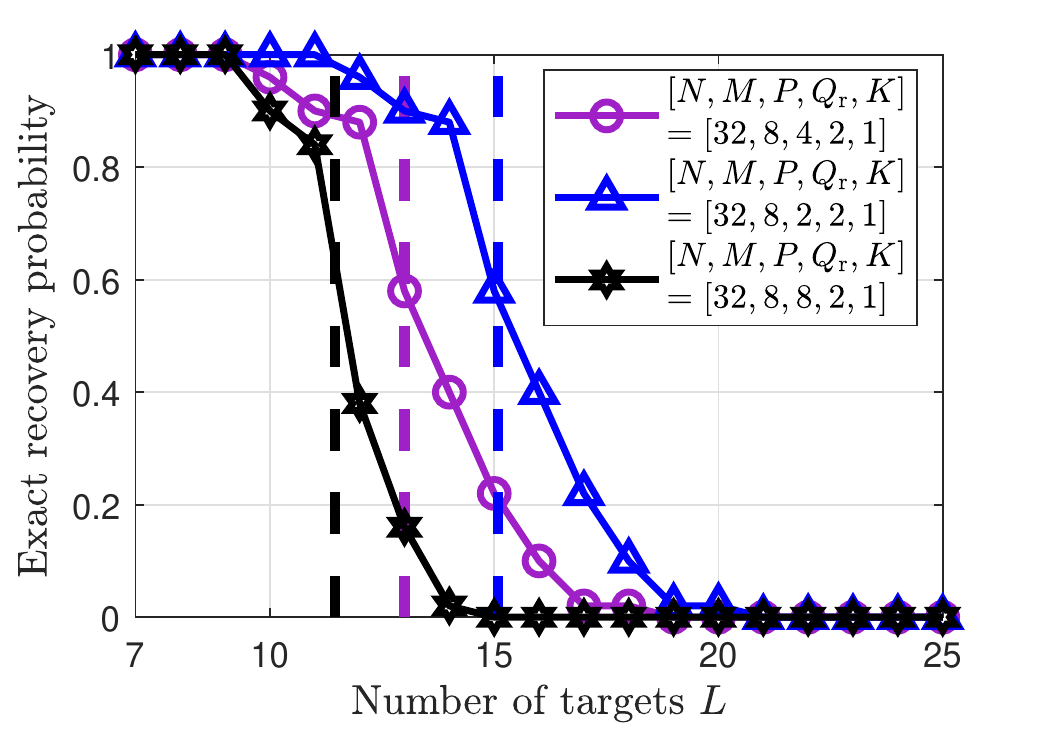}}
			\centerline{(c) Varying $P$}	 
		\end{minipage}	
		\begin{minipage}[b]{0.485\columnwidth}
			\centering
			\centerline{\includegraphics[width= \columnwidth]{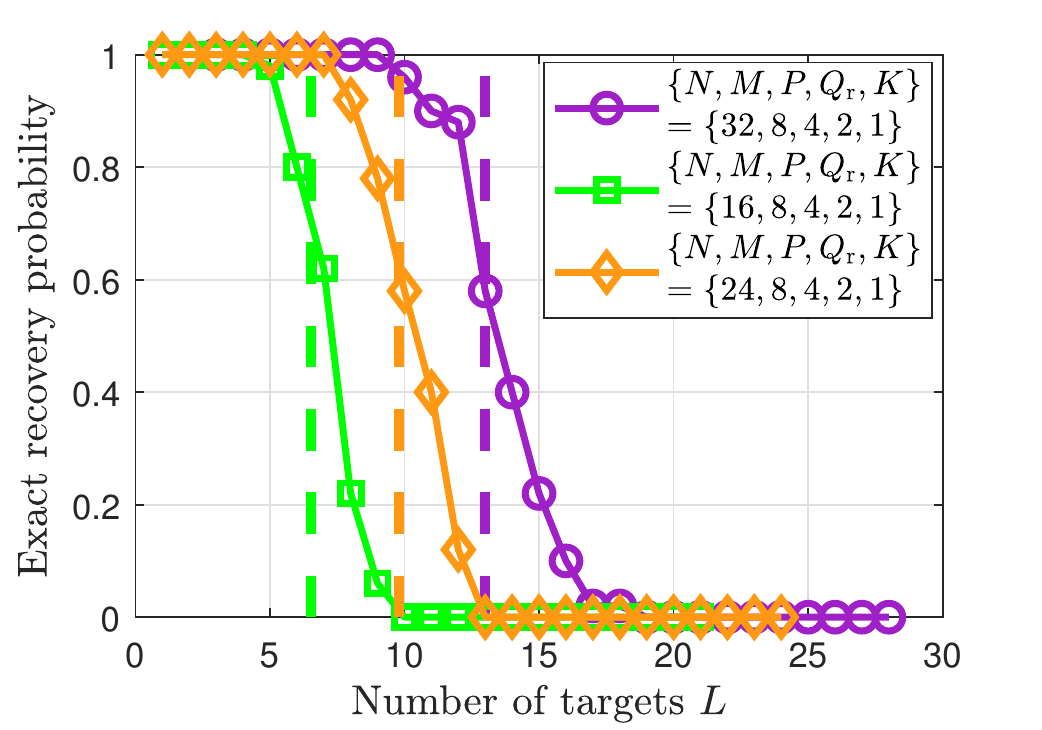}}		
			\centerline{(d) Varying $N$} 
		\end{minipage}
		\vspace{-0.22cm}
		\caption{Exact recovery probabilities versus the number of targets. In each subfigure, only one waveform parameter varies, i.e. varying $K$ in (a), varying $M$ in (b), varying $P$ in (c) and varying $N$ in (d). The vertical dash lines describe the locations of the  theoretical phase transition thresholds of the corresponding waveform parameters indicated by different colors.} 
		\label{fig:PhaseTransition}
	\end{figure}

	\begin{table}  
		\caption{Waveform parameters simulations of phase transition}
		\vspace{-0.6cm}
		\footnotesize
		\begin{center}  
			\begin{tabular}{|c||c||c||c|c||c|c||c|c|}  
				\hline 
				\hline 
				\textbf{} & \textbf{Base} & \textbf{Vary $K$} &  \multicolumn{2}{c||}{\textbf{Vary $M$}} & \multicolumn{2}{c||}{\textbf{Vary $P$}} &\multicolumn{2}{c|}{\textbf{Vary $N$}} \\
				\hline 
				$N$ & 32 & 32 & 32 & 32 & 32 & 32 & 16 & 24\\
				\hline
				$M$ & 8 & 8 & 4 & 16 & 8 & 8 & 8 & 8 \\
				\hline
				$P$ & 4 & 4 & 4 & 4 & 2 & 8 & 4 & 4 \\
				\hline
				$Q_{\rm r}$ & 2 & 2 & 2 & 2 & 2 & 2 & 2 & 2 \\
				\hline 
				$K$ & 1 & 2 & 1 & 1 & 1 & 1 & 1 & 1 \\
				\hline
				$L^{\ast}$ & 13 & 30.2 & 15.1 & 11.4 & 15.1 & 11.4 & 6.5 & 9.8 \\
				\hline
				\hline		
			\end{tabular}  
		\end{center} 
		\vspace{-0.2cm}
		\label{tab:PhaseTransition} 
	\end{table}
	To reveal the relationship between the phase transition threshold and the waveform parameters, we classify the simulations into four categories, depending on the parameter which is modified: $1)$  the number of transmit sub-carriers $K$; $2)$ the cardinality of the carrier set $M$; $3)$ thr total number of transmit antennas $P$; and $4)$ the number of pulses $N$ in one \ac{cpi}. The waveform parameters and the corresponding theoretical phase transition threshold for each parameter setting are listed in TABLE~\ref{tab:PhaseTransition}. In each category, the simulated phase transition threshold is compared to that of the waveform with the baseline parameters, as well as  the theoretical phase transition threshold  calculated via \eqref{eqn:PhaseTransition1}.  	The recovery probability curves are depicted in Fig.~\ref{fig:PhaseTransition}, where solid curves with different colors represent empirically evaluated recovery probabilities for different waveform parameters, while the vertical dashed curves are the  theoretical phase transition  of the waveform setting with the corresponding color. We observe in  Fig.~\ref{fig:PhaseTransition} that, for each waveform parameter setting, the  exact recovery probability drastically decreases around the theoretical   threshold of \eqref{eqn:PhaseTransition1}. Thus, the phase transition can be obtained by theoretical calculation, while accurately reflecting the empirical performance,  thus facilitating  the configuration of \ac{frac}.
	
	In particular, we observe in Fig.~\ref{fig:PhaseTransition}(a) and Fig.~\ref{fig:PhaseTransition}(d) that more targets can be detected when $K$ and $N$ are increased. This is because when $K$ and $N$ increase, the number of measurement increases, which leads to an increase in the maximum number of recoverable targets. Observing Fig.~\ref{fig:PhaseTransition}(b) and Fig.~\ref{fig:PhaseTransition}(c), we note that the phase transition threshold decreases with the increase of $M$ and $P$ for a fixed $K$. This is because the transmit waveform  utilizes a smaller portion of the bandwidth and uses less antennas, when increasing the cardinality of the carrier set and the number of total transmit antennas, which decreases the maximum number of recoverable targets. These numerical results are in line with the analysis in Subsection~\ref{subsec:PhaseThreshold}, indicating that conclusions made out based on the quantitative approximation in Corollary~\ref{cor:PhaseTransition1} also reflect on scenarios of practical interest. % which  provides an approximation to calculate the phase transition threshold, and reveals a obvious relationship between the number of recoverable targets and the waveform parameters in Corollary~\ref{cor:PhaseTransition1}. 
	
	\subsubsection{Comparison of Hardware Complexity}
	From the analysis in Section~\ref{sec:PerformanceRadar} and the simulations in this subsection, we conclude that \ac{frac} achieves the same radar resolution performance as a wideband \ac{fmcw} \ac{mimo} radar which utilizes the full available bandwidth and antenna elements. The fact that \ac{frac} utilizes a subset of the available band and transmit elements at each \ac{pri}, which is exploited to convey information in the form of \ac{im} and reduces the hardware complexity,  results in a minor performance loss in hit rate, and a reduction in the number of maximum recoverable targets. In the following we quantify the hardware complexity gains of \ac{frac} compared to the wideband benchmark, while the communication gains are evaluated in the following subsection. 
	
	%	Nonetheless, when the number of targets in the scenario is less than the maximum recoverable targets indicated by the phase transition threshold, \ac{frac} leads to significant reductions to hardware complexity, such as decreasing the  sampling rate, reducing the data volume, and utilizing less \ac{rf} modules, by transmitting narrowband waveforms from selected active antenna elements. Here, we compare the hardware complexities of \ac{frac} and the wideband radar which transmits a wideband waveform and using a large receive array. 
	
	The hardware complexity reduction of \ac{frac} compared to wideband   \ac{mimo} radar is expressed in its number of \ac{rf} modules, the required sample rate, and the volume of the data being processed at the radar receiver.
	To quantify the reduction in  \ac{rf} modules, we note that the \ac{dfrc} transmitter of \ac{frac} utilizes   $K$ \ac{rf} modules, while  $Q_{\rm r}$ \ac{rf} modules are used in reception. Thus the total number of \ac{rf} modules required in \ac{frac} is $K + Q_{\rm r}$. \ReviseReview{For comparison, a wideband \ac{mimo} radar uses an overall of $PQ_{\rm r}$ \ac{rf} modules  to achieve the same angle resolution. 
		Furthermore, to achieve the same radar resolutions as \ac{frac}, the waveform bandwidth of wideband \ac{mimo} radar is  $B_{\rm W} = MB_{\rm sub}$, and the number of elements in receive antenna is set to $PQ_{\rm r}$. Substituting $B_{\rm W}$ into \eqref{eqn:SampRateFMCW}, we note that the sampling rate of the wideband benchmark is $M$ times larger than that of \ac{frac}.\label{txt:SamplingRateSimu}}
	%Denote the sampling rate of \ac{frac} and the compared radar by $F_{s\rm N}^{{\left(\rm r\right)}}$ and $F_{s\rm W}^{\left(\rm r\right)}$, respectively. Therefore, for a determined maximum range $r_{\rm max}$, we obtain the relationship between the sampling rate of \ac{frac} and that of the compared radar system, which are given by  $F_{s\rm W}^{\left(\rm r\right)} = M F_{s\rm N}^{\left(\rm r\right)}$. 
	
	To quantify the data volume reduction,  we note that \ac{frac} receives $Q_{\rm r}$ signals, and uses them to generate $KQ_{\rm r}$ virtual channels. Assuming each virtual channel is acquired using an \ac{adc} with rate $F_{s}\Radar$, the number of samples gathered by the \ac{frac} receiver in each pulse  is $ KQ_{\rm r}T_{0}F_{s}\Radar$. For comparison, a wideband \ac{mimo} receiver acquires  $ P Q_{\rm r}T_{0}F_{s}\Radar$ samples per pulse. As the sampling rate of the wideband radar is $M$ times larger than that  of \ac{frac}, it acquires $PM/K$ more samples per pulse than \ac{frac} does.
	
	The hardware overhead of \ac{frac} using the parameters in TABLE~\ref{tab:ParameterValue} compared to wideband \ac{fmcw} \ac{mimo} radar is summarized in TABLE~\ref{tab:HardComplexity}. Observing this table, we find that there is a significant reduction to the hardware complexity by \ac{frac}, indicating its attractiveness to vehicular systems operating with tight constraints in size, power, and cost.
	\begin{table}[t]
		\caption{Comparison of Hardware Complexity} 
		\vspace{-0.3cm}
		\label{tab:HardComplexity}
		\small
		\begin{center}			
			\begin{tabular}{|c|c|c|c|}
				\hline\hline
				& \textbf{RF modules} & \textbf{Sampling rate}  & { \textbf{Data volume}} \\
				\hline
				\ac{frac} & 3 & 416.68 kHz & 42 Samples\\
				\hline
				Wideband  & 8 & 2.74 MHz & 1333 Samples\\
				\hline\hline		
			\end{tabular}  
		\end{center}
		\vspace{-0.26cm}
	\end{table}
	
	\vspace{-0.2cm}
	\subsection{Communications Subsystem Evaluation}
	\vspace{-0.1cm}
	\label{subsec:CommSimu}
	We next evaluate the communication capabilities of \ac{frac}. To that aim, we first compare it with the \ac{majorcom} and a similar \ac{dfrc} system which only exploits \ac{pm} for data conveying, utilizing the achievable rate and uncoded \ac{ber} as the performance measures.  Then, the performances of \ac{ml} and \ac{sod} decoders are compared. 
	In the experiments, the number of taps for the channel filter is set to $I  = 8$, and the channel filter taps $\{h_{p, q_{\rm c}}^i\}$ are modeled as i.i.d. zero-mean proper-complex Gaussian random variables with variance $e^{-{\left|i\right|}}$.
	The \ac{snr} is defined here as $\mathrm{SNR}^{\left(\rm c\right)}:= \frac{KQ_{\rm c}U}{\sigma_{\rm c}^{2}}\sum_{i}^{I-1}e^{-\left|i\right|}$. % Here, the achievable rate is calculated via empirical averaging over a large number of realizations of the communications channel and additive noises, which results in an extremely long simulation duration for large dimensional signals. 
	To reduce the computational burden, the total bandwidth of the \ac{dfrc} waveform is set to $B = 200$ kHz when simulating the achievable rate, whose computation involves exhaustive empirical averaging, while the total bandwidth is still set to $100$ MHz in the \ac{ber} evaluations as in Subsection~\ref{subsec:RadarSimu}. Unless stated otherwise, the remaining parameters are set according to TABLE~\ref{tab:ParameterValue}.  
	
	\color{black}
	\subsubsection{Achievable Rate}
	\label{txt:SchemeCompareComm}
	We numerically compare the achievable rates of \ac{frac}, \ac{majorcom} and the wideband \ac{fmcw} with \ac{pm}.  Following \cite{Younis2018Information}, the achievable rate is computed via 
	\vspace{-0.3cm}
	\begin{align} 
	&\!\!I\left(\myVec{e}_n;\myVec{y}_n\Comm\right) = -Q_{\rm c}\log_2 e \notag  \\
	&\!\!  - \mathbb{E}_{\myMat{\Psi}}\left\{\mathbb{E}_{\myVec{y}_n\Comm|\myMat{\Psi}}\left\{\log_2\left(\frac{1}{|\mySet{E}|}\mathop{\sum}\limits_{\myVec{e}_n \in \mySet{E}}e^{\frac{-\Vert\myVec{y}_n\Comm-\myMat{\Psi}\myVec{e}_n\Vert_{2}^{2}}{\sigma_{\rm c}^{2}}}\right)\right\}\right\}.
	\vspace{-0.2cm}
	\label{eqn:AchieavableRate} 
	\end{align}
	The stochastic expectations are computed via empirical averaging. Similarly, the achievable rates of \ac{majorcom} and the compared \ac{pm} scheme are obtained by substituting the corresponding input vector into \eqref{eqn:AchieavableRate}.
	
	To compare the achievable rates of \ac{frac} and the \ac{pm}-only based scheme, we keep the data rates of \ac{frac} and the compared \ac{pm} system to be the same, using $6$ bits/pulse and $7$ bits/pulse. The waveform parameters are configured as $\left\{M , P, Q_{\rm c}, K\right\} = \left\{8, 4, 4, 1\right\}$. As \ac{frac} uses \ac{im} to convey $N_{\rm IM}=\lfloor\log_2\binom{M}{K}\rfloor  + \lfloor  \log_2\binom{P}{K}\rfloor +  \lfloor \log_{2}K!\rfloor = 5$ bits per pulse, \ac{bpsk} and \ac{qpsk} are utilized to realize the data rates of $6$ bits/pulse and $7$ bits/pulse, respectively. The \ac{pm}-only based benchmark thus  utilizes the constellations of orders $64$  and $128$,  to convey the same amounts of $6$ bits/pulse and $7$ bits/pulse, respectively. 
	To compare the achievable rates of \ac{frac} and  \ac{majorcom} in a fair manner, the number of  \ac{rf} modules in the transmitter are set as the same in \ac{frac} and \ac{majorcom}. To meet this condition, we also simulate \ac{frac} with the waveform parameters of \ac{frac}  configured as $\left\{M, P, Q_{\rm c}, K, J\right\} = \left\{8, 4, 4, 2, 2\right\}$, while the waveform parameters of \ac{majorcom} are set as $\left\{M, P, Q_{\rm c}, K, P_K\right\} = \left\{8, 2, 4, 2, 1\right\}$. With this parameter setting, according  to TABLE~\ref{tab:CompareSchemes}, the data rates of \ac{frac} and \ac{majorcom} are $\lfloor\log_2\binom{M}{K}\rfloor  + \lfloor  \log_2\binom{P}{K}\rfloor +  \lfloor \log_{2}K!\rfloor + \log_2 J = 9$ bits/pulse and $\lfloor \log_{2}\binom{P}{K}\rfloor + \lfloor \log_{2} \frac{P!}{\left(P_K !\right)^{K}} \rfloor = 5$ bits/pulse, respectively.
	
\begin{figure}
		%    \vspace{-0.3cm}
		\centerline{\includegraphics[width= 0.8\columnwidth]{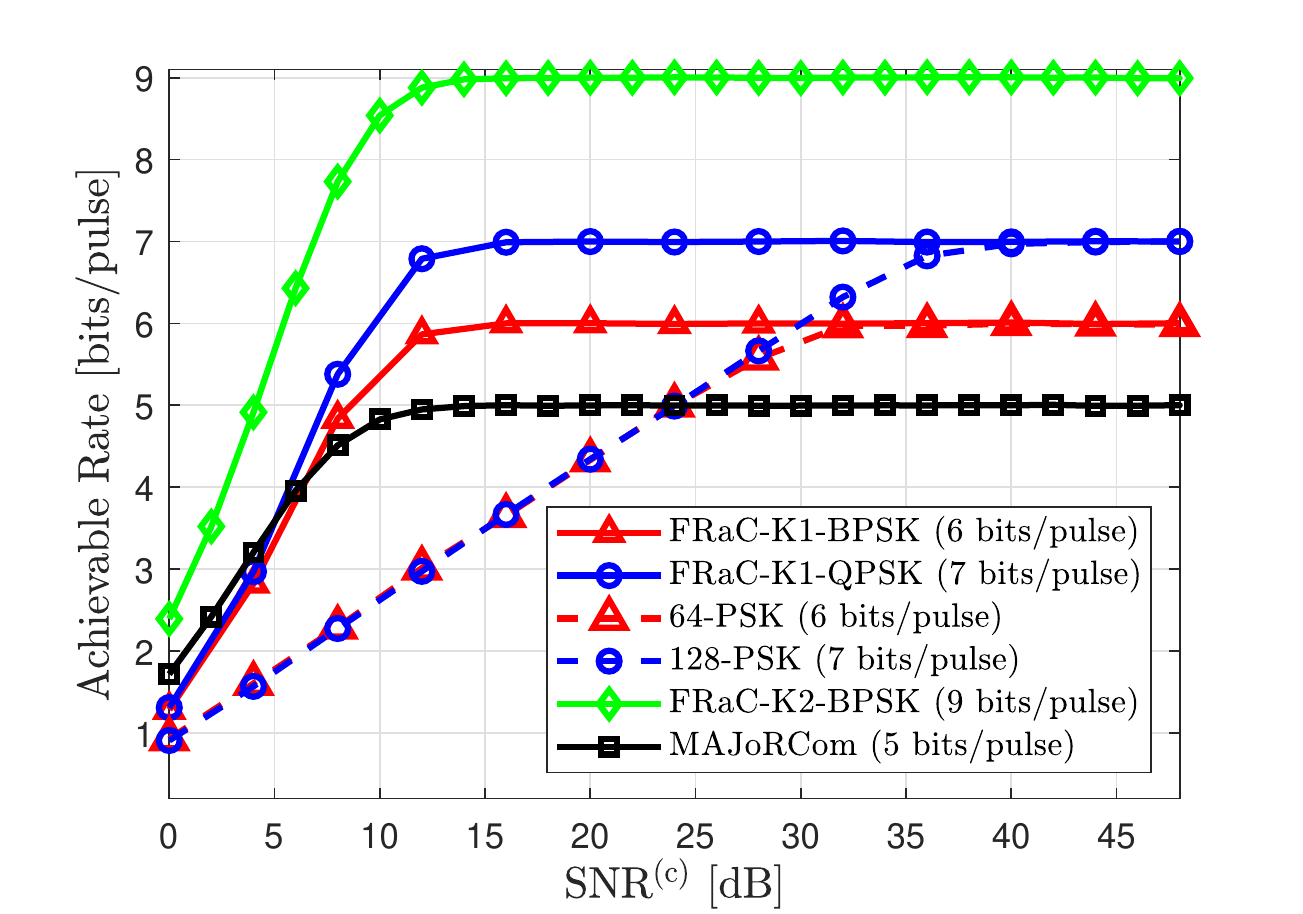}}
		\vspace{-0.2cm}
		\caption{Achievable rate comparison.} 
		\label{fig:AchievableRate1}
	\end{figure}

	The evaluated achievable rates are depicted in  Fig.~\ref{fig:AchievableRate1}. 
	Observing  Fig.~\ref{fig:AchievableRate1}, we note that, as expected, the achievable rate does not exceed the cardinality of  $\mySet{E}$, i.e., the maximal rates of FRaC-K1-BPSK and 64-BPSK are $6$ bits/pulse, while the maximal rates of FRaC-K1-QPSK and 128-BPSK are $7$ bits/pulse. However, these rates are only achieved at high \ac{snr} values. In low \acp{snr}, the achievable rates of \ac{frac} outperform that of the \ac{pm}  wideband \ac{fmcw}, indicating the improved spectral efficiency of combining \ac{im} with \ac{pm} in \ac{dfrc} signalling. We also observe in Fig.~\ref{fig:AchievableRate1}  that the rates achieved in FRaC-K2-BPSK are higher than that of the \ac{majorcom} in all considered \acp{snr}. As expected, in high \acp{snr}, the maximal achievable rates of FRaC-K2-BPSK and \ac{majorcom} approaches $9$ bits/pulse and $5$ bits/pulse, respectively. This is because more bits can be embedded in \ac{frac} by utilizing the sparse array and combining the \ac{pm}. These results indicate \ac{frac} can convey more information than \ac{majorcom} while utilizing the same number of \ac{rf} modules.
	
	\subsubsection{Bit Error Rate}	
\begin{figure}
		%    \vspace{-0.3cm}
		\centerline{\includegraphics[width= 0.8\columnwidth]{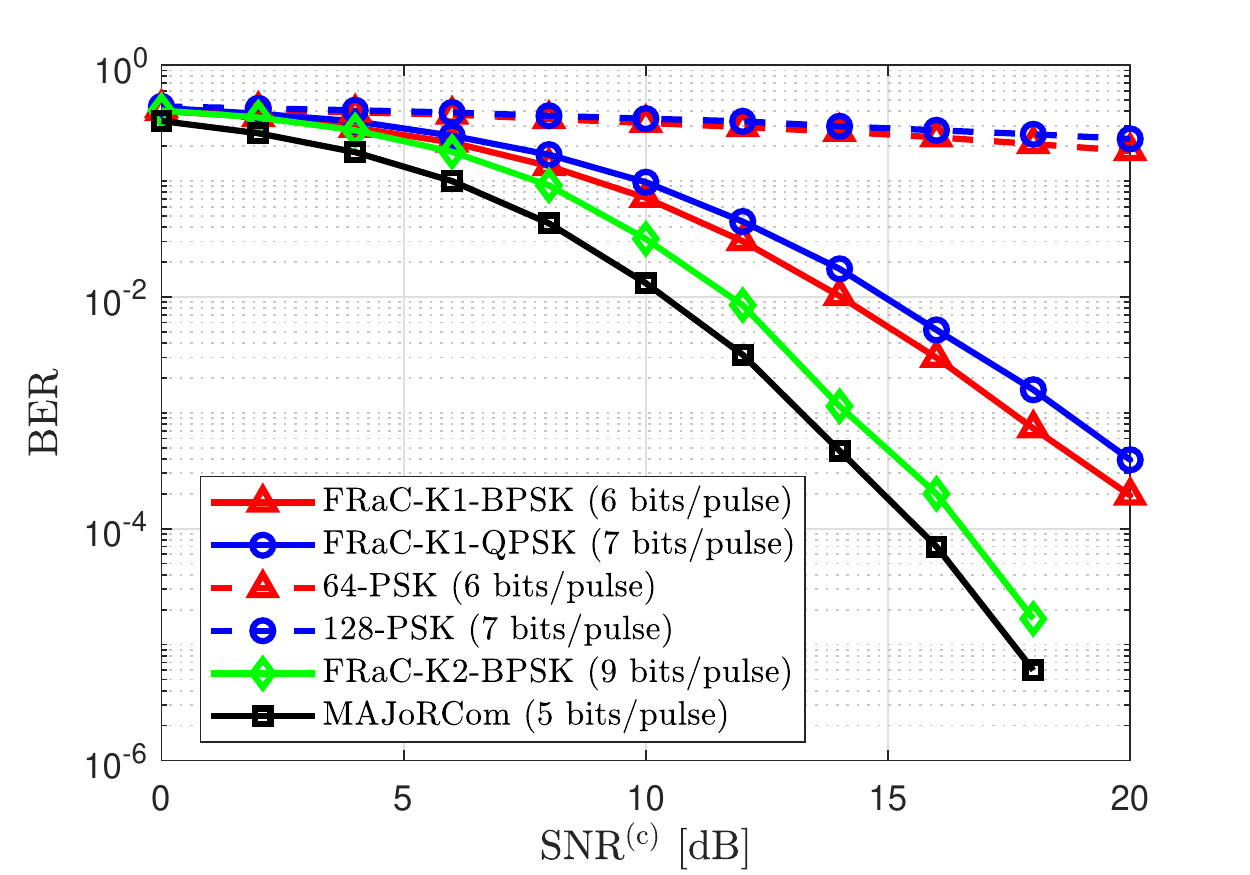}}
		\vspace{-0.2cm}
		\caption{\acs{ber} comparison.} 
		\label{fig:BER1}
	\end{figure}

	We next evaluate the uncoded \ac{ber} performance of the communications subsystem using the same setups in the achievable rate evaluations. In this experiment, a total of $10^5$ \ac{dfrc} waveforms are generated and decoded for each \ac{snr} value. 
	The numerically evaluated \ac{ber} results of \ac{frac}, the \ac{pm} scheme and \ac{majorcom} are shown in  Fig.~\ref{fig:BER1}. As in the achievable rate study,  we compare the \ac{ber} curves of FRaC-K1-BPSK, FRaC-K1-QPSK, and FRaC-K2-BPSK  with that  of 64-PSK, 128-PSK, and \ac{majorcom}, respectively. Observing  Fig.~\ref{fig:BER1}, we find that the \ac{ber} performances of \ac{frac} significantly outperform that of the \ac{pm} system. These improvements stem from the fact that \ac{frac} uses less dense \ac{pm} constellations, since it conveys additional bits through \ac{im}. Comparing the \ac{ber} curves of FRaC-K2-BPSK with \ac{majorcom}, we note that FRaC-K2-BPSK, which conveys almost twice the amount of bits as that of \ac{majorcom} for the considered setup, achieves \ac{ber} within an \ac{snr} gap of merely $1$dB, indicating on its ability to achieve higher data rates using coded transmissions. 
	
	\color{black}
	\subsubsection{Detector Comparison}
\begin{figure}
		\vspace{-0.3cm}
		\centerline{\includegraphics[width= 0.8\columnwidth]{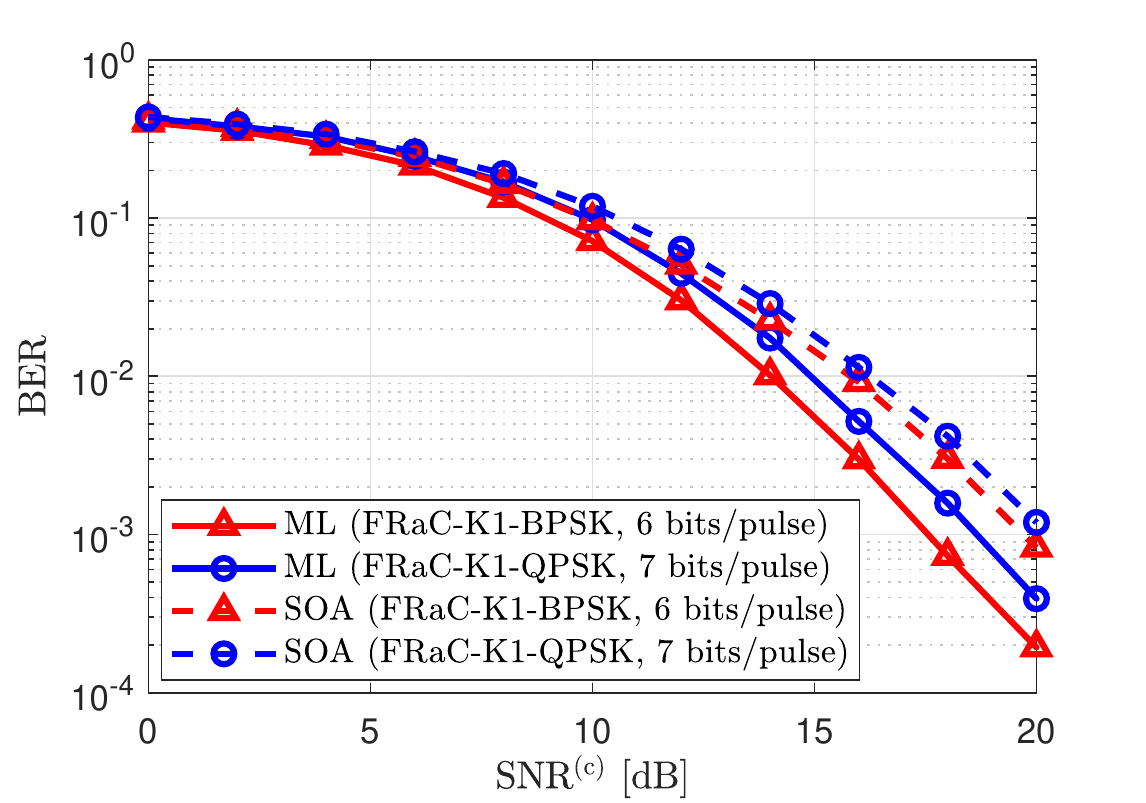}}
		\vspace{-0.2cm}
		\caption{BER comparison of the ML decoder and the SOA decoder.} 
		\label{fig:BER2}
	\end{figure}

	Finally, we compare the \ac{ber} of the \ac{ml} decoder and the reduced complexity \ac{sod} decoder. The \ac{ber} curves obtained utilizing different decoders are depicted in Fig.~\ref{fig:BER2}. In this experiment, the waveform parameters are set as the same with the parameters for  Fig.~\ref{fig:BER1}. Observing Fig.~\ref{fig:BER2},  we note that the complexity reduction of the \ac{sod} algorithm results in a performance loss of approximately $2 \mathrm{dB}$ in \ac{snr} compared to the \ac{ml} algorithm. To quantify the complexity of the decoders, recall that for the \ac{sod} decoder, the transmit sub-carriers are first detected, followed by searching over a subset corresponding to the detected sub-carriers.  Taking the setting $M = 8$ and $K = 1$ for example,   \ac{sod} involves searching over a set whose cardinality is ${2^{-\lfloor \log_{2}\binom{M}{K}\rfloor}} = \frac{1}{8}$  of that examined by the \ac{ml} algorithm. This indicates that the \ac{sod} algorithm allows to balance   complexity at the cost of a relatively minor performance loss compared to the \ac{ml} decoder. 
	\iffalse	
	We next simulate the performance of the \ac{sod} decoder in terms of the waveform parameters. Thus, the performances of the \ac{sod} decoder is directly influenced by the performances of the sub-carrier detection. In this experiment, the cardinality of the sub-carrier set is set to be $16$ and $32$  for a fixed total bandwidth of $160$ MHz. The other waveform parameters are set as $\left\{P, Q_{\rm c}, K\right\} = \left\{4, 4, 1\right\}$. The 
	
	\begin{figure}
		%    \vspace{-0.3cm}
		\centerline{\includegraphics[width= 0.75\columnwidth]{Fig1/CER.eps}}
		%	\vspace{-0.4cm}
		\caption{sub-carrier error rate versus the number of sub-carriers $M$ using \ac{sod} algorithm.} 
		\label{fig:CER}
	\end{figure}
	\fi		
	%----------------------------------------------------------------------------------------
	%	CONCLUSIONS
	%----------------------------------------------------------------------------------------
	\vspace{-0.1cm}
	\section{Conclusions}
	\vspace{-0.1cm}
	\label{sec:Conclusions}
	%\ac{dfrc} design is an attracting strategy for future self-driving cars which leads to substantial gains in size, cost, power consumption, and performance. 
	In this work we proposed \ac{frac}, which is a \ac{dfrc} system based on \ac{fmcw} signaling with \ac{im}, utilizing sparse arrays and narrowband waveforms. \ac{frac} conveys its message in the combinations of carrier selection, antenna selection, waveform permutation, and \ac{pm}. We presented the signal models and detection algorithms for both radar and communications subsystems. For the radar subsystem, we analyzed the ambiguity function, showing that the resolution of \ac{frac} is similar to that of a wideband \ac{mimo} radar. We also characterized the relationship between the maximum number of recoverable targets and the waveform parameters via phase transition theory, and numerically demonstrated that the theoretical analysis matches the empirical performance. Furthermore, we numerically showed that the achievable rate and \ac{ber} performances of \ac{frac} outperform a system which only exploits \ac{pm}.

	%----------------------------------------------------------------------------------------
	%	APPENDICES
	%----------------------------------------------------------------------------------------
	\numberwithin{proposition}{subsection} 
	\numberwithin{lemma}{subsection} 
	\numberwithin{corollary}{subsection} 
	\numberwithin{remark}{subsection} 
	\numberwithin{equation}{subsection} 
	
	\vspace{-0.3cm}
	\begin{appendix}	
		\vspace{-0.15cm}	
		%----------------------------------------------------------------------------------------
		%	APPENDIX A
		%----------------------------------------------------------------------------------------
		\subsection{Proof of Theorem \ref{thm:ExpctatedAmbFun}}
		\label{app:Proof1}
		\vspace{-0.1cm}
		For a given radar pulse index $n$, the indices of the active antenna elements, denoted by  ${\mathbf{P}}_n$, and the carrier frequencies, denoted by ${\mathbf{M}}_n$, are randomized from the radar antenna combination set and the sub-carrier combination set, respectively. The random vector ${\mathbf{P}}_n$ thus obeys a discrete uniform distribution following $\Pr\left({\mathbf{P}}_n \right)=1/\Omega_{p}$, where $\Omega_{p} := \binom{P}{K}$ is the total number of possible antenna index combinations. Similarly, the random vector ${\mathbf{M}}_n$ obeys a discrete uniform distribution following $\Pr\left({\mathbf{M}}_n \right)=1/\Omega_{m}$, where $\Omega_{m} := \binom{M}{K}$ is the total number of possible sub-carrier index combinations.
		The expected ambiguity function can be calculated as follows: 
		\begin{align}
		&\mathbb{E}\big\{\chi\left(\delta \! f_r, \delta \! f_v, \delta \! f_\theta\right) \big\} = \sum_{n=0}^{N-1}\sum_{q=0}^{Q_{\rm r}-1} e^{- j2\pi n \delta \! f_v} \cdot e^{ - j2\pi q_{\rm r} \delta \! f_\theta} \notag \\ 
		& \qquad \qquad \times \mathbb{E}\big\{\sum_{k=0}^{K-1} e^{-j2\pi m_{n,k} \delta \! f_r}\cdot e^{ - j2\pi Q_{\rm r}p_{n,k} \delta \! f_\theta}\big\}.	
		\label{eqn:proofexpectation1}		
		\end{align}
		
		The expected value in \eqref{eqn:proofexpectation1} is
		\begin{align}
		& \mathbb{E}\big\{\sum_{k=0}^{K-1} e^{-j2\pi m_{n,k} \delta \! f_r}\cdot e^{ - j2\pi Q_{\rm r}p_{n,k} \delta \! f_\theta}\big\} \notag \\
		& \stackrel{(a)}{=} \frac{1}{\Omega_{m} \Omega_{p}} \sum_{i_1 = 0}^{\Omega_{m} - 1} \sum_{i_2 = 0}^{\Omega_{p} - 1} \sum_{k = 0}^{K-1 } e^{-j 2 \pi m_{n,k}^{\left(i_1\right)}\delta \! f_r} \cdot e^{-j 2 \pi Q_{\rm r} p_{n,k}^{\left(i_2\right)}\delta \! f_\theta} \notag \\
		& \stackrel{(b)}{=} \frac{1}{\Omega_{m} \Omega_{p}} \frac{K\Omega_{m} \Omega_{p}}{MP} \sum_{m=0}^{M-1}\sum_{p = 0}^{P-1} e^{-j2\pi m \delta \! f_r} e^{-j2\pi Q_{\rm r} p \delta \! f_\theta}, 
		\label{eqn:proofexpectation2}	
		\end{align}
		where $(a)$ follows since $\mathbf{M}_n$ and $\mathbf{P}_n$ are uniformly distributed, and $(b)$ holds  as there are $K\Omega_{m}\Omega_{p}$ summands, where each index in the  summation over $m$ and $n$ occurs $K\Omega_{m}\Omega_{p}/\left(MP\right)$ times. Substituting \eqref{eqn:proofexpectation2} into \eqref{eqn:proofexpectation1}, we obtain that 
		$\mathbb{E}\big\{\chi\left(\delta \! f_r, \delta \! f_v, \delta \! f_\theta\right) \big\}  = \frac{K}{MP}\sum_{m=0}^{M-1}e^{-j2\pi m \delta \! f_r} \sum_{n=0}^{N-1}e^{-j2\pi n \delta \! f_v} \sum_{q' = 0}^{PQ_{\rm r} -1 }e^{-j2\pi q' \delta \! f_\theta}$,
		and thus
		\begin{align}
		& \mathbb{E}\big\{\chi\left(\delta \! f_r, \delta \! f_v, \delta \! f_\theta\right) \big\} = \frac{K}{MP} e^{-j \pi \left(\delta \! f_r + \delta \! f_v + \delta \! f_\theta\right)}, 
		\notag \\
		& \quad \quad \times \frac{\sin\left(M\pi \delta \! f_r\right)}{\sin\left(\pi \delta \! f_r\right)} \cdot \frac{\sin\left(N\pi \delta \! f_v\right)}{\sin\left(\pi \delta \! f_v\right)}\cdot \frac{\sin\left(PQ_{\rm r}\pi \delta \! f_\theta\right)}{\sin\left(\pi \delta \! f_\theta\right)}.
		\label{eqn:proofexpectation3}
		\end{align}
		Taking the absolute value of \eqref{eqn:proofexpectation3} proves \eqref{eqn:ExpctatedAmbFun}. \qed 

		\subsection{Proof of Corollary \ref{cor:PhaseTransition1}}
		\label{app:Proof4}
		When $\frac{NMPQ_{\rm r}}{L^{\ast}}\gg 1$,  $\beta_{\ast}$ is sufficiently larger than one\cite{Li2021Phase}, thus $\frac{\beta_{\ast}^2}{2} \gg \ln \left(\beta_{\ast}^2 + 1\right)$. Under these conditions, we ignore the terms of $\ln\left(\beta_{\ast}^2 + 1\right)$ and $L^{\ast}$ in \eqref{eqn:betaast2}, which yields  
		\begin{align}
		\beta_{\ast}^2 = \mathcal{O}\left(\ln\left({NMPQ_{\rm r}}\right)\right). 
		\label{eqn:proofCorollary1_1}
		\end{align}	
		According to \eqref{pro:phasethreshold}, $L^{\ast}$ is approximated by  
		\begin{align}
		L^{\ast} \approx \frac{NKQ_{\rm r}}{2 + \beta_{\ast}^2/2}.
		\label{eqn:eqn:proofCorollary1_2}
		\end{align}
		Substituting \eqref{eqn:proofCorollary1_1} into \eqref{eqn:eqn:proofCorollary1_2} yields $L^{\ast}=\mathcal{O}\left(\frac{NKQ_{\rm r}}{\ln \left(NMPQ_{\rm r}\right)}\right)$, completing the proof.  \qed 
	\end{appendix}
	
	%----------------------------------------------------------------------------------------
	%	BIBLIOGRAPHY
	%----------------------------------------------------------------------------------------
	\begin{spacing}{1}
		\bibliographystyle{IEEEtran}
		\bibliography{IEEEabrv,references1} 
	\end{spacing}
\end{document}